\documentclass[journal,onecolumn]{IEEEtran}
\IEEEoverridecommandlockouts \makeatletter
\def\ps@headings{%
\def\@oddhead{\mbox{}\scriptsize\rightmark \hfil \thepage}%
\def\@evenhead{\scriptsize\thepage \hfil \leftmark\mbox{}}%
\def\@oddfoot{}%
\def\@evenfoot{}}

\makeatother
\usepackage{xr-hyper} 

\usepackage{algorithm}
\usepackage{algpseudocode}

\usepackage{longtable}
\usepackage{nicematrix}
\usepackage[english]{babel}
\usepackage{fontenc}
\usepackage{placeins}
\usepackage{tikz}
\usetikzlibrary{patterns}
\usepackage{amssymb}
\usepackage{float}
\usepackage{booktabs}
\usepackage{amsmath,amsthm}
\usepackage{cleveref}
\usepackage{colortbl}
\usepackage{makecell}
\usepackage{hhline}
\usepackage{color}
\usepackage{caption}
\usepackage{subcaption}
\usepackage{diagbox}
\usetikzlibrary{matrix}
\usepackage{threeparttable}
\usepackage{array}
\usepackage{arydshln}
\usepackage{pgfplots}
\usepackage[margin=2cm]{geometry}
\usepackage{booktabs}
\usepackage{array}
\usepackage{adjustbox}
\usepackage{makecell}
\usepackage{cuted}
\usepackage{stfloats}

\renewcommand{\arraystretch}{1.2} 

\usepackage{multirow}
\usepackage{bm}
\usepackage{soul}
\usepackage[normalem]{ulem}
\usepackage{ifthen}

\usepackage{afterpage}
\newcommand{\mr}{\mathrm}
\newcommand{\mb}{\mathbf}
\newcommand{\mc}{\mathcal}

\newcommand{\h}{\mb{h}}

\newcommand{\showrev}{0} 
\newcommand{\rv}[1]{%
  \ifnum\showrev=1
    \hl{#1}%
  \else
    #1%
  \fi
}
\newcommand{\rem}[1]{%
	\ifthenelse{\equal{\showrev}{1}}%
	{\textcolor{red}{\sout{#1}}}
	{}%
}
\newcommand{\mrv}[1]{%
	\ifthenelse{\equal{\showrev}{1}}%
	{\colorbox{yellow}{$\displaystyle #1$}}
	{#1}%
}
\newcommand{\br}{\mb{r}}

\newcommand{\pref}[2]{Part #1~Thm.\ref{#2}}

\newcommand{\SetR}{\mathbb{R}}
\newcommand{\SetZ}{\mathbb{Z}}

\newcommand{\oa}{\mr{OA}}
\newcommand{\voa}{\mr{VOA}}
\newcommand{\mvoa}{\mr{MVOA}}
\makeatletter

\newcommand{\Rmnum}[1]{\expandafter\@slowromancap\romannumeral #1@}
\makeatother
\newtheorem{theorem}{Theorem}[subsection]
\newtheorem{lemma}{Lemma}

\newtheorem{definition}{Definition}
\newtheorem{example}{Example}

\newcommand{\tridiagbox}[3]{%
 \begin{tikzpicture}[baseline=(current bounding box.center)]
  \draw (0,0.7) -- (2,0);        
  \draw (1.1,1) -- (2,0);        
  \node at (0.4,0.2) {#1};     
  \node at (0.8,0.7) {#2};    
  \node at (1.7,0.8) {#3};    
 \end{tikzpicture}%
}

\title{ Entropy Functions on Two-Dimensional
	Faces of Polymatroidal Region of Degree Four: Part \Rmnum{2}: Information Theoretic Constraints Breed New Combinatorial Structures}
\author{Shaocheng Liu$^{*}$, Qi Chen$^{*}$, and Minquan Cheng$^{\dagger}$\\$^{*}$Xidian University, Xi'an 710071, China,\\
lsc@stu.xidian.edu.cn, qichen@xidian.edu.cn 
\\$^{\dagger}$Guangxi Normal University, chengqinshi@hotmail.com}

\makeatletter

\@addtoreset{equation}{subsection}
\makeatother

\begin{document}

\maketitle

\begin{abstract}
The characterization of entropy functions is of fundamental importance in
information theory. By imposing constraints on their
Shannon outer bound, i.e., the polymatroidal region, one obtains the faces
of the region and entropy functions on them with special
structures. In this series of two papers, we characterize entropy functions on the $2$-dimensional
faces of the polymatroidal region $\Gamma_4$. 
In Part I, we formulated the problem, enumerated all $59$ types
of $2$-dimensional faces of $\Gamma_4$ by an algorithm, and fully characterized entropy
functions on $49$ types of them. In this paper, i.e., Part II, we will characterize
entropy functions on the remaining $10$ types of
faces, among which $8$ types are
fully characterized, and $2$ types are partially characterized. To
characterize these types of faces, we introduce some new combinatorial
design structures that are interesting \rv{in themselves}.

Index terms-entropy function, polymatroid, information inequalities,
orthogonal array, multi-level variable-strength orthogonal array
\end{abstract}

\section{Introduction}
\label{int}

Let $N_n=\{1,2,\ldots,n\}$ and $\mb{X}\triangleq(X_i,i\in N_n)$ be a random vector indexed by
$N_n$. The set function $\h: 2^{N_n}\to \SetR$ defined by
\begin{equation*}
  \h(A)=H(X_A), \quad A\subseteq N_n
\end{equation*}
is called the \emph{entropy function} of $\mb{X}$, while $\mb{X}$ is called a \emph{characterizing random vector} of $\h$.
The Euclidean space $\mc{H}_n\triangleq\SetR^{2^{N_n}}$ where entropy functions live is called the
\emph{entropy space} of degree $n$.
The set of all entropy functions, denoted by $\Gamma^*_n$, is called
the \emph{entropy region} of degree $n$. 
The characterization of entropy functions, i.e., determining whether an
$\h\in \mc{H}_n$ is in $\Gamma^*_n$, is of fundamental importance in information theory.

Entropy functions are (the rank functions of) polymatroids, i.e., they satisfy polymatroidal
axioms, that is, for all $A,B\subseteq N_n$,
\begin{align}
  \h(A) &\ge 0,   \label{a1}\tag{I.1}\\
  \h(A)&\le \h(B),\quad \text{ if }  A\subseteq B,\tag{I.2}\\
  \h(A)+\h(B)&\ge \h(A\cap B)+\h(A\cup B).   \label{a3}\tag{I.3}
\end{align}
The region in $\mc{H}_n$ bounded by such inequalities, denoted by $\Gamma_n$ is called the
\emph{polymatroidal region} of degree $n$. Thus, $\Gamma_n$ is an
outer bound on $\Gamma^*_n$. For more about entropy functions, we
referred the readers to \cite[Chapter 13-15]{yeung2008information}.


Traditionally, entropy functions are characterized by information
inequalities. Those inequalities derived by polymatroidal axioms are called Shannon-type,
as they correspond to the non-negativity of Shannon information
measures. 
 Since 1998, a series of non-Shannon-type information
inequalities, among which Zhang-Yeung inequality is the first one \cite{ZY98},
were discovered \cite{zhang2002new}\cite{makarychev2002new}\cite{dougherty2011nonshannon}. Thus $\overline{\Gamma^*_n}$, the closure of $\Gamma^*_n$, is strictly included in $\Gamma_n$ when $n\ge 4$.  
In 2007, Mat{\'u}{\v{s}} showed that for $n\ge 4$, there exist infinitely non-redundant non-Shannon-type inequalities, indicating that $\overline{\Gamma_n^*}$ is not polyhedral \cite{matusinfinite}.
Each information inequality determines an outer bound on $\Gamma^*_n$, as those set functions in $\mc{H}_n$ that dissatisfy it must be located outside $\Gamma^*_n$. In this series of two papers, we develop a system of entropy function characterization from the perspective of faces of $\Gamma_n$, which covers the traditional inequality characterization.

By definition, $\Gamma_n$ is a polyhedral cone in $\mc{H}_n$.
Thus, each Shannon-type information inequality determines a face $F$ of
$\Gamma_n$. It is natural to
characterize entropy functions on the specific $F$ of
$\Gamma_n$ (See Subsection \ref{fpc} for details on the faces of a
polyhedral cone). Let $F^*\triangleq F\cap \Gamma^*_n$ be the set of all entropy
functions in $F$. In the following, to determine the entropy
functions on $F$, or the region $F^*$, we will call it characterize $F$ for
short. 
A non-Shannon-type information inequality 
can be considered as an outer bound on $F^*$ when
$F=\Gamma_n$ itself, the improper face of $\Gamma_n$. 
A constrained non-Shannon-type information inequality is an outer bound on $F^*$
when $F$ is the face determined by the constraints that are equalities
obtained by setting some Shannon-type inequalities as equalities. When $F$
is an extreme ray, i.e., a $1$-dimensional face of $\Gamma_n$, if it
contains a matroid, entropy functions on $F$ are called matroidal
entropy functions, and they can be fully characterized by the
probabilistically characteristic set of the matroid \cite{CCB21}\cite{CCB22}\cite{CCB24}. Mat{\'u}{\v{s}}
fully characterized the first non-trivial 2-dimensional face of $\Gamma_n$
in 2006 \cite{matus2005piecewise}. It is a $2$-dimensional face of $\Gamma_3$. In
2012, Chen and Yeung  characterized another $2$-dimensional face of $\Gamma_3$ \cite{chen2012characterizing}. They are the
only two types of non-trivial faces of $\Gamma_3$ that  need to be
characterized. To the best of our knowledge, so far, there is no fully characterized
no-trivial $3$-dimensional faces. However,
partial characterizations of $3$-dimensional faces of $\Gamma_3$ can
be found in \cite{HCG12}\cite{11003168}.

Many information-theoretic problems can be considered  as applications
of entropy function characterizations on faces of $\Gamma_n$. In a
series of three papers \cite{matuvs1995a, matuvs1995b, matu1999conditional},
Mat{\'u}{\v{s}} and Studne{\'y} solved the probabilistic conditional independence problem
for four random variables. Note each class of conditional independence
constraints, which is called a semimatroid in their papers, 
determines a face $F$ of $\Gamma_n$.  
The solution to this problem, i.e., the
probabilistic-representability of a semimatroid 
determines whether the relative interior of the corresponding face $F$ intersects with
$\Gamma^*_n$.
This problem thus can be considered as partial characterizations of
the faces of $\Gamma_n$. 
In \cite{yan2012implicit},
Yan, Yeung, and Zhang proved a formula involving $\Gamma_n^*$ for the capacity of multi-source
multi-sink network coding. Those constraints in the formula induced
by network topology and source independency form a face of $\Gamma_n$,
which shows that this holy-grail network coding problem corresponds to
the entropy function characterizations on this face. For the secret-sharing problem, see \cite{beimel2011secret} for example, the perfect
secrecy and decoding correctness conditions of an access structure
determine a face $F$ on
$\Gamma_n$, and the information ratio of the secret-sharing problem
can be considered as an optimization problem whose feasible region
is $F\cap \Gamma^*_n$. Other problems, such as distributed data storage
\cite{tian2014characterizing}, coded caching \cite{tian2018symmetry},
Markov random fields \cite{yeung2018information}, and relational
database \cite{Dan2023} are also related to the
entropy regions on the faces of $\Gamma_n$.

Though the information theory problems discussed above usually involve more
  than four random variables, and the corresponding faces are of
  dimension higher than $2$, following the characterizations of
  extreme rays. i.e., $1$-dimensional faces containing matroid
\cite{CCB21}\cite{CCB22}\cite{CCB24}, and
$2$-dimensional faces of $\Gamma_3$
\cite{matus2005piecewise}\cite{chen2012characterizing},
in this series of two papers, we characterize entropy functions on the $2$-dimensional
faces of $\Gamma_4$, which may serve as stepping stones to the
  general cases of this problem. In Part I \cite{LCC2024part1}, we enumerated all $59$ types $2$-dimensional faces
of $\Gamma_4$ by an algorithm and completely characterized $49$ types of
them. In this part, we characterize the remaining $10$ types of faces,
among which $8$ types are
fully characterized and $2$ types are partially characterized. To
characterize these types of faces, we adopt two sets of combinatorial structures, that is, mixed-level variable-strength
  orthogonal arrays and orthogonal Latin hypercubes, and the
  characterizations breed 
some new combinatorial
design structures that are interesting in themselves.

The rest of this paper is organized as follows. In Section \ref{pre},
for self-contained, we give the preliminaries on integer polymatroids
and matroids, polyhedral cones, mixed-level variable-strength
orthogonal arrays and their relationship to orthogonal Latin hypercubes. We then list the
results of the characterization of extreme rays of $\Gamma_4$, which has
been done in Part I, without proofs. 
In Section \ref{af}, we complete the characterization of the remaining
$10$ types of $2$-dimensional 
faces of $\Gamma_4$, and the results are
summarized in Table \ref{table:2} together with those in Part I.
The correspondences between the faces and the theorems
characterizing them in two papers are summarized in
Table \ref{table:1}.

\section{Preliminaries}
\label{pre}
\subsection{Integer polymatroids and matroids}
For a polymatroid $\h\in \Gamma_n$, if $\h(A)\in \SetZ$ for all
$A\subseteq N_n$, $\h$ is called \emph{integer}.
An ordered pair $M=(N_n,\br)$ is called a \emph{matroid} with \emph{rank function}
$\br$ if $\br$ is an integer polymatroid with $\br(A)\le |A|$ for all
$A\subseteq N_n$. Like polymatroids, in this paper, we do not distinguish a matroid and
its rank function unless otherwise specified. 

A \emph{uniform matroid} $U_{k,n}$ is a matroid with rank function
$\br(A)=\min\{k,|A|\}$ for all $A\subseteq N_n$.

For a matroid $M$ and
$e\in N_n$, if $\br(e)=0$, $e$ is called a \emph{loop} of $M$. For $e, e'\in
N_n$, if $\br(\{e,e'\})=1$, then $e$ and $e'$ are called
\emph{parallel}.

For more
about matroid theory, readers
are referred to \cite{oxley2006matroid}.

\subsection{Faces of a polyhedral cone}
\label{fpc}
Let $C\subseteq \SetR^d$ be a full-dimensional polyhedral cone. For a
hyperplane $P$ containing the origin $O$ in $\SetR^d$, if $C\subseteq P^+$, where
$P^+$ is one of the two halfspaces corresponding to $P$,
$F\triangleq C\cap P$ is called a (proper) face of $C$, while $C$
itself is its improper face. When $\dim F=d-1$, $F$ is called a
\emph{facet} of $C$, and when $\dim F=1$, $F$ is an extreme ray of
$C$. Either the set of all facets or the set of all extreme rays of $C$
uniquely determines the cone, and they are called the H-representation and
$V$-representation of the cone, respectively. For each face $F$ of the
cone, it can be written as the intersection of the facets of the cone
that contain $F$, 
or the convex hull of the extreme rays contained in $F$. We also call
them the H-representation and V-representation of $F$,
respectively. More details on this topic are referred to \cite{ziegler2012lectures}.

As we discussed in Section \ref{int}, $\Gamma_n$ is a polyhedral cone
in $\mc{H}_n$ determined by polymatroidal axioms in \eqref{a1}-\eqref{a3}.
They are equivalent to the following 
elemental inequalities
\begin{align}
  \label{eq:1}
 &\h(N_n)\ge \h({N_n\setminus \{i\}})\quad\quad  i\in N_n;\\
  &\h(K)+\h(K\cup ij)\le \h(K\cup i)+\h(K\cup j) ,  \label{eq:1a}\\
                         & \quad\quad i,j\in N_n, K\subseteq N_n\setminus\{i,j\}\nonumber
\end{align}
each of which determines a facet of $\Gamma_n$\cite[Chapter 14]{yeung2008information}.  
When $n=4$, it can be checked that there are $28$ elemental
inequalities, or $28$ facets of $\Gamma_4$.

It can be seen in \cite{matuvs1994extreme} that 
there are $41$ extreme
rays of $\Gamma_4$. Note that each extreme ray $E$ of $\Gamma_4$ can be written in the form
\begin{equation}
\label{eray}
  E=\{a\br: a\ge 0\}
\end{equation}
where $\br$ is the minimal integer polymatroid in the ray , that is,
   an integer polymatroid such that $\br/t$ is not integer for any
   integer $t>1$. Therefore, in
this paper, when we say a minimal integer polymatroid, we
mean the extreme ray containing it unless otherwise specified. Note
that some of these integer polymatroids are matroids.
The
$41$ extreme rays can be classified into the following $11$
types.
\begin{itemize}
\item $U^{i}_{1,1}$, $i\in N_4$;
\item $U^\alpha_{1,2}$, $\alpha\subseteq N_4$, $|\alpha|=2$;
\item $U^\alpha_{1,3}$, $\alpha\subseteq N_4$, $|\alpha|=3$;
\item $U^\alpha_{2,3}$, $\alpha\subseteq N_4$, $|\alpha|=3$;
\end{itemize}
for $U^\alpha_{k,m}$ with $\alpha\subseteq N_4$ and $|\alpha|=m$, we mean a
matroid on $N_4$ whose restriction on $\alpha$ is a $U_{k,m}$ and
$e\in N_4\setminus \alpha$ being loops; 
\begin{itemize}
\item $U_{1,4}$;
\item $\mc{W}^\alpha_2$, $\alpha\subseteq N_4$, $|\alpha|=2$;
\end{itemize}
for $\mc{W}^\alpha_2$ with
$\alpha\subseteq N_4$, $|\alpha|=2$, it is called a wheel matroid with
order $2$,\footnote{We adopt the notation and terminology in
  \cite[Section 8.4]{oxley2006matroid}} and it is a matroid with two parallel elements in $\alpha$, and
each element in $\alpha$ and the other two elements in $N_4$ form a
$U_{2,3}$;
\begin{itemize}
\item $U_{2,4}$;
\item $U_{3,4}$;
\item $\hat{U}^i_{2,5}$, $i\in N_4$;
   \end{itemize}
for $\hat{U}^i_{2,5}$ with $i\in N_4$, it is a polymatroid whose free
expansion is a $U_{2,5}$,\footnote{See \cite[Theorem
  1.3.6]{nguyen1978semimodular} and \cite[Theorem 4]{matuvs2007}  for
  the definition of free expansion.} and its rank function is defined by 
\begin{equation*}
  \br(A)=
  \begin{cases}
    \min \{2, |A|\},\quad &A\neq \{i\}\\
    2,\quad & A=\{i\}
  \end{cases}
\end{equation*}
for any $A\subseteq N_4$, 
\begin{itemize}
    \item $\hat{U}^i_{3,5}$;
\end{itemize}
for $\hat{U}^i_{3,5}$ with $i\in N_4$, it is a polymatroid whose free
expansion is a $U_{3,5}$, and its rank function is defined by 
\begin{equation*}
  \br(A)=
  \begin{cases}
    \min \{3, |A|+1\},\quad & i\in A,\\
    |A|,\quad & i\not\in A,
  \end{cases}
\end{equation*}
for any $A\subseteq N_4$; 
\begin{itemize}
  \item $V^\alpha_8$, $\alpha\subseteq N_4$, $|\alpha|=2$;
  \end{itemize}
for $V^\alpha_8$ with $\alpha\subseteq N_4$ and $|\alpha|=2$, it is a polymatroid whose free
expansion is the V\'amos matroid, and its rank function is defined by 
\begin{equation*}
  \br(A)=
  \begin{cases}
   3,\quad & |A|=2 \text{ and } A\neq \alpha, \\
    \min \{4, 2|A|\},\quad & \text{ o.w.}
  \end{cases}
\end{equation*}

It can be seen that for an extreme ray in the form $E^\alpha$ with
$\alpha\subseteq N_4$, it is in a type with $\binom{4}{|\alpha|}$
extreme rays, and each extreme ray in the type can be obtained from
each other by permuting the indices in $N_4$. 
For a specific extreme ray, say $U^{\{1,2\}}_{1,2}$, for
simplicity, we will drop the bracket and comma of the set in the
superscript, and write it as $U^{12}_{1,2}$. To facilitate the readers,
 the $11$ types of extreme rays are presented in Table
\ref{tablevector} by the rank functions of their minimal
  integer polymatroid in the form of $15$-dimensional vectors, i.e.,
  $\br=(\br(A),\emptyset\neq A\subseteq N_4)$.

In the same manner, we
denote the $8$ extreme rays of $\Gamma_3$ and classify them in the following $4$
types.
\begin{itemize}
\item $U^{i;3}_{1,1}$, $i\in N_3$;
\item $U^{\alpha;3}_{1,2}$, $\alpha\subseteq N_3$, $|\alpha|=2$;
\item $U_{1,3}$;
\item $U_{2,3}$.
\end{itemize}
Here we put second superscript $3$ to $U^{i;3}_{1,1}$ and
$U^{\alpha;3}_{1,1}$ to distinguish them from $U^{i}_{1,1}$ and
$U^{\alpha}_{1,1}$, the extreme rays of $\Gamma_4$, respectively.

\begin{table*}[h]
	\centering
	\caption{Extreme rays of $\Gamma_4$ and their rank functions}
	\renewcommand{\arraystretch}{2}
	{\setlength{\tabcolsep}{4pt} 
	     \begin{tabular}{@{}c@{}|cccc|cccccc|cccc|c}
			\hline \tridiagbox{$E_{M}$}{$\mb{r}_M(A)$}{$A$} & $1$ &$2$& $3$& $4$
			& $12$& $13$& $14$ &$23$& $24$ &$34$ &$123$ &$124$& $134$& $234$& $1234$  \\
			\hline\hline  $U^{1}_{1,1}$& $1$& $0$ &$0$ &$0$ & $1$& $1$ &$1$ &$0$& $0$& $0$& $1$& $1$& $1$& $1$&$1$\\
			\hline $U^{12}_{1,2}$& $1$& $1$ &$0$ &$0$ & $1$& $1$ &$1$ &$1$& $1$& $0$&$1$& $1$& $1$& $1$&$1$\\
			\hline $U^{123}_{1,3}$& $1$& $1$ &$1$ &$0$ & $1$& $1$ &$1$ &$1$& $1$& $1$& $1$&$ 1$& $1$& $1$&$1$\\
			\hline $U_{1,4}$& $1$& $1$ &$1$ &$1$ &$ 1$& $1$ &$1$ &$1$& $1$& $1$& $1$& $1$& $1$& $1$&$1$\\
			\hline $U^{123}_{2,3}$& $1$& $1$ &$1$ &$0$ & $2$& $2$ &$1 $&$2$& $1$& $1$& $2$& $2$& $2$& $2$&$2$\\
			\hline $\mc{W}^{12}_2$& $1$& $1$ &$1$ &$1$ & $1$& $2$ &$2$ &$2$&$ 2$& $2$& $2$& $2$&$ 2$& $2$&$2$\\
			\hline $U_{2,4}$& $1$& $1$ &$1$ &$1$ & $2$& $2$ &$2$ &$2$& $2$&$ 2$& $2$&$ 2$& $2$& $2$&$2$\\
			\hline $U_{3,4}$& $1$& $1$ &$1$ &$1$ & $2$& $2$ &$2$ &$2$& $2$&$ 2$& $3$&$ 3$& $3$& $3$&$3$\\
			\hline $\hat{U}^1_{2,5}$& $2$& $1$ &$1$ &$1$ & $2$& $2$ &$2$ &$2$& $2$& $2$& $2$& $2$&$ 2$& $2$&$2$\\
			\hline $\hat{U}^1_{3,5}$& $2$& $1$ &$1$ &$1$ &$ 3$& $3$ &$3$ &$2$& $2$&$ 2$& $3$& $3$&$ 3$& $3$&$3$\\
			\hline $V^{12}_8$  & $2$&$ 2$ &$2$ &$2$ & $4$& $3$ &$3$ &$3$& $3$& $3$&$ 4$& $4$& $4$&$ 4$&$4$\\
			\hline
		\end{tabular}
	}	\label{tablevector}
\end{table*}

\subsection{Mixed-level variable-strength orthogonal arrays and
  orthogonal  Latin hypercubes}
\label{jlvoaolh}

To characterize entropy functions on the extreme rays of $\Gamma_4$,
we adopt two sets of combinatorial structures, that is, mixed-level variable-strength
  orthogonal arrays and orthogonal Latin hypercubes, which are
  equivalent in some sense, but each has its own advantages. For
  mixed-level variable-strength
  orthogonal arrays, it will be generalized more easily to the
  cases with more random variables and higher dimensions, and the
  symmetries between random variables are more straightforward.
In contrast, the language of orthogonal Latin hypercubes will be
better visualized for the cases of four random variables. 
Furthermore,
they will also play important roles in characterizing
$2$-dimensional faces of $\Gamma_4$, especially those with two extreme rays containing integer
polymatroids of rank exceeding $1$.


\begin{definition}[MVOA,\rv{ }\cite{CCB24}]
  Given an integer polymatroid $P=(N_n,\mb{r})$ with $\br(N_n)\ge 2$ and
  an integer $v>1$, a $v^{\br(N_n)}\times n$
  array $\mb{T}$ with columns indexed by $N_n$,
  whose entries of a column indexed by $i\in N_n$
 are from a set $I_i$ with cardinality $v^{\br(i)}-1$, is called a \emph{mixed-level variable-strength
  orthogonal array}(MVOA) induced by $P$
  with \emph{base level} $v$ if for each
 $A\subseteq N_n$, $v^{\mb{r}(N_n)}\times |A|$ subarray $\mb{T}(A)$ of $\mb{T}$ consisting of
 columns indexed by $A$ satisfies the following condition:
each row of $\mb{T}(A)$ occurs in $\mb{T}(A)$ exactly $v^{\br(N_n)-\br(A)}$ times.
 We also call such $\mb{T}$ a $\mvoa(P,v)$.
\end{definition}

Usually, we set $I_i=\mathbb{I}_{ v^{\br(i)}-1}$, unless otherwise specified, where
$\mathbb{I}_u\triangleq \{0,1,\ldots, u-1\}$ for any integer $u\ge
1$.
When the integer polymatroid is a matroid $M$, its induced MVOA is reduced to
a \emph{variable-strength orthogonal array} $\voa(M,v)$\mrv{\cite{CCB21},\cite{CCB22},\cite{CCB24}}. 
When a matroid is a uniform matroid $U_{t,n}$, the induced $\voa(U_{t,n},v)$ is an
orthogonal array with index unity, usually denoted by $\oa(t,n,v)$
\cite{CCB21, CCB22}.

To \rv{describe} MVOAs, we introduce some notations which will be used
throughout this paper.
The $v^{\br(N_n)}\times |A|$ subarray of $\mb{T}$ consisting of columns indexed by $A$ is denoted by $\mb{T}(A)$. 
For simplicity, we drop the brackets of $\mb{T}(A)$, e.g., we write
$\mb{T}(\{1,2,3\})$ as $\mb{T}(1,2,3)$. Let $\mb{T}(A;j)$ denote the $j$-th row of $\mb{T}(A)$.

For the orthogonal Latin hypercubes, we will not give a general
definition in this paper. Because, on the one hand, for some extreme rays containing an
integer polymatroid other than a uniform matroid, the variants of traditional
orthogonal Latin hypercubes are utilized. On the other hand, in this
paper, we only consider the cases corresponding
to integer polymatroids in $\Gamma_4$, \rv{and so, a case-by-case discussion is enough.}
For the general form of these variants, it can be studied in future research.

We begin our discussion from Latin squares, the non-trivial orthogonal Latin
hypercubes with the smallest parameters, \rv{which corresponds to}  the
smallest connected matroid with rank exceeding $1$.
For uniform matroid
$U_{2,3}$,  a $\voa(U_{2,3},v)$ corresponds to a Latin square $S$ of
order $v$ with row and column indices and symbols in $\mathbb{I}_v$, that is,
a $v\times v$ square $S$ with cells indexed by $(i,j)\in \mathbb{I}^2_v$ and symbols in
the cells of each row $\{S(i,j),j\in \mathbb{I}_v\}$ and each column $\{S(i,j),i\in \mathbb{I}_v\}$ all distinct. For each row $(i,j,k)$ of a $\voa(U_{2,3},v)$ $\mb{T}$, it
corresponds to the cell $S(i,j)$ with symbol $k$. Here, we also define Latin squares of the
zeroth-kind and the second-kind,\footnote{The definition of Latin squares of
  the zeroth-kind and the second-kind are in the spirit of the definition
  of the ``Latin cubes of first order and second order'' in
  \cite{OrthogonalDesigns}. ``Order'' is usually used for the size of
  a Latin square, so in our paper, we use ``kind'' instead. In the
  following, Latin cubes of the first kind and the second kind will be
  defined similarly.} which are trivial themselves,
but can be considered as building blocks in construction of other
combinatorial structures in this paper, while for a traditional Latin
square, we call it of \emph{the first kind}. For a $v\times v$ square $S$
with symbols in all cells, each identical to $0$, we call it a \emph{Latin square of
the zeroth kind}. It corresponds to a $\voa(U^{1,2;3}_{1,1},v)$, where
$U_{1,1}^{1,2;3}$ is a matroid whose rank function is the \rv{summation} of the
rank functions of $U^{1;3}_{1,1}$ and $U^{2;3}_{1,1}$.
For a $v\times v$ square $S$ with symbols in all cells
distinct from $\mathbb{I}_{v^2}$, we call it a \emph{Latin square of the second kind}. It corresponds to a $\mvoa(U^{13,23;3}_{1,2},v)$, where
$U^{13,23;3}_{1,2}$ is an integer polymatroid whose rank function is
the summation of the rank functions of $U^{13;3}_{1,2}$ and
$U^{23;3}_{1,2}$. For a traditional Latin square, or a Latin square of
the first kind, we call it a Latin square and omit ``of the first
kind'' if there is no ambiguity.   

\begin{example}
In Table \ref{tab:example1}, we list  Latin squares of the zeroth kind, the first kind,
and  the second kind,
and their corresponding MVOAs for $v=3$.
\begin{table*}[ht]
	\centering
	\caption{Three kinds of Latin squares and their corresponding (M)VOAs}
	\label{tab:example1}
	\renewcommand{\arraystretch}{1.1}
	
	\begin{subtable}[t]{0.3\linewidth}
		\centering
		\subcaption*{\textbf{(a) The zeroth kind}}
		\vspace{0.5em}
		$\begin{array}{c|ccc}
			& 0 & 1 & 2 \\ \hline
			0 & 0 & 0 & 0 \\ 
			1 & 0 & 0 & 0 \\ 
			2 & 0 & 0 & 0
		\end{array}$\\[0.8em]
		
		\hspace{-1.7cm}$\voa(U^{1,2;3}_{1,1},3):$
		$\begin{array}{ccc}
			0 & 0 & 0 \\ 
			0 & 1 & 0 \\ 
			0 & 2 & 0 \\ 
			1 & 0 & 0 \\ 
			1 & 1 & 0 \\ 
			1 & 2 & 0 \\ 
			2 & 0 & 0 \\ 
			2 & 1 & 0 \\ 
			2 & 2 & 0
		\end{array}$
	\end{subtable}
	\hfill\hspace{-0.5  cm}
	\begin{subtable}[t]{0.15\linewidth}
		\centering
		\subcaption*{\textbf{(b) The first kind}}
		\vspace{0.5em}
		$\begin{array}{c|ccc}
			& 0 & 1 & 2 \\ \hline
			0 & 0 & 1 & 2 \\ 
			1 & 2 & 0 & 1 \\ 
			2 & 1 & 2 & 0
		\end{array}$\\[0.8em]
		
		\hspace{-1.5cm}$\voa(U_{2,3},3):$
		$\begin{array}{ccc}
			0 & 0 & 0 \\ 
			0 & 1 & 1 \\ 
			0 & 2 & 2 \\ 
			1 & 0 & 2 \\ 
			1 & 1 & 0 \\ 
			1 & 2 & 1 \\ 
			2 & 0 & 1 \\ 
			2 & 1 & 2 \\ 
			2 & 2 & 0
		\end{array}$
	\end{subtable}
	\hfill
	\begin{subtable}[t]{0.3\linewidth}
		\centering
		\subcaption*{\textbf{(c) The second kind}}
		\vspace{0.5em}
		$\begin{array}{c|ccc}
			& 0 & 1 & 2 \\ \hline
			0 & 0 & 1 & 2 \\ 
			1 & 3 & 4 & 5 \\ 
			2 & 6 & 7 & 8
		\end{array}$\\[0.8em]
		
		\hspace{-2.2cm}$\mvoa(U^{13,23;3}_{1,2},3):$
		$\begin{array}{ccc}
			0 & 0 & 0 \\ 
			0 & 1 & 1 \\ 
			0 & 2 & 2 \\ 
			1 & 0 & 3 \\ 
			1 & 1 & 4 \\ 
			1 & 2 & 5 \\ 
			2 & 0 & 6 \\ 
			2 & 1 & 7 \\ 
			2 & 2 & 8
		\end{array}$
	\end{subtable}
\end{table*}
\end{example}
In the following, we list the mixed-level variable-strength orthogonal array induced by
  integer polymatroids on the extreme rays of $\Gamma_4$ with rank
  exceeding $1$, and their
  corresponding variants of orthogonal Latin hypercubes. For each type
  of extreme rays, we only consider one of its representatives.

\begin{itemize}
\item For a $\voa(U^{123}_{2,3},v)$ $\mb{T}$, $\mb{T}(1,2,3)$ forms a 
 $\voa(U_{2,3},v)$, and all the entries of the column
 $\mb{T}(4)$ can be a constant symbol $0$. It corresponds to a pair
 of Latin squares sharing the same row and column indices and one of which is
 the first-kind, and the other of which is the zeroth-kind, where $\mb{T}(1)$ and
 $\mb{T}(2)$, $\mb{T}(3)$, $\mb{T}(4)$ corresponds to the indices of
 rows and columns, the symbols of the first square and symbols of the
 second square, respectively.

 \item For a $\voa(\mc{W}^{34},v)$ $\mb{T}$, $\mb{T}((1,2,3)$ forms a
   $\voa(U_{2,3},k)$ and $\mb{T}(4)=\mb{T}(3)$. Thus $\mb{T}$ corresponds
   to a pair of $v\times v$ identical Latin squares.

\item For a $\voa(U_{2,4},v)$ $\mb{T}$, it correspond to a pair of
  mutual orthogonal $v\times v$ Latin squares. A pair of Latin squares
  (of the first kind) $S_1$ and $S_2$ are called orthogonal if the pairs
  of symbols in the pairs of cells
  $\{(S_1(i,j),S_2(i,j)):i,j\in \mathbb{I}_v\}$ are all distinct, or exactly those in $\mathbb{I}^2_v$.

\item For a $\voa(U_{3,4},v)$ $\mb{T}$, it corresponds to a Latin cube
  of order $v$ (type 2 and the first kind, permutation cube \cite[VI.22.33]{colbourn2006handbook}), where entries in $\mb{T}(1)$,
  $\mb{T}(2)$ and $\mb{T}(3)$ correspond to the indices of three
  dimensions, respectively, and $\mb{T}(4)$ corresponds to the symbols.
  A \emph{Latin cube of order $v$ and the first kind} is a $v\times
  v\times v$ cube with symbols in each line
  $\{C(i,j,k), i\in \mathbb{I}_v\}$ for fixed $j,k\in \mathbb{I}_v$, $\{C(i,j,k),
  j\in \mathbb{I}_v\}$ for fixed $i,k\in \mathbb{I}_v$, and $\{C(i,j,k),
  i\in \mathbb{I}_v\}$ for fixed $i,j\in \mathbb{I}_v$ all distinct. 


\item For a $\mvoa(\hat{U}^4_{2,5},v)$ $\mb{T}$, it corresponds to a pair of
$v\times v$ squares $(S_1,S_2)$, where $S_1$ is a Latin square of the 
first kind, and
$S_2$ is a Latin square of the second kind. 

\item For a $\mvoa(\hat{U}^4_{3,5},v)$ $\mb{T}$, it corresponds a
Latin cube $C$ of order $v$ and the second kind. A \emph{Latin cube of order
$v$ and the second kind} is  
$v\times v \times v$ cube with
each layer $\{C(i,j,k),i,j\in \mathbb{I}_v\}$ for fixed $k\in \mathbb{I}_v$,
$\{C(i,j,k),i,k\in \mathbb{I}_v\}$ for fixed $j\in \mathbb{I}_v$, and $\{C(i,j,k),j,k\in
\mathbb{I}_v\}$ for fixed $i\in \mathbb{I}_v$ all Latin squares of order $v$ and the second kind. 

\end{itemize}

It can be seen that MVOAs induced by rank $2$ integer ploymatroids
correspond to a pair of Latin
squares of the same kind superimposed in different ways, or of
different kinds. While  MVOAs induced by $2$ rank $3$ integer 
ploymatroids correspond to Latin cubes of two different kinds.

\begin{example}
In Table \ref{tab:example2}, we list orthogonal Latin hypercubes and their
corresponding MVOAs discussed above for $v=3$.

\begin{table*}[htbp]
	\centering
	\caption{Orthogonal Latin hypercubes and their corresponding MVOAs}
	\label{tab:example2}
	\renewcommand{\arraystretch}{1.1}
	
	\begin{subtable}[t]{0.45\linewidth}
		\centering
		\subcaption*{\textbf{(a) Latin squares of the first and zeroth kinds}}
		\vspace{0.5em}
		$\begin{array}{c|ccc}
			& 0 & 1 & 2 \\ \hline
			0 & (0,0) & (1,0) & (2,0) \\ 
			1 & (2,0) & (0,0) & (1,0) \\ 
			2 & (1,0) & (2,0) & (0,0)
		\end{array}$\\[0.4em]
		$\voa(U^{123}_{2,3},3):$
		$\begin{array}{cccc}
			0 & 0 & 0 & 0 \\ 
			0 & 1 & 1 & 0 \\ 
			0 & 2 & 2 & 0 \\ 
			1 & 0 & 2 & 0 \\ 
			1 & 1 & 0 & 0 \\ 
			1 & 2 & 1 & 0 \\ 
			2 & 0 & 1 & 0 \\ 
			2 & 1 & 2 & 0 \\ 
			2 & 2 & 0 & 0
		\end{array}$
	\end{subtable}\hfill
	\begin{subtable}[t]{0.45\linewidth}
		\centering
		\subcaption*{\textbf{(b) Two orthogonal Latin squares}}
		\vspace{0.5em}
		$\begin{array}{c|ccc}
			& 0 & 1 & 2 \\ \hline
			0 & (0,0) & (1,1) & (2,2) \\ 
			1 & (2,1) & (0,2) & (1,0) \\ 
			2 & (1,2) & (2,0) & (0,1)
		\end{array}$\\[0.4em]
		$\voa(U_{2,4},3):$
		$\begin{array}{cccc}
			0 & 0 & 0 & 0 \\ 
			0 & 1 & 1 & 1 \\ 
			0 & 2 & 2 & 2 \\ 
			1 & 0 & 2 & 1 \\ 
			1 & 1 & 0 & 2 \\ 
			1 & 2 & 1 & 0 \\ 
			2 & 0 & 1 & 2 \\ 
			2 & 1 & 2 & 0 \\ 
			2 & 2 & 0 & 1
		\end{array}$
	\end{subtable}
	
	\vspace{1em}
	\begin{subtable}[t]{0.45\linewidth}
		\centering
		\subcaption*{\textbf{(c) Two identical Latin squares}}
		\vspace{0.5em}
		$\begin{array}{c|ccc}
			& 0 & 1 & 2 \\ \hline
			0 & (0,0) & (1,1) & (2,2) \\ 
			1 & (2,2) & (0,0) & (1,1) \\ 
			2 & (1,1) & (2,2) & (0,0)
		\end{array}$\\[0.4em]
		$\voa(\mathcal{W}^{34}_2,3):$
		$\begin{array}{cccc}
			0 & 0 & 0 & 0 \\ 
			0 & 1 & 1 & 1 \\ 
			0 & 2 & 2 & 2 \\ 
			1 & 0 & 2 & 2 \\ 
			1 & 1 & 0 & 0 \\ 
			1 & 2 & 1 & 1 \\ 
			2 & 0 & 1 & 1 \\ 
			2 & 1 & 2 & 2 \\ 
			2 & 2 & 0 & 0
		\end{array}$
	\end{subtable}\hfill
	\begin{subtable}[t]{0.45\linewidth}
		\centering
		\subcaption*{\textbf{(d) Latin squares of the first and second kinds}}
		\vspace{0.5em}
		$\begin{array}{c|ccc}
			& 0 & 1 & 2 \\ \hline
			0 & (0,0) & (1,1) & (2,2) \\ 
			1 & (2,3) & (0,4) & (1,5) \\ 
			2 & (1,6) & (2,7) & (0,8)
		\end{array}$\\[0.4em]
		$\mvoa(\hat{U}^4_{2,5},3):$
		$\begin{array}{cccc}
			0 & 0 & 0 & 0 \\ 
			0 & 1 & 1 & 1 \\ 
			0 & 2 & 2 & 2 \\ 
			1 & 0 & 2 & 3 \\ 
			1 & 1 & 0 & 4 \\ 
			1 & 2 & 1 & 5 \\ 
			2 & 0 & 1 & 6 \\ 
			2 & 1 & 2 & 7 \\ 
			2 & 2 & 0 & 8
		\end{array}$
	\end{subtable}
	
	\vspace{1.5em}
	\begin{subtable}[t]{\linewidth}
		\centering
		\subcaption*{\textbf{(e) Latin cube of the first kind}}
		$\begin{array}{c|ccc|ccc|ccc}
			& 0 & 1 & 2 & 0 & 1 & 2 & 0 & 1 & 2 \\ \hline
			0 & (0,0)&(0,1)&(0,2)&(1,1)&(1,2)&(1,0)&(2,2)&(2,0)&(2,1)\\
			1 & (0,1)&(0,2)&(0,0)&(1,2)&(1,0)&(1,1)&(2,0)&(2,1)&(2,2)\\
			2 & (0,2)&(0,0)&(0,1)&(1,0)&(1,1)&(1,2)&(2,1)&(2,2)&(2,0)
		\end{array}$\\[0.8em]

$\voa(U_{3,4},3)$ (transposed):\\[0.5em]
	\resizebox{0.8\linewidth}{!}{$
		\begin{array}{cccccccccccccccccccccccccccc}
				0 & 0 & 0 & 1 & 1 & 1 & 2 & 2 & 2 & 0 & 0 & 0 & 1 & 1 & 1 & 2 & 2 & 2 & 0 & 0 & 0 & 1 & 1 & 1 & 2 & 2 & 2 &  \\
				0 & 1 & 2 & 0 & 1 & 2 & 0 & 1 & 2 & 0 & 1 & 2 & 0 & 1 & 2 & 0 & 1 & 2 & 0 & 1 & 2 & 0 & 1 & 2 & 0 & 1 & 2 &  \\
				0 & 0 & 0 & 0 & 0 & 0 & 0 & 0 & 0 & 1 & 1 & 1 & 1 & 1 & 1 & 1 & 1 & 1 & 2 & 2 & 2 & 2 & 2 & 2 & 2 & 2 & 2 &  \\
				0 & 1 & 2 & 1 & 2 & 0 & 2 & 0 & 1 & 1 & 2 & 0 & 2 & 0 & 1 & 0 & 1 & 2 & 2 & 0 & 1 & 0 & 1 & 2 & 1 & 2 & 0 & 
			\end{array}
		$}
	\end{subtable}\\[2em]

\begin{subtable}[t]{\linewidth}
	\centering
	\subcaption*{\textbf{(f) Latin cube of the second kind}}
	$\begin{array}{c|ccc|ccc|ccc}
		& 0 & 1 & 2 & 0 & 1 & 2 & 0 & 1 & 2 \\ \hline
		0 & (0,0)&(0,1)&(0,2)&(1,4)&(1,5)&(1,6)&(2,8)&(2,3)&(2,7)\\
		1 & (0,3)&(0,4)&(0,5)&(1,7)&(1,8)&(1,0)&(2,2)&(2,6)&(2,1)\\
		2 & (0,6)&(0,7)&(0,8)&(1,1)&(1,2)&(1,3)&(2,5)&(2,0)&(2,4)
	\end{array}$\\[1em]
	
     $\mvoa(\hat{U}^4_{3,5},3)$ (transposed):\\[0.5em]
	\resizebox{0.8\linewidth}{!}{$
	\begin{array}{@{}*{27}{c}@{}}
				0 & 0 & 0 & 1 & 1 & 1 & 2 & 2 & 2 & 0 & 0 & 0 & 1 & 1 & 1 & 2 & 2 & 2 & 0 & 0 & 0 & 1 & 1 & 1 & 2 & 2 & 2 \\
				0 & 1 & 2 & 0 & 1 & 2 & 0 & 1 & 2 & 0 & 1 & 2 & 0 & 1 & 2 & 0 & 1 & 2 & 0 & 1 & 2 & 0 & 1 & 2 & 0 & 1 & 2 \\
				0 & 0 & 0 & 0 & 0 & 0 & 0 & 0 & 0 & 1 & 1 & 1 & 1 & 1 & 1 & 1 & 1 & 1 & 2 & 2 & 2 & 2 & 2 & 2 & 2 & 2 & 2 \\
				0 & 1 & 2 & 3 & 4 & 5 & 6 & 7 & 8 & 4 & 5 & 6 & 7 & 8 & 0 & 1 & 2 & 3 & 8 & 3 & 7 & 2 & 6 & 1 & 5 & 0 & 4
			\end{array}
		$}
\end{subtable}
\end{table*}
\end{example}

\subsection{Entropy functions on the extreme rays of $\Gamma_4$}
\label{ray}

As we discussed in Subsection \ref{fpc}, for the V-representation of a
face, it can be written as the convex hull of the extreme rays it
contains. Therefore, to characterize entropy functions on
$2$-dimensional faces of $\Gamma_4$, we need first to characterize the
entropy functions on the extreme rays of $\Gamma_4$. For the eleven
types of extreme rays, the entropy functions on seven of them are
matroidal and have been characterized in \cite{CCB21}, while those
containing $\hat{U}^i_{2,5}$ or $\hat{U}^i_{3,5}$ are characterized in
Part I of this series \cite{LCC2024part1}, and those containing $V^\alpha_8$ are
non-entropic. Theorems that characterize these entropy functions are listed
below without proof.


\begin{theorem}
	\label{ex1}
  For $E=U^1_{1,1}, U^{12}_{1,2}, U^{123}_{1,3}, U_{1,4}$, i.e.,
  extreme rays containing a matroid with rank $1$, $a\br\in E$ is entropic
  for all $a\ge 0$.
\end{theorem}

The characterization of the following four types of extreme rays follows
immediately from matroidal entropy functions in \cite{CCB21} and \cite{CCB24}.

\begin{theorem}
	\label{ex2}
  For $E=U^{123}_{2,3}, \mc{W}^{14}_2, U_{3,4}$, $a\br\in E$ is entropic if and only if
  $a=\log k$ for integer $k\ge 1$.
\end{theorem}

\begin{theorem}(\cite[Proposition 2]{CCB21})
	\label{ex3}
  For $E=U_{2,4}$, $a\br\in E$ is entropic if and only if
  $a=\log k$ for positive integer $k\neq 2,6$.
\end{theorem}

\begin{theorem}(\cite[Theorem II-C.4]{LCC2024part1})
	\label{rk16}
	For the rank function $\br$ of $\hat{U}^i_{2,5}$, $\h=a\br$ is entropic if and only if $a=\log k$ for integer $k>0$.
\end{theorem}


\begin{theorem}(\cite[Theorem II-C.5]{LCC2024part1})
	\label{rk17}
	For the rank function $\br$ of $\hat{U}^i_{3,5}$, $\h=a\br$ is entropic if and only if $a=\log k$ for integer $k>0$.
\end{theorem}

\begin{theorem}
	\label{exv8}
  For $E=V^\alpha$, $a\h\in E$ is entropic if and only if $a=0$.
\end{theorem}

\noindent \textbf{Remark} In Part I \cite{LCC2024part1}, Theorem \ref{rk16}(resp. \ref{rk17}) on extreme 
ray containing $\hat{U}^i_{2,5}$(resp. $\hat{U}^i_{3,5}$) was proved by
the construction of a specific $\mvoa(\hat{U}^i_{2,5},v)$(resp.
$\mvoa(\hat{U}^i_{3,5},v)$) for all $v\ge 1$. However, the correspondence
between a pair of Latin squares of the first kind and the second kind
 and an $\mvoa(\hat{U}^i_{2,5},v)$(resp.  a Latin cube of the second kind
and an $\mvoa(\hat{U}^i_{3,5},v)$) setting up in Subsection
\ref{jlvoaolh} provides a general construction of the problem. Such
construction also sheds some light on the open question
\cite[Question 1]{CCB24}.

\begin{table}[t] 
	\renewcommand{\arraystretch}{2}
	\begin{center}   		
		\huge
		\caption{Two-dimensional faces of $\Gamma_4$.}  
		\label{table:1}
		\resizebox{1\columnwidth}{!}{
			\begin{tabular}{|c|c|c|c|c|c|c|c|c|c|c|c|}  
				\hline   &$U^i_{1,1}$ & $U^\alpha_{1,2}$&$U^\alpha_{1,3}$ &$U_{1,4}$ &	$U^\alpha_{2,3}$&	$\mc{W}^\alpha_{2}$& 	$U_{2,4}$ &	$U_{3,4}$ &$\hat{U}^i_{2,5}$&	$\hat{U}^i_{3,5}$&	$V^\alpha_8$ \\
				\hline  \multirow{4}{*}{$U^j_{1,1}$ }     &    &   $(U^{12}_{1,2},U^1_{1,1}),12$&  	$(U^{123}_{1,3},U^1_{1,1}),12$ & &{  $(U_{2,3}^{123},U_{1,1}^{1}),12$}  &	{  $(\mc{W}_{2}^{14},U_{1,1}^{1}),12$}&	&	&  	$(\hat{U}^1_{2,5},U_{1,1}^{1}),4$ &  $(\hat{U}^{1}_{3,5},U^{1}_{1,1}),4$ &  $(V^{12}_8,U^{1}_{1,1}),12$	\\
				&  $(U^1_{1,1},U^2_{1,1}),6$ &\pref{1}{part1-rk1}&\pref{1}{part1-rk1}    &$(U_{1,4},U^{1}_{1,1}),4$&\pref{1}{part1-rk3}  & \pref{1}{part1-rk3}  &{$(U_{2,4},U^{1}_{1,1}),4$} &  $(U_{3,4},U^{1}_{1,1}),4$ & \pref{1}{part1-rk18} & \pref{1}{part1-rk21}  &  \pref{1}{part1-rk31}  \\
				\cline{3-4}\cline{6-7} \cline{10-12} 
				& \pref{1}{part1-rk1}  & $(U^{12}_{1,2},U^3_{1,1}),12$  & $(U^{123}_{1,3},U^4_{1,1}),4$ &  \pref{1}{part1-rk1} & {  $(U_{2,3}^{123},U_{1,1}^{4}),4$} &
				{ $(\mc{W}_{2}^{34},U_{1,1}^{1}),12$} & \pref{2}{rk25} &\pref{1}{part1-rk8} &  	$(\hat{U}^1_{2,5},U_{1,1}^{2}),12$&  	$(\hat{U}^{1}_{3,5},U^{2}_{1,1}),12$ &  $(V^{12}_8,U^{3}_{1,1}),12$\\
				& & \pref{1}{part1-rk1} &\pref{1}{part1-rk1} & &{ \pref{1}{part1-rk3}} &{  \pref{1}{part1-rk3}} & & & \pref{1}{part1-rk18} &  \pref{1}{part1-rk21}& \pref{1}{part1-rk31} \\
				\hline  \multirow{6}{*}{$U^\beta_{1,2}$} &	\multirow{6}{*}{$\backslash$ }& & &	& &	{ $(\mc{W}_{2}^{14},U_{1,2}^{14}),6$}&	& & & & \\
				& & $(U^{12}_{1,2},U^{13}_{1,2}),12$ & $(U^{123}_{1,3},U^{12}_{1,2}),12$ & &{ 	$(U_{2,3}^{123},U_{1,2}^{12}),12$} &\pref{1}{part1-rk3} & & & & & \\
				\cline{7-7}
				& & \pref{1}{part1-rk1} & \pref{1}{part1-rk1} & $ (U_{1,4},U^{12}_{1,2}),6$&\pref{1}{part1-rk2} &{$(\mc{W}_{2}^{24},U_{1,2}^{14}),24$}&$(U_{2,4},U^{12}_{1,2}),6$ & $(U_{3,4},U^{12}_{1,2}),6$ & $(\hat{U}^1_{2,5},U_{1,2}^{12}),12$ & $(\hat{U}^{1}_{3,5},U^{12}_{1,2}),12$ &  $(V^{12}_8,U^{13}_{1,2}),24$\\
				\cline{3-4} \cline{6-6}
				& & $(U^{12}_{1,2},U^{34}_{1,2}),3$ & $(U^{123}_{1,3},U^{14}_{1,2}),12$ & \pref{1}{part1-rk1} & { $(U_{2,3}^{123},U_{1,2}^{14}),12$} &\pref{1}{part1-rk3} &  \pref{2}{rk15} &  \pref{1}{part1-rk9}& \pref{1}{part1-rk18} &  \pref{1}{part1-rk21} &  \pref{1}{part1-rk31}\\
				\cline{7-7}
				& & \pref{1}{part1-rk1} &\pref{1}{part1-rk1} & & \pref{1}{part1-rk3}  & {$(\mc{W}_{2}^{34},U_{1,2}^{12}),6$} & & & & & \\
				& & & & & &\pref{1}{part1-rk2}& & & & & \\
				\hline \multirow{2}{*}{$U^\beta_{1,3}$}  & \multirow{2}{*}{ $\backslash$} &	 \multirow{2}{*}{ $\backslash$} & 	$(U^{123}_{1,3},U^{124}_{1,3}),6$& 	$(U_{1,4},U^{123}_{1,3}),4$&	{ $(U_{2,3}^{123},U_{1,3}^{124}),12$}&{ 	$(\mc{W}_{2}^{14},U_{1,3}^{124}),12$}&	$(U_{2,4},U^{123}_{1,3}),4$& 	$(U_{3,4},U^{123}_{1,3}),4$&	 $(\hat{U}^1_{2,5},U_{1,3}^{123}),12$& 	$(\hat{U}^{1}_{3,5},U^{234}_{1,3}),4$&	 $(V^{12}_8,U^{134}_{1,3}),12$\\
				& & & \pref{1}{part1-rk1} &\pref{1}{part1-rk1} & \pref{1}{part1-rk4} & \pref{1}{part1-rk2} & \pref{2}{rk13} &\pref{1}{part1-rk11} & \pref{1}{part1-rk19} & \pref{1}{part1-rk22} &\pref{1}{part1-rk31} \\
				\hline \multirow{2}{*}{$U_{1,4}$ }        &	\multirow{2}{*}{$\backslash$} &	\multirow{2}{*}{$\backslash$} &	\multirow{2}{*}{$\backslash$}&	\multirow{2}{*}{$\backslash$}&	{ 	$(  U_{2,3}^{123  } , U_{1,4} ),4     $}&	\multirow{2}{*}{ 0}&	\multirow{2}{*}{ 0}&	 $(U_{3,4},U_{1,4}),1$& \multirow{2}{*}{ 0}&	\multirow{2}{*}{ 0}&	 $(V^{12}_8,U_{1,4}),6$\\
				& & & & &\pref{1}{part1-rk7} & & &\pref{1}{part1-rk12} & & &\pref{1}{part1-rk31} \\
				\hline \multirow{2}{*}{$U^\beta_{2,3}$}  &	\multirow{2}{*}{$\backslash$} &\multirow{2}{*}{	$\backslash$} &	\multirow{2}{*}{$\backslash$}&\multirow{2}{*}{	$\backslash$}&	$ (  U_{2,3}^{123  } , U_{2,3}^{124  } ),6$ &  {	$(  \mc{W}_{2}^{12  } , U_{2,3}^{134  } ),12$}&	$(U_{2,4},U^{123}_{2,3}),4$&	 $(U_{3,4},U^{123}_{2,3}),4$&$(\hat{U}^1_{2,5},U_{2,3}^{234}),4$& 	$(\hat{U}^{1}_{3,5},U^{123}_{2,3}),12$& $(V^{12}_8,U^{123}_{2,3}),12$\\
				& & & & &\pref{1}{part1-rk5} &\pref{1}{part1-rk6} & \pref{2}{rk14} &\pref{1}{part1-rk10} &\pref{2}{rk27} &\pref{1}{part1-rk23} & \pref{1}{part1-rk31}\\
				\hline \multirow{2}{*}{$\mc{W}^\beta_{2}$}	   &  \multirow{2}{*}{$\backslash$} &	\multirow{2}{*}{$\backslash$} &	\multirow{2}{*}{$\backslash$} &	\multirow{2}{*}{$\backslash$} &\multirow{2}{*}{$\backslash$}&{	$(\mc{W}_{2}^{12 },\mc{W}_{2}^{13 }),12$}&{		$(U_{2,4},\mc{W}^{12}_{2}),6$}&	\multirow{2}{*}{ 0}&{	{$(\hat{U}^1_{2,5},\mc{W}_{2}^{12}),12$}}&{	$ (\hat{U}^{1}_{3,5},\mc{W}^{23}_{2}),12$}&	\multirow{2}{*}{ 0} \\
				& & & & & & \pref{2}{rk28} & \pref{2}{rk26} & & \pref{2}{rk20} &\pref{1}{part1-rk24} &
				\\
				\hline  \multirow{2}{*}{$U_{2,4}$}       &\multirow{2}{*}{$\backslash$} &	\multirow{2}{*}{$\backslash$} &	\multirow{2}{*}{$\backslash$}&	\multirow{2}{*}{$\backslash$}&	\multirow{2}{*}{$\backslash$}&	\multirow{2}{*}{$\backslash$}& 	\multirow{2}{*}{$\backslash$}	& 	\multirow{2}{*}{ 0}&{$(\hat{U}^1_{2,5},U_{2,4}),4$}&{	$(\hat{U}^{1}_{3,5},U_{2,4}),4$}&	\multirow{2}{*}{ 0}\\
				& & & & & & & & &\pref{2}{rk29}  & \pref{2}{rk30}  & \\
				\hline \multirow{2}{*}{$U_{3,4}$}     &  \multirow{2}{*}{$\backslash$} &\multirow{2}{*}{$\backslash$}&	\multirow{2}{*}{$\backslash$}&	\multirow{2}{*}{$\backslash$}&	\multirow{2}{*}{$\backslash$} &	\multirow{2}{*}{$\backslash$} &	\multirow{2}{*}{$\backslash$} &	\multirow{2}{*}{$\backslash$}	&	\multirow{2}{*}{ 0}&	\multirow{2}{*}{ 0}&	{$ (V^{12}_8,U_{3,4}),6$}\\
				& & & & & & & & & & &\pref{1}{part1-rk31} \\
				\hline $\hat{U}^j_{2,5}$      &	$\backslash$ &	$\backslash$ &	$\backslash$ &	$\backslash$&	$\backslash$&	$\backslash$&$\backslash$&	$\backslash$&	 0&	 0&	 0\\
				\hline $\hat{U}^j_{3,5}$  &  $\backslash$ &$\backslash$&	$\backslash$&	$\backslash$&	$\backslash$&	$\backslash$&	$\backslash$&	$\backslash$&$\backslash$& 	0&	 0\\
				\hline  $V^\beta_8$    &  $\backslash$ &	$\backslash$ &	$\backslash$ &	$\backslash$&	$\backslash$&	$\backslash$&$\backslash$&$\backslash$&	$\backslash$&	$\backslash$& 	0  \\
				\hline   
			\end{tabular}  
		}
	\end{center}  
 We label the rows and
columns with the $11$ types of extreme rays of $\Gamma_4$.
For simplicity, we
denote the face $F=\mr{cone}(E_1, E_2)$ by $(E_1,E_2)$. In each cell
with ``$(E_1, E_2)\  n$, Part $k$ Thm. $m$'', $k=1$ or $2$, ``$(E_1, E_2)$'' denote a representative of the
type of $2$-dimensional faces, where ``$E_1$ $(E_2)$'' is a representative of
the type the extrem rays in the column (row),
``$n$'' is the number of the faces in this type and this face is
characterized in ``Thm. $m$'' of ``Part $k$'' of this series. For the cell with
``$0$'', the convex hull of the two extreme rays in each type forms no
$2$-dimensional faces of $\Gamma_4$.
\end{table}

\section{\rv{Characterization of entropy functions on Two-dimensional  faces of $\Gamma_4$} }
\label{af}

In this section, we characterize the entropy functions on
$2$-dimensional faces of $\Gamma_4$. We embed each face $F=(E_1, E_2)$
in the first octant of a $2$-dimensional Cartesian coordinate system
whose axes are labeled by $a$ and $b$. Thus, for each $(a,b)$, $a,b\ge
0$, it represents the polymatroid $a\br_1+b\br_2$, where $\br_i, i=1,2$, is
the rank function of the minimal integer polymatroid in $E_i$, respectively. Throughout this paper, for a random vector $(X_i,i\in N_4)$ or its subvectors, 
we assume each $X_i$ is distributed on a finite set $\mathcal{X}_{i}$, and for each $x_i\in X_i$, $p(x_i)>0$.
\begin{lemma}(\cite[Lemma 1]{LCC2024part1})
	\label{lem1}
	  If $X_1$ and $X_2$ are independent and 
	for any $p(x_1,x_2,x_3)>0$, $p(x_1)=p(x_2)$, then $X_1$ and $X_2$ are uniformly distributed on $\mathcal{X}_{1}$ and $\mathcal{X}_{2}$, respectively, $|\mathcal{X}_{1}|=|\mathcal{X}_{2}|$ and $H(X_1)=H(X_2)$.
\end{lemma}

\begin{lemma}(\cite[Lemma 15.3]{yeung2008information})
  \label{lem}
  For any $\h_1,\h_2\in \Gamma^*_n$, $\h_1+\h_2\in \Gamma^*_n$.
\end{lemma}
\subsection{Entropy functions on faces with extreme rays containing $U_{2,4}$ and one rank-1 matroid}
\label{u24rank1}
In this subsection, we characterize entropy functions on the $2$-dim
faces of $\Gamma_4$ with extreme rays containing $U_{2,4}$ and one rank-1 matroid.

\begin{theorem}
	\label{rk13}
	For $F=(U_{2,4},U^{123}_{1,3})$, $\mathbf{h}=(a,b) \in F
		$ is entropic if and only if 	
		\begin{itemize}
			\item   $	a+b\geq \log v   $ and  $ \log(v-1)< a \leq \log v $ for positive integer $  v\neq 2,6$; or
			\item  	$a+b\geq \log{(v+1)} $ and  $\log(v-1)< a \leq \log v $  for $ v = 2,6.$
		\end{itemize}
\end{theorem}
\begin{figure}[H]
	\centering
	\begin{tikzpicture}[scale = 2]
		\draw [->,densely dashed] (0,0)--(2.5,0) node[below right] { $a$};
		\draw [->] (0,0)--(0,1.3) node[above left] {$b$};
		\node[below left] at (0,0) {O};  		
		\fill (2,0.6) circle (0.4pt);  
		\fill (2.1,0.6) circle (0.4pt);  
		\fill (2.2,0.6) circle (0.4pt);  
		
		\draw [gray!40,fill=gray!40] (0,1.098)--(1.098,0)--(1.098,1.386-1.098)--(1.386,0)--(1.386,1.609-1.386)--(1.609,0)--(1.609,1.945-1.609)--(1.945,0)--(1.945,1.2)--(0.006,1.2)--(0.006,0.693);
		\draw[black] (0,1.098)--(1.098,0)--(1.098,1.386-1.098)--(1.386,0)--(1.386,1.609-1.386)--(1.609,0)--(1.609,1.945-1.609)--(1.945,0)--(1.945,2.079-1.945);
		\foreach \x in{0,1.098,1.386,1.609,1.945}
		{
			\draw[fill] (\x,0) circle (.02 );  
		}
		\draw[fill] (0,1.098) circle (.02); 
		\node[below,font=\fontsize{7}{6}\selectfont] at (1.098,0) {log3};  
		\node[below,font=\fontsize{7}{6}\selectfont] at (1.386,0) {log4};
		\node[below,font=\fontsize{7}{6}\selectfont] at (1.65,0) {log5};
		\node[below,font=\fontsize{7}{6}\selectfont] at (1.945,0) {log7};
		\node[left,font=\fontsize{7}{6}\selectfont] at (0,1.098) {log3};		
	\end{tikzpicture}
	\captionsetup{justification=centering}
	\caption{The face $(U_{2,4},U_{1,3}^{123})$}
	\label{fig8}
\end{figure}
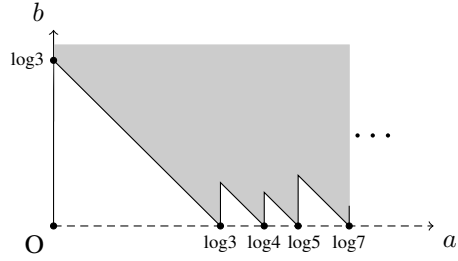

\afterpage{
	\clearpage
	\begin{longtable}{c|c|c|c}
         \caption{Entropy functions on two-dimensional faces of $\Gamma_4$ }
         \label{table:2} \\ 
		\hline  Theorem & Two-dimensional face $F$ &Entropy region $F^*=F\cap \Gamma^*_4$ & Figure \\
		\hline
		\hline
		\endfirsthead
		\hline  Theorem & Two-dimensional face $F$ &Entropy region $F^*=F\cap \Gamma^*_4$ & Figure \\
		\hline
		\endhead
        	   Part 1.Thm.IV-A.1   & \makecell{ \\ $(U^{1}_{1,1},U^{2}_{1,1})$, $(U^{12}_{1,2},U^{1}_{1,1})$, \\$(U^{12}_{1,2},U^{3}_{1,1})$,  $(U^{12}_{1,2},U^{13}_{1,2})$,\\ $(U^{12}_{1,2},U^{34}_{1,2})$, $(U^{123}_{1,3},U^{1}_{1,1})$, \\ $(U^{123}_{1,3},U^{4}_{1,1})$, $(U^{123}_{1,3},U^{12}_{1,2})$,\\ $(U^{123}_{1,3},U^{14}_{1,2})$,  $(U^{123}_{1,3},U^{124}_{1,3})$,\\$(U_{1,4},U^{1}_{1,1})$, $(U_{1,4},U^{12}_{1,2})$, \\ $(U_{1,4},U^{123}_{1,3})$.} &\makecell{ $\{a\br_1+b\br_2:$\\$a\geq 0, b\geq0\}$ }  & \makecell[l]{\\ \includegraphics[trim=10 10 0 10,scale=0.9]{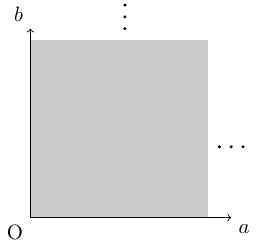}}  \\
		& &     & \\
		\hline
		 Part 1.Thm.IV-B.1 &\makecell{$(U^{123}_{2,3},U^{12}_{1,2})$,\\ $(\mc{W}_{2}^{34},U_{1,2}^{12})$,\\ $(\mc{W}_{2}^{14},U_{1,3}^{124})$. } &\makecell{ $\{a\br_1+b\br_2:$\\$ a+b\geq \log v   $ and \\ $ \log(v-1)\leq a \leq \log v$  \\
			for positive integer $ v$\} }& \makecell[l]{\includegraphics[trim=10 0 0 0,scale=0.9]{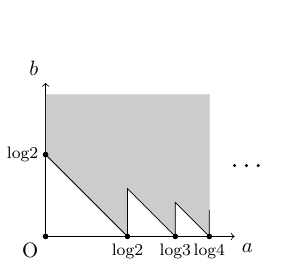}}
		\\
		\hline
		\makecell{\\  Part 1.Thm.IV-B.2} & \makecell{ \\ $(U_{2,3}^{123},U_{1,1}^{1})$,$(U_{2,3}^{123},U_{1,1}^{4})$, \\$,
			(U_{2,3}^{123},U_{1,2}^{14})$, $  (\mc{W}_{2}^{14},U_{1,1}^{1})$,\\$
			(\mc{W}_{2}^{34},U_{1,1}^{1}), (\mc{W}_{2}^{14},U_{1,2}^{14})$,\\
			$(\mc{W}_{2}^{24},U_{1,2}^{14})$.}  &\makecell{$\{a\br_1+b\br_2:$\\$a=\log v$ for \\some positive \\integer $v$, $b\geq 0$\}  }& \makecell[l]{\includegraphics[trim=3 0 0 0,scale=1]{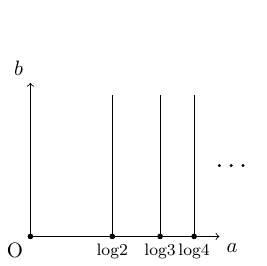}} \\
		\hline
		\makecell{\\  Part 1.Thm.IV-B.3 } &\makecell{ $(\hat{U}_{2,5}^{1},U_{1,1}^{1})$,\\ $(\hat{U}_{2,5}^{1},U_{1,1}^{2})$, \\ $(\hat{U}_{2,5}^{1},U_{1,2}^{12})$ }&\makecell{$\{a\br_1+b\br_2:$\\$a=\log v$ for \\some positive \\integer $v$, $b\geq 0$\}  }& \makecell[l]{\includegraphics[trim=3 0 0 0,scale=0.9]{fig_rk4.pdf}} 
		\\         
		\hline
		 \makecell{ Part 1.Thm.IV-B.4 }& \makecell{$(\hat{U}_{2,5}^{1},U_{1,3}^{123})$} &\makecell{ $\{a\br_1+b\br_2:$\\$ a+b\geq \log v   $ and \\ $ \log(v-1)\leq a \leq \log v$  \\
			for positive integer $ v$\} }& \makecell[l]{\includegraphics[trim=10 0 0 0,scale=0.9]{fig_rk2.pdf}} \\ 
		\hline
		 \makecell{ Part 1.Thm.IV-B.5 }& \makecell{$(\hat{U}_{3,5}^{1},U_{1,1}^{1})$, \\ $(\hat{U}_{3,5}^{1},U_{1,1}^{2})$, \\  $(\hat{U}_{3,5}^{1},U_{1,2}^{12})$} &\makecell{$\{a\br_1+b\br_2:$\\$a=\log v$ for \\some positive \\integer $v$, $b\geq 0$\}  }& \makecell[l]{\includegraphics[trim=3 0 0 0,scale=0.8]{fig_rk4.pdf}}\\
		\hline
	 \makecell{ Part 1.Thm.IV-C.1} &\makecell{ $( U_{2,3}^{123} , U_{1,3}^{124})$ }& \makecell{$\{a\br_1+b\br_2:$\\$a=\log v$ for \\some positive \\integer $v$, $b\geq 0$\}  }  &\makecell[l]{\includegraphics[trim=3 0 0 0,scale=1]{fig_rk4.pdf}} \\
		\hline
		 \makecell{Part 1.Thm.IV-C.2} & \makecell{$(U_{2,3}^{123  } , U_{2,3}^{124  } ) $}& \makecell{$\{a\br_1+b\br_2:$\\$a=\log v_1, b=\log v_2$    \\for some positive\\ integer $v_1$, $v_2$\} }& \makecell[l]{\\ \includegraphics[trim=10 0 0 0,scale=1]{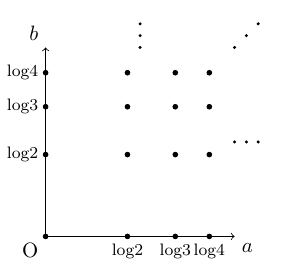}} \\
		\hline
		 \makecell{ Part 1.Thm.IV-C.3} & \makecell{$  ( U_{2,3}^{123} , U_{1,4}  ) $}  &\makecell{ $\{a\br_1+b\br_2:$\\$a\geq 0,b>0$ or\\ $(a,b)=(\log{v},0)$\\
			for  positive \\integer $k$\} } &  \makecell[l]{\\ \includegraphics[trim=5 0 0 0,scale=1]{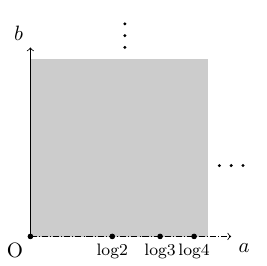}}\\
		\hline
		 \makecell{ Part 1.Thm.IV-D.1} &\makecell{$ ( \mc{W}_{2}^{12  } , U_{2,3}^{134  })$ }&\makecell{ $\{a\br_1+b\br_2:$\\$ a+b=\log{v}$,
			\\$a=H(\bm{\alpha})
			$, where \\ integer $v\ge 1$ and\\ $\bm{\alpha}$ is
			a  partition of $v$\}   } &  \makecell[l]{\\ \includegraphics[trim=10 0 0 0,scale=0.77]{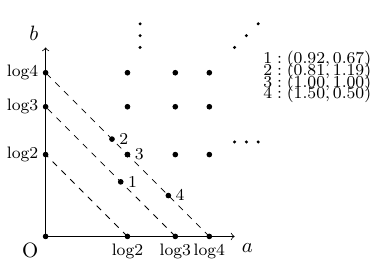}}\\
		\hline
		 \makecell{ Part 1.Thm.IV-E.1} &$(  U_{3,4}  ,U^1_{1,1} ) $&\makecell{$\{a\br_1+b\br_2:$\\$a=\log v$ for \\some positive \\integer $v$, $b\geq 0$\}  } & \makecell[l]{\includegraphics[trim=3 0 0 0,scale=1]{fig_rk4.pdf}}  \\
		\hline
		  \makecell{ Part 1.Thm.IV-E.2} &$(  U_{3,4}  ,U^{12}_{1,2} )$ & \makecell{$\{a\br_1+b\br_2:$\\$a=\log v$ for \\some positive \\integer $v$, $b\geq 0$\}  } &\makecell[l]{\includegraphics[trim=3 0 0 0,scale=1]{fig_rk4.pdf}}  \\
		\hline
		 \makecell{ Part 1.Thm.IV-E.3} &$(  U_{3,4}  ,U^{123}_{2,3} )$ & \makecell{$\{a\br_1+b\br_2:$\\$a+b=\log v$ for \\some positive\\ integer $v$\}  } & \makecell[l]{\\ \includegraphics[trim=10 0 0 0,scale=1]{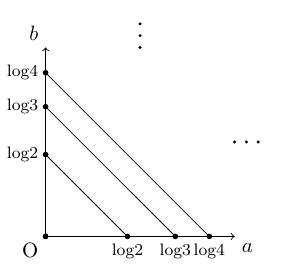}}  \\
		\hline
		  \makecell{ Part 1.Thm.IV-E.4} &$(  U_{3,4}  ,U^{123}_{1,3} )$ &\makecell{$\{a\br_1+b\br_2:$\\$a=\log v$ for \\some positive\\ integer $v$, $b\geq 0$\}  } &\makecell[l]{\includegraphics[trim=3 0 0 0,scale=1]{fig_rk4.pdf}}  \\
		\hline
		 \makecell{ Part 1.Thm.IV-E.5} &$(  U_{3,4}  ,U_{1,4} )$ &\makecell{$\{a\br_1+b\br_2:$\\$a\geq 0,b>0$ or\\ $(a,b)=(\log{v},0)$\\
			for  positive integer $v$\} } & \makecell[l]{\\ \includegraphics[trim=5 0 0 0,scale=1]{fig_rk7.pdf}}\\
		\hline
		  \makecell{ Part 1.Thm.IV-F.1} &$(\hat{U}_{3,5}^{1},U_{1,3}^{234})$ &\makecell{$\{a\br_1+b\br_2:$\\$a=\log v$ for \\some positive \\ integer $v$, $b\geq 0$\}  } & \makecell[l]{\includegraphics[trim=3 0 0 20,scale=1]{fig_rk4.pdf}}  \\
		\hline
		 \makecell{ Part 1.Thm.IV-F.2} &$ (\hat{U}_{3,5}^{1},U_{2,3}^{123})$ &\makecell{$\{a\br_1+b\br_2:$\\$ a=\log v_1, b=\log v_2$\\ for some positive \\integer $v_1$, $v_2$\} }  & \makecell[l]{\\ \includegraphics[trim=0 0 0 10,scale=1]{fig_rk5.pdf}} \\
		\hline
		 \makecell{Part 1.Thm.IV-F.3} &$(\hat{U}_{3,5}^{1},\mc{W}_{2}^{23})$ & \makecell{ $\{a\br_1+b\br_2:$\\$a+b=\log{v}$ for \\some integer $v\ge 1$\} }& \makecell[l]{\\ \includegraphics[trim=0 0 0 0,scale=1]{fig_rk10.pdf}}   \\
		\hline 
		 \makecell{ Part 1.Thm.IV-G.1}&  \makecell{\\ $(V^{12}_8, U^{1}_{1,1})$, $(V^{12}_8,  U^{3}_{1,1})$, \\$(V^{12}_8,  U^{12}_{1,2})$, $(V^{12}_8,  U^{134}_{1,3})$, \\$(V^{12}_8,  U_{1,4})$ \\ \quad}  &\makecell{ $\{a\br_1+b\br_2: $\\$ a= 0, b\geq 0$}  &\makecell[l]{\\ \includegraphics[trim=0 0 0 0,scale=1]{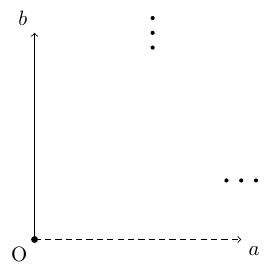}} \\
		\hline
		\makecell{ Part 1.Thm.IV-G.1}&  \makecell{\\ $(V^{12}_8,  U^{123}_{2,3})$,\\ $(V^{12}_8,  U_{3,4})$ \\ \quad}  &\makecell{ $\{a\br_1+b\br_2: $\\$ a= 0, b=\log v$ for\\ some integer $v\ge 1\}$ }  &\makecell[l]{\\ \includegraphics[trim=0 0 0 0,scale=1]{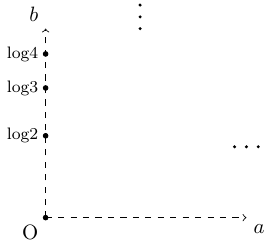}} \\
        \hline
		  \makecell{Part 2.Thm.\ref{rk13}} &\makecell{$ (  U_{2,4}  ,U^{123}_{1,3} )$ }&  \makecell{ $\{a\br_1+b\br_2:$\\$a+b\geq \log v $\\ and  $ \log(v-1)< a \leq \log v $\\ for positive integer $  v\neq 2,6\}$; or\\
		 	$a+b\geq \log{(v+1)} $ \\and  $\log(v-1)< a \leq \log v $ \\ for $ v = 2,6.$}
		 &  \makecell[l]{\\ \includegraphics[trim=5 0 0 0,scale=0.9]{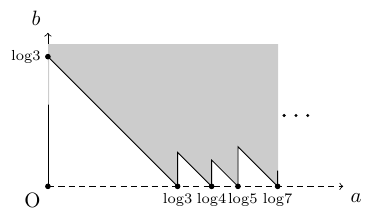}} \\
	 \hline 
	 \makecell{Part 2.Thm.\ref{rk15}} & \makecell{$(  U_{2,4}  ,U^{12}_{1,2} )$} & \makecell{$\{a\br_1+b\br_2:$\\$a=\log v$ for\\  positive integer $v\neq 2,6$; or\\
		 $a=\log2$, $b\geq \log2$; or\\
		$a=\log6$, $b\geq \log2.\}$} &\makecell[l]{\\ \includegraphics[trim=15 0 0 0,scale=0.9]{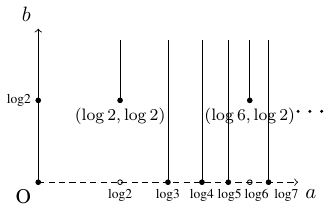}} \\
	\hline 
	\makecell{Part 2.Thm.\ref{rk25}} &$(U_{2,4}, U^{4}_{1,1})$ & \makecell{  $\{a\br_1+b\br_2:$\\$a=\log v$ \\for integer $v\neq 2,6$ or\\  $a=\log6,b\geq \log2\} \subseteq F^{*}$   and\\ 
		$\{a\br_1+b\br_2: a\neq \log k$  \\for some integer $k>0$ or\\ $a=\log2\}\cap F^{*}=\emptyset$  }& \makecell[l]{\\ \includegraphics[trim=5 0 0 5,scale=0.9]{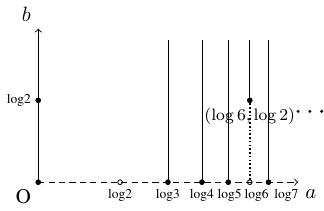}} \\
        \hline
         \makecell{ Part 2.Thm.\ref{rk26}} & \makecell{  $  (U_{2,4},  \mc{W}^{12}_{2}) $} & \makecell{ $\{a\br_1+b\br_2:a+b=\log v$\\ for integer $v>0$, and\\ there  exists a  $v^2\times 4 $ \\array $\mb{T}$ such that \\$\mb{T}(1,3,4)$ and $\mb{T}(2,3,4)$ are \\$\voa(U_{2,3},v)$, and \\$a=H(\bm\alpha) -\log v $, where \\ $\bm\alpha=(\alpha_{x_1,x_2}>0:x_1,x_2\in \mathbb{I}_v)$\\ and $\alpha_{x_1,x_2}$ denotes the times\\ of the row $(x_1,x_2)$ \\that occurs in $\mb{T}(1,2)$\} } & \makecell[l]{\\ \includegraphics[trim=5 0 0 0,scale=0.9]{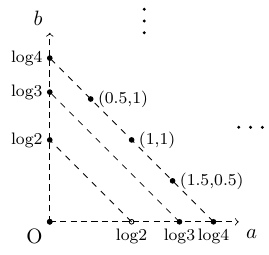}}\\
        \hline
         \makecell{Part 2.Thm.\ref{rk14}} & \makecell{$(  U_{2,4}  ,U^{123}_{2,3} ) $} &\makecell{ $\{a\br_1+b\br_2:$\\ $a+b=\log{v},
        	a=H(\bm{\alpha})
        	$ and\\ $(a,b)\neq (\log 2,0),(\log 6,0)$, \\ where integer $v>0$ and \\$\bm{\alpha}$ is
        	a  partition of $v$\} }&  \makecell[l]{\\ \includegraphics[trim=5 0 0 0,scale=0.9]{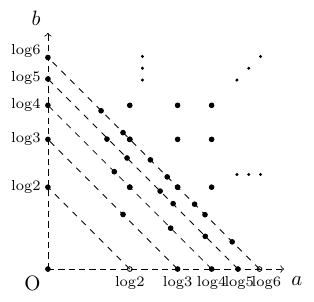}}\\
        \hline
          \makecell{ Part 2.Thm.\ref{rk20}} & \makecell{  $  (\hat{U}_{2,5}^{1},\mc{W}_{2}^{12}) $}  & \makecell{$\{a\br_1+b\br_2:$ \\ $a+b= \log v$\\ for some positive $v$  and\\  $a=\frac{1}{v}\sum_{i=0}^{v-1} H(\bm{\alpha_i})$, where\\ $ \bm{\alpha_i}\in\mathcal{P}(v),i\in \mathbb{I}_v$\} } & \makecell[l]{\\ \includegraphics[trim=5 0 0 0,scale=1]{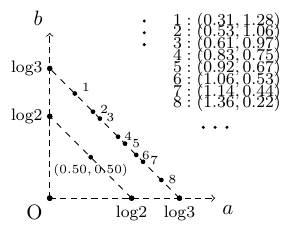}} \\
        \hline
         \makecell{Part 2.Thm.\ref{rk28}} &$(\mc{W}_{2}^{12}, \mc{W}_{2}^{13})$&\makecell{$\{a\br_1+b\br_2:$\\ there exists  a uniform \\ decomposition $\{ \mb{T}_0,\dots,\mb{T}_{v-1}\}$ \\of  $\voa(U_{2,3},v)$ $\mb{T}$ such that \\
                 	$a =\log v -\frac{1}{k}\sum_{i=0}^{v-1}\log |B_i|$,\\
                 	$ b =\log v -\frac{1}{v}\sum_{i=0}^{v-1}\log |A_i|$, \\
                 	where the subarray $\mb{T}_i$ of $\mb{T}$ are\\ induced by $A_i$ and $B_i$ \\ for $i\in \mathbb{I}_v\}$ } & \makecell[l]{\\ \includegraphics[trim=5 0 0 0,scale=1.0]{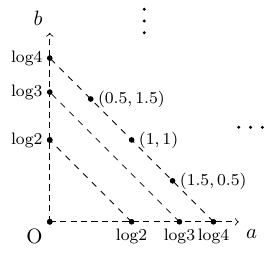}} \\ 
                 \hline
                   \makecell{Part 2.Thm.\ref{rk27}}&   $(\hat{U}_{2,5}^{1},  U^{234}_{2,3})$& \makecell{ $\{a\br_1+b\br_2:$\\$ a+b=\log v$ for some positive $v$ \\and there exists a suborder\\ decomposition $\{\mb{T}_0,\mb{T}_1,\dots,$\\$\mb{T}_{t-1}\}$ of a $\voa(U_{2,3},v)$  $\mb{T}$ \\such that \\
                 	$a=\frac{1}{2} H(\dfrac{|A_i|^2}{v^2}:i\in \mathbb{I}_v)$,\\
                 	where  subarray $\mb{T}_i$ of $\mb{T}$\\ are induced by $A_i$ and $B_i$\\ for $i\in \mathbb{I}_v$\} }  & \makecell[l]{\\ \includegraphics[trim=5 0 0 0,scale=1]{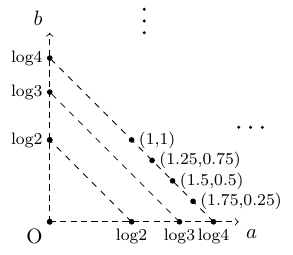}}    \\ 
                \hline
                   \makecell{Part 2.Thm.\ref{rk29}}&  $(\hat{U}_{2,5}^{1},  U_{2,4})$  & \makecell{ $\{a\br_1+b\br_2:$\\$ a+b=\log v$  and\\ there exists a $\{1\}$-partial\\
                 	$\voa(U_{2,3},v)$ $\mb{T}$ such that \\
                 	$a=H(\dfrac{\alpha_0}{v^2},\dfrac{\alpha_1}{v^2},\dots,$\\ $\dfrac{\alpha_{t-1}}{v^2})-\log v$, \\
                 	where $\alpha_i$ denotes the times\\ of the entry $i\in\mathbb{I}_v$ that\\ occurs in $\mb{T}_1$\} }  & \makecell[l]{\\ \includegraphics[trim=0 0 0 0,scale=0.9]{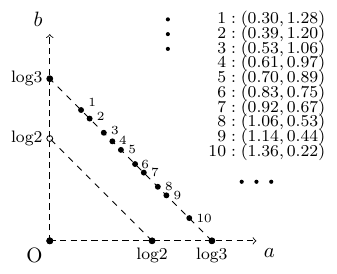}}  \\
                  \hline
                 \makecell{Part 2.Thm.\ref{rk30}} &$(\hat{U}_{3,5}^{4},  U_{2,4})$ & \makecell{ 	 
                 	$\{a\br_1+b\br_2:$\\$a+b=\log v$ for\\ integer $v\neq2,6$;\\
                 	$(a,b)=(\log2,0)$; or\\
                 	$a+b=\log6,a\geq\log2$\}$\subseteq F^{*}$ \\and
                 	$\{a\br_1+b\br_2:$\\
                 	$a+b\neq \log v$ \\for some integer $v>0$;\\
                 	$a+b=\log2$, $a<\log2$; or\\
                 	$(a,b)=(0,\log6)$\} $\cap F^{*}=\emptyset$. 
                 }& \makecell[l]{\\ \includegraphics[trim=5 0 0 5,scale=1.0]{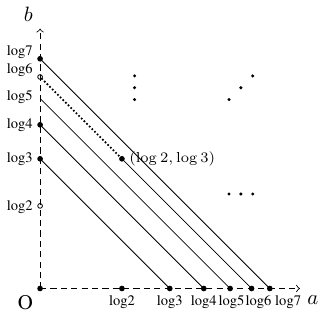}}  \\
                  \hline
	\end{longtable} }
 \begin{proof}
	If $\mathbf{ h} \in  F $ is entropic, its characterizing random vector
	$(X_i,i\in N_4)$ satisfies the following information equalities,
	\begin{align}
		H(  X_{N_4})&=H(X_{N_4-i})  ,\ i\in N_4, \nonumber \\
		H(X_{i4 })&=H(X_{i})+H(X_{4}) ,   i \in \{1,2,3\},  \nonumber \\
		H(X_{i\cup K})+H(X_{  j\cup K})&=H(X_{ K})+H(X_{  ij
			\cup K }), |K|=2.   \nonumber
	\end{align}
  For $ (x_i,i\in N_4) \in
	\mathcal{X}_{N_4}$ with $p(x_{1234})>0$, above information equalities
	imply that the probability mass function satisfies
	\begin{align}
		p(x_{1234})&=p(x_{1},x_{2},x_{3}) \label{c350}\\
		&=p(x_{1},x_{2},x_{4}) \label{c351} \\ 
		&=p(x_{1},x_{3},x_{4}) \label{c352}\\
		&= p(x_{2},x_{3},x_{4}), \label{c353}\\
		p(x_1,x_4)&=p(x_1)p(x_4), \label{c354}\\
		p(x_2,x_4)&=p(x_2)p(x_4), \label{c355}\\
		p(x_3,x_4)&=p(x_3)p(x_4), \label{c356}\\
		p(x_{1},x_{2},x_{3})p(x_{1},x_{2},x_{4})&=p(x_{1},x_{2})p(x_{1234}),\label{c357}\\
		p(x_{1},x_{2},x_{3})p(x_{1},x_{3},x_{4})&=p(x_{1},x_{3})p(x_{1234}),\label{c358}\\
		p(x_{1},x_{2},x_{3})p(x_{1},x_{3},x_{4})&=p(x_{1},x_{3})p(x_{1234}),\label{c359}
		\\
		p(x_{1},x_{2},x_{3})p(x_{2},x_{3},x_{4})&=p(x_{2},x_{3})p(x_{1234}),\label{c360}\\
		p(x_{1},x_{2},x_{4})p(x_{2},x_{3},x_{4})&=p(x_{2},x_{4})p(x_{1234}),\label{c361}\\
		p(x_{1},x_{3},x_{4})p(x_{2},x_{3},x_{4})&=p(x_{3},x_{4})p(x_{1234}).\label{c362}
	\end{align}
	By \eqref{c350}-\eqref{c353} and \eqref{c357}-\eqref{c362}, we have
	\begin{align}
		&p(x_{1234})=p(x_1,x_2,x_3)=p(x_1,x_2,x_4) \label{c363} \\ 
		&=p(x_1,x_3,x_4)=p(x_2,x_3,x_4)   \label{c364}\\
		&=p(x_1,x_2)=p(x_1,x_3)=p(x_1,x_4)   \label{c365}\\
		&=p(x_2,x_3)=p(x_2,x_4)=p(x_3,x_4).\label{c366}
	\end{align}
	Then by \eqref{c354}-\eqref{c356} and \eqref{c365}-\eqref{c366}, we obtain
	\begin{align}
		p(x_1)p(x_4)=p(x_2)p(x_4)=p(x_3)p(x_4).\label{c367}
	\end{align}
	Canceling $p(x_4)$ in the above equation,
	\begin{equation}
		p(x_1)=p(x_2)=p(x_3).  \label{c368}
	\end{equation}
	
	Let $\h=a\br_1+b\br_2$, where $\br_1$ and $\br_2$ are the rank functions of the matroids on
	the two extreme rays of the face, respectively. Restricting $\h$ on
	$\{1,2,4\}$, we obtain $\h'=a\br'_1+b\br'_2$, where $\br'_i, i=1,2$ are the restriction
	of $\br_i$ on $\{1,2,4\}$. It can be checked that they are 
	the rank functions of $U_{2,3}$ and $U^{12,3}_{1,2}$,
	respectively. Thus, $\h'\in (U_{2,3}, U^{12,3}_{1,2})$ and 
	\begin{equation}
          a+b \ge \log \lceil 2^a \rceil
	\end{equation}
 by \cite[Theorem 1]{matus2005piecewise}, which is an outer bound on the entropy region of $F$. 
 
 Now we show that 
	\begin{itemize}
	\item  $a+b\geq \log v   $ and  $ \log(v-1)< a \leq \log v $ for positive integer $  v\neq 2,6$; or
	\item  $a+b\geq \log{(v+1)} $ and  $\log(v-1)< a \leq \log v $  for $ v= 2,6$
\end{itemize}
  form an inner bound on the entropy region on $F$. Let $\mb{T}$ be a $\voa(U_{2,4},v)$. Let $X_4$ be distributed on  $\mathbb{I}_v$ such that $H(X_4)=a$. Let $(X_i,i\in N_4)$ be distributed on the rows of $\mb{T}$ with $p(x_1,x_2,x_3,x_4)=\frac{p(x_4)}{v}$. It can be checked that the entropy function of such constructed $(X_i,i\in N_4)$ is $(a,\log v-a)$. Then by Lemma \ref{lem}, all $\h$ in the inner bound are entropic.
	\begin{figure}[H]
		\centering
		\begin{tikzpicture}[scale = 2]
			\draw [->,,densely dashed] (0,0)--(2.5,0) node[below right] { $a$};
			\draw [->] (0,0)--(0,1.3) node[above left] {$b$};
			\node[below left] at (0,0) {O};  		
			\fill (2,0.6) circle (0.4pt);  
			\fill (2.1,0.6) circle (0.4pt);  
			\fill (2.2,0.6) circle (0.4pt);  
			
			\draw [black,fill=green,pattern=north east lines] (0,0.693)--(0.693,0)--(0.693,1.098-0.693)--(1.098,0)--(1.098,1.386-1.098)--(1.386,0)--(1.386,1.609-1.386)--(1.609,0)--(1.609,1.791-1.609)--(1.791,0)--(1.791,1.945-1.791)--(1.945,0)--(1.945,1.2)--(0.004,1.2)--(0.004,0.693);

			\draw [gray!40,fill=gray!40] (0,1.098)--(1.098,0)--(1.098,1.386-1.098)--(1.386,0)--(1.386,1.609-1.386)--(1.609,0)--(1.609,1.945-1.609)--(1.945,0)--(1.945,1.2)--(0.006,1.2)--(0.006,1.386);
			\draw[black] (0,1.098)--(1.098,0)--(1.098,1.386-1.098)--(1.386,0)--(1.386,1.609-1.386)--(1.609,0)--(1.609,1.945-1.609)--(1.945,0)--(1.945,2.079-1.945);
			\foreach \x in{0,1.098,1.386,1.609,1.945}
			{
				\draw[fill] (\x,0) circle (.02 );  
			}
			\draw[fill] (0,1.098) circle (.02); 
			
			\node[below,font=\fontsize{7}{6}\selectfont] at (0.693,0) {log2}; 
			\node[below,font=\fontsize{7}{6}\selectfont] at (1.098,0) {log3};  
			\node[below,font=\fontsize{7}{6}\selectfont] at (1.386,0) {log4};
			\node[below,font=\fontsize{7}{6}\selectfont] at (1.65,0) {log5};
			\node[below,font=\fontsize{7}{6}\selectfont] at (1.945,0) {log7};
			\node[left,font=\fontsize{7}{6}\selectfont] at (0,1.098) {log3};
			\node[left,font=\fontsize{7}{6}\selectfont] at (0,0.693) {log2};		
		\end{tikzpicture}
		\captionsetup{justification=centering}
		\caption{Inner and outer bounds on $F^*$}
		\label{fig11}
	\end{figure}

	It can be seen that there exists a gap between the inner and outer bounds,
	\begin{equation}
		\log{v} \leq a+b < \log{(v+1)}
	\end{equation}
    and
    \begin{equation}
        \log{(v-1)} < a\leq \log{v} \text{ for } v=2,6,
    \end{equation}
which is the slash region in Fig. \ref{fig11}. In the following we
prove that polymatroids in this gap are all non-entropic.

	Consider the bipartite graph $G=(V,E)$ with $V=\mc{X}_1\cup \mc{X}_2$ and   $(x_1,x_2)\in E$  if and only if $p(x_{1},x_{2})>0$. By $p(x_1,x_2,x_4)=p(x_1,x_2)$, each edge can be colored by a unique $x_4$. By $p(x_1,x_2,x_4)=p(x_1,x_4)$,  any two edges incident to $x_1$ are
 colored differently. Due to \eqref{c354}, $X_1$ and  $X_4$ are independent, thus all colors will occur at least once on the edges incident to $x_1$. Hence, for each vertex $x_1$, it is incident to $k$ edges where 	$v=|\mathcal{X}_{4}|$. It holds for each $x_2$ as well by symmetry. We denote the number of the vertices of $\mc{X}_{i}$ in the connected component $C_{j}$ by $n_{i}^{(j)}, i=1,2, j=1,2,\cdots, t$  and  the  probability mass of $ C_{j}$ by $p_{j}$, that is, the  probability of the event that the random vector takes a tuple in $C_j$. In a connected component $C_{j}$,  the number of edges is $n_{1}^{(j)}v=n_{2}^{(j)}v$, which implies that $n_{1}^{(j)}=n_{2}^{(j)}$, and so it can be simplified to $n^{(j)}$. Since each vertex is incident to $v$ edges, we have $n^{(j)}\geq v$. As $ p(x_1,x_2)$ =$\sum_{x_3,x_4} p(x_1,x_2,x_3,x_4)$, there exist  $x_3,x_4$ such that $p(x_1,x_2,x_3,x_4)>0$. By \eqref{c368}, $p(x_1)=p(x_2)$, which implies that the probability mass of the two adjacent vertices are the same, and so are the vertices in a connected compoent as well. Since $p(x_1,x_2,x_3)=p(x_1,x_2)$, we color each $(x_1,x_2)$ by $x_3\in \mc{X}_3$. As $p(x_1)=p(x_2)=p(x_3)$ by \eqref{c368}, the probability mass of the color $x_3$ in a connected  component in $G$ are the same.
	We classify the connected components of $G$ into $t_1$ equivalence class such that for any two components in an equivalence class, they share a common color in $\mc{X}_3$. Let $A_i,i=1,2,\dots,t_1$ be index set of the components in each equivalence class $i$.
	 Thus, the probability mass of vertices in the same equivalence class in $G$ are equal. Then
	\begin{align}
		H(X_1)&=-\sum_{j=1}^{t_1} n^{(j)'} \dfrac{p'_j}{n^{(j)'}} \log \dfrac{p'_j}{n^{(j)'}}  \label{c372} \\
		&=H(p'_1,\dots,p'_{t_1}) +   \sum_{j=1}^{t_1}    p'_j\log{n^{(j)'}},                      \label{c373}
	\end{align}
	where $p'_j \triangleq  \sum_{i\in A_j}p_i$, $n^{(j)'}\triangleq  \sum_{i\in A_j }n^{(i)}$.

	As $\h\in F$ and $(X_i, i\in N_4)$ is
	its characterizing random vector, we have 
	\begin{align}
		H(X_1)&=H(X_2)=	H(X_3) \label{c369} \\
		&=a+b, \label{c370} \\
		H(X_4)&=a. \label{c371}
	\end{align}
	By \eqref{c373} and $n^{(j)}\geq v$,
	\begin{align}
		a+b&=H(p'_1,\dots,p'_{t_1}) +   \sum_{j=1}^{t_1}    p'_j\log{n^{(j)'}}  \label{c374}\\
		&\geq  H(p'_1,\dots,p'_{t_1}) +   \sum_{j=1}^{t_1}    p'_j\log v  \label{c375}\\
		&\geq \log v. \label{c376}
	\end{align}
	Note that $|\mathcal{X}_{4}|=v$, we have $a\leq \log v$. Then
	\begin{equation}
		a+b\geq \log v \geq a. \label{c377}
	\end{equation}
 Assume there exists one equivalence  class that contains only one connected component $C_j$ and satisfies $n^{(j)}=v$ for $v=2,6$. Then for the  connected component $C_j$, as $p(x_1)=p(x_2)=p(x_3)$, 
	\begin{equation}
		n_3^{(j)}=\dfrac{p'_j}{p(x_3)}=\dfrac{p_j}{p(x_1)}=n^{(j)},  \label{c378}
	\end{equation}
	where $n_3^{(j)}$ denotes the number of the colors $x_3$ in $\mathcal{X}_{3}$ in $C_j$.
	Let $\mb{T}$ be a  $v^2 \times 4$ array, and for each row of $\mb{T}$, the four entries correspond to the two ends of an edge $(x_1,x_2)$ in $C_j$, and the color in $\mathcal{X}_{3}$ and $\mathcal{X}_{4}$ of the edge, respectively.
	It is easy to check that both $\mb{T}(1,2,3)$ and $\mb{T}(1,2,4)$ are $\voa(U_{2,3},v)$s. Since $X_3$ and $X_4$ are independent, each pair $(x_3,x_4)\in  \mathbb{I}^2_v$ appears in $\mb{T}(3,4)$ as a row exactly once. Therefore, $\mb{T}$ is a
	$\voa(U_{2,4},v)$, which contradicts the non-existence of $\voa(U_{2,4},2)$ or $\voa(U_{2,3},6)$. Hence, each equivalence  class either contains multiple connected components or contains only one connected component satisfying $n^{(j)}>v$.
	 Thus,   
	\begin{equation}
		n^{(j)'}= \sum_{i\in A_j }n^{(i)} \geq v+1. \label{c379}
	\end{equation}
	By \eqref{c373} and \eqref{c370},
	\begin{align}
		a+b&=H(X_1)=H(p'_1,\dots,p'_{t_1}) +  \sum_{j=1}^{t_1}    p'_j\log{n^{(j)'}}  \label{c380}\\
		&\geq  H(p'_1,\dots,p'_{t_1}) +   \sum_{j=1}^{t_1}    p'_j\log{(v+1)}  \label{c381}\\
		&\geq \log{(v+1)} \label{c382},
	\end{align}
	which implies the gap is non-entropic.
\end{proof}

\begin{theorem}
	\label{rk15}
	For $F=(U_{2,4}, U^{12}_{1,2} )$, $\mathbf{h}=(a,b) \in F$ is entropic if and only if 
	\begin{itemize}
		\item  $a=\log v$ for  positive integer $v\neq 2,6$;
		\item $a=\log2$, $b\geq \log2$; or
		\item $a=\log6$, $b\geq \log2$.
	\end{itemize}
\end{theorem}

\begin{figure}[H]
    \centering\mrv{\includegraphics{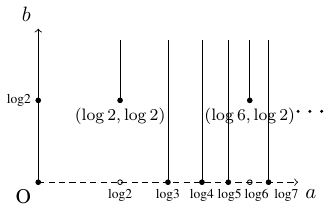}}
	\captionsetup{justification=centering}
	\caption{The face $( U_{2,4},U^{12}_{1,2})$}
	\label{fig10}
\end{figure}
 \begin{proof}
	If $\mathbf{ h} \in  F $ is entropic, its characterizing random vector
	$(X_i,i\in N_4)$ satisfies the following information equalities,
	\begin{align}
		H(  X_{N_4})&=H(X_{N_4-i})  ,\ i\in N_4 \nonumber \\
		H(X_{ij })&=H(X_{i})+H(X_{j}),\{i,j\}\neq \{1,2\}  \nonumber \\
		H(X_{i\cup K})+H(X_{  j\cup K})&=H(X_{ K})+H(X_{  ij
			\cup K }),\nonumber\\& \qquad\qquad\qquad |K|=2, K\neq\{3,4\}.   \nonumber
	\end{align}
  For $ (x_i,i\in N_4) \in
\mathcal{X}_{N_4}$ with $p(x_{1234})>0$, above information equalties
imply that the probability mass function satisfies
	\begin{align}
		p(x_{1234})&=p(x_{1},x_{2},x_{3}) \label{c200}\\
		&=p(x_{1},x_{2},x_{4}) \label{c201} \\ 
		&=p(x_{1},x_{3},x_{4}) \label{c202}\\
		&= p(x_{2},x_{3},x_{4}), \label{c203}\\
		p(x_1,x_3)&=p(x_1)p(x_3), \label{c204}\\
		p(x_1,x_4)&=p(x_1)p(x_4), \label{c205}\\
		p(x_2,x_3)&=p(x_2)p(x_3), \label{c206}\\
		p(x_2,x_4)&=p(x_2)p(x_4), \label{c207}\\
		p(x_3,x_4)&=p(x_4)p(x_4), \label{c208}\\
		p(x_{1},x_{2},x_{3})p(x_{1},x_{2},x_{4})&=p(x_{1},x_{2})p(x_{1234}),\label{c209}\\
		p(x_{1},x_{2},x_{3})p(x_{1},x_{3},x_{4})&=p(x_{1},x_{3})p(x_{1234}),\label{c210}\\
		p(x_{1},x_{2},x_{4})p(x_{1},x_{3},x_{4})&=p(x_{1},x_{4})p(x_{1234}).\label{c211}\\
		p(x_{1},x_{2},x_{3})p(x_{2},x_{3},x_{4})&=p(x_{2},x_{3})p(x_{1234}),\label{c212}\\
		p(x_{1},x_{2},x_{4})p(x_{2},x_{3},x_{4})&=p(x_{2},x_{4})p(x_{1234}).\label{c213}
	\end{align}
According to \eqref{c200}, canceling $p(x_1,x_2,x_3)$ and $p(x_1,x_2,x_3,x_4)$  on either side of \eqref{c209}, we have 
	\begin{equation}
		p(x_{1},x_{2},x_{4})=p(x_1,x_2).   \label{c214}
	\end{equation}
Together with \eqref{c201}, we obtain
     \begin{equation}
     	p(x_{1},x_{2},x_3,x_{4})=p(x_1,x_2).   \label{c800}
     \end{equation}
	By the same argument, we have
	\begin{align}
		p(x_{1234})&=p(x_1,x_3)=p(x_1,x_4) \label{c215} \\
		&=p(x_2,x_3)=p(x_2,x_4).   \label{c216}    
	\end{align} 
	By \eqref{c204} and \eqref{c205}, replacing $p(x_1,x_3)$ and $p(x_1,x_4)$ by $p(x_1)p(x_3)$ and $p(x_1)p(x_4)$  in \eqref{c215}, we obtain
	\begin{equation}
		p(x_3)=p(x_4) . \label{c217}
	\end{equation}
	Since $X_3$ and $X_4$ are independent, by Lemma \ref{lem1}, $X_3$ and $X_4$ are uniformly distributed on $\mathcal{X}_{3}$ and $\mathcal{X}_{4}$, respectively, and $H(X_3)=H(X_4)=\log v$ where $v=|\mathcal{X}_{3}|=|\mathcal{X}_{4}|$.
  As $\h\in F$ and $(X_i, i\in N_4)$ is
	its characterizing random vector, we have 
	\begin{align}
		H(X_3)&=H(X_4)=a, \label{c219}  
	\end{align}
	which implies that $a$ can only take the value of $\log v$. By
        Lemma \ref{lem} and the fact that $a=\log v$ for $v\neq 2,6$
        on the ray $U_{2,4}$ and the whole ray $U^{12}_{1,2}$ are entropic, all $\h=(a,b)\in F$ are entropic when $a=\log v $ for positive integer $v\neq 2,6$, and $b\geq 0$.
	
	Now, we show that when $a=\log 2$ or $\log 6$, 
	\begin{align}
	 b\ge \log 2.\label{l1}
	\end{align}
	Consider the bipartite graph $G=(V,E)$ with $V=\mc{X}_1\cup
        \mc{X}_2$ and  $(x_1,x_2)\in E$  if and only if
        $p(x_{1},x_{2})>0$. Assume $G$ has $t$ connected components
        and   $|\mathcal{X}_{3}|=|\mathcal{X}_{4}|=v$ for $v=2,6$. By
        \eqref{c200}, \eqref{c201} and \eqref{c214}, we have
        $p(x_1,x_2)=p(x_1,x_2,x_3)=p(x_1,x_2,x_4)$, which implies that
        $X_i$ is a function of $X_1$ and $X_2$, $i=3,4$. Then each
        edge $(x_1,x_2) $ can be colored by a unique
        $x_3\in\mathcal{X}_{3}$ and a unique $x_4\in\mathcal{X}_{4}$.
        By \eqref{c200} and \eqref{c215}, we have
        $p(x_1,x_3)=p(x_1,x_2,x_3)$. Thus, any two edges incident to
        $x_1$ are colored by different $x_3\in\mathcal{X}_{3}$. Since
        $X_1$ and $X_3$ are independent, all colors $x_3\in
        \mathcal{X}_{3} $ will occur at least once on the edges
        adjacent to $x_1$. Hence, each $x_1$ is incident to $k$
        edges. It holds for each $x_2$ as well by symmetry. We denote
        the number of the vertices in $\mc{X}_{i}$ by $n_{i}^{(j)}, i=1,2, j=1,2,\cdots,t$ in the connected component $C_{j}$ and  the  probability mass of $ C_{j}$ by $p_{j}$.
In each connected component  $C_{j}$, the number of edges is $n_{1}^{(j)}k=n_{2}^{(j)}k$, which implies $n_{1}^{(j)}=n_{2}^{(j)}$ and so it can be simplified to $n^{(j)}$. 
As $p(x_1,x_2)=\sum_{x_3,x_4} p(x_1,x_2,x_3,x_4)$,
there exist $x_3,x_4$ such that $p(x_1,x_2,x_3,x_4)>0$.  Due to \eqref{c215} and \eqref{c216}, we have $p(x_1,x_3)=p(x_2,x_3)$. By \eqref{c204} and \eqref{c206}, replacing $p(x_1,x_3)$ and $p(x_2,x_3)$ by $p(x_1)p(x_3)$ and $p(x_2)p(x_3)$, we obtain 
\begin{equation}
	p(x_1)=p(x_2),
\end{equation} 
which implies  the probability mass of two adjacent vertices are the same, and so are the vertices in a connected compoent as well. In each connected component $C_j$, the probability mass of the vertices  are equal to
\begin{equation}
	p(x_1)=p(x_2)=\dfrac{p_j}{n^{(j)}}.
\end{equation} 
By \eqref{c204}, \eqref{c800} and \eqref{c215}, the probability mass of the edges are equal to
	\begin{align}
		p(x_{12})&=p(x_{1234})=p(x_1,x_3)=p(x_1)p(x_3). \label{c223}   
	\end{align}
Note that $X_3$ is uniformly distributed on $\mathcal{X}_{3}$ and $|\mathcal{X}_{3}|=v$. Replacing $p(x_3)$ by $\frac{1}{v}$ in \eqref{c223}, we have
\begin{align}
	p(x_1,x_2)=p(x_1)p(x_3)=\dfrac{p(x_1)}{v}=\dfrac{p_j}{n^{(j)}v}.\label{c1000}  
\end{align}
 Hence,
	\begin{align}
		H(X_1)&=-\sum_{j=1}^{t} n^{(j)}\dfrac{p_j}{n^{(j)}} \log \dfrac{p_j}{n^{(j)}} \label{c224}\\
		&=H(p_1,...,p_t)+\sum_{j=1}^{t} p_j \log n^{(j)}. \label{c225} 
        \end{align}
    \begin{align}
		&H(X_1,X_2)=-\sum_{j=1}^{t} n^{(j)}v\dfrac{p_j}{n^{(j)}v}\log \dfrac{p_j}{n^{(j)}v} \label{c226}\\
		&=H(p_1,...,p_t)+\sum_{j=1}^{t} p_j \log n^{(j)}v \label{c227}\\
		&=H(p_1,...,p_t)+\sum_{j=1}^{t} p_j \log n^{(j)}+ \sum_{j=1}^{t} p_j \log v  \label{c228}\\
		&=H(p_1,...,p_t)+\sum_{j=1}^{t} p_j \log n^{(j)}+\log v.  \label{c229} 
	\end{align}
 As $\h\in F$ and $(X_i, i\in N_4)$ is
its characterizing random vector, we have 
\begin{align}
	H(X_1)&=a+b, \label{c750}  \\
 H(X_1,X_2)&=2a+b. \label{c751} 
\end{align}
	Equating the above equations with \eqref{c225} and \eqref{c229}, we obtain
	\begin{align}
		a&=\log v,  \\
		b&=H(p_1,...,p_t)+\sum_{j=1}^{t} p_j \log n^{(j)} -\log v. \label{c231}
	\end{align}
	Now we show that for each connected component, there exists an $(x_3, x_4)$ that occurs at least twice. By \eqref{c208}, $X_3$ and $X_4$ are independent. Since $|\mathcal{X}_{3}|=|\mathcal{X}_{4}|=v$, the number of the pairs $(x_3,x_4)$ is $v^2$. The probability mass of each $(x_3,x_4)$ is
	\begin{equation}
		p(x_3,x_4)=p(x_3)p(x_4)=\dfrac{1}{v^2}. \label{c232}
	\end{equation}
	 In each connected component $C_j$, when $n^{(j)}>v$, there are $n^{(j)}v$ edges.  Due to the pigeonhole principle, there  exists an $(x_3,x_4)$ that occurs at least twice.  When $n^{(j)}=v$, let $\mb{T}$ be a  $v^2 \times 4$ array, and for each row of $\mb{T}$, the four entries correspond to the two ends of an edge $(x_1,x_2)$ in $C_j$, and the colors in $\mathcal{X}_{3}$ and $\mathcal{X}_{4}$ of the edge, respectively.
	 It is easy to check that both $\mb{T}(1,2,3)$ and $\mb{T}(1,2,4)$ are $\voa(U_{2,3},v)$s. The non-existence of $\voa(U_{2,4}, 2)$ or $\voa(U_{2,4}, 6)$ implies that the number of different entries $(x_3,x_4)$ on the rows of $\mb{T}(3,4)$ is less than $v^2$, so there exists an $(x_3, x_4)$ that occurs at least twice.
	 
	
	 The probability mass of the pair $(x_3,x_4)$ that occurs at least twice in $C_j$ is
	\begin{equation}
		p(x_3,x_4)=\sum_{x_1,x_2} p(x_1,x_2,x_3,x_4). \label{c233} 
	\end{equation} 
  By \eqref{c223} and \eqref{c1000}, 
  \begin{equation}
  	p(x_1,x_2,x_3,x_4)=\dfrac{p_j}{n^{(j)}v},
  \end{equation}
which is independent of $(x_1,x_2)$ and so
	\begin{align}
		p(x_3,x_4)&=\sum_{x_1,x_2} p(x_1,x_2,x_3,x_4)\geq\dfrac{2p_j}{n^{(j)}v}.  \label{c236}
	\end{align}
Together with \eqref{c232}, we obtain
	\begin{align}
	\dfrac{1}{v^2}\geq 	\dfrac{2p_j}{n^{(j)}v},
	\end{align}
	which implies 
	\begin{equation}
		n^{(j)}\geq 2vp_j .  \label{c237}
	\end{equation}
	Substituting \eqref{c237} into \eqref{c231}, 
	\begin{align}
		b&=H(p_1,...,p_t)+\sum_{j=1}^{t} p_j \log n^{(j)} -\log v  \label{c238}\\
		&\geq H(p_1,...,p_t)+\sum_{j=1}^{t} p_j \log2vp_j  -\log v  \label{c239}\\
		&=H(p_1,...,p_t)-H(p_1,...,p_t)+\log2v -\log v  \label{c241}\\
		&=\log2.   \label{c242}
	\end{align}

        To prove the theorem, it remains to show that all $(a,b)\in
        F$ are entropic if $a=\log 2$ or $\log 6$, \rv{$b\ge \log2$}, which can
        be implied by the fact that polymatroids $\h=(\log2,\log2)$
        and $(\log6,\log2)$ are entropic. For $v=2,6$, let $\mb{T}_v$
        be a $2v^2\times 4$ array with entries in $ \mathbb{I}_{2v}$ such that
        \begin{itemize}
        \item each pair in $ \mathbb{I}^2_v\cup \mathbb{I}'^2_v$ occurs exactly
          once in $\mb{T}_v(1,2)$, where $\mathbb{I}'_v=\mathbb{I}_{2v}\setminus
          \mathbb{I}_{v}$;
       \item each pair in $\mathbb{I}^2_{v}$ occurs exactly twice in
         $\mb{T}_v(3,4)$; and
       \item $\mb{T}'_v(1,2,3)$,  $\mb{T}'_v(1,2,4)$,
         $\mb{T}''_v(1,2,3)$ and $\mb{T}''_v(1,2,4)$ are all
         $\voa(U_{2,3},v)$s, where $\mb{T}'_v$ is a $v^2\times 4$
         subarray of $\mb{T}_v$  formed by the rows with first two
         entries in $\mathbb{I}_{v}$, and  $\mb{T}''_v$ is a $v^2\times 4$
         subarray of $\mb{T}_v$ formed by the rows with first two
         entries in $\mathbb{I}'_{v}$. 
        \end{itemize}
        It can be seen that the following $\mb{T}_2$ and $\mb{T}_6$
        are such constructed.

			 \begin{equation*}
		\mb{T}_2\ = \
		\begin{matrix}
			0 & 0 & 0 & 0 \\
			0 & 1 & 1 & 1 \\     
			1 & 0 & 1 & 1 \\
			1 & 1 & 0 & 0 \\ 
			2 & 2 & 0 & 1 \\
			2 & 3 & 1 & 0 \\
			3 & 2 & 1 & 0 \\
			3 & 3 & 0 & 1 
		\end{matrix}
	\end{equation*} 
\begin{equation*}
\mrv{\mb{T}_6}
=
\begin{array}{@{}c@{\qquad\qquad}c@{}}
\mrv{\mb{T}'_6} & \mrv{\mb{T}''_6}\\[2pt]
\begin{matrix}
  0 & 0 & 0 & 0 \\
  0 & 1 & 1 & 1 \\
  0 & 2 & 2 & 2 \\
  0 & 3 & 3 & 3 \\
  0 & 4 & 4 & 4 \\
  0 & 5 & 5 & 5 \\
  1 & 0 & 5 & 1 \\
  1 & 1 & 0 & 2 \\
  1 & 2 & 1 & 3 \\
  1 & 3 & 2 & 4 \\
  1 & 4 & 3 & 5 \\
  1 & 5 & 4 & 0 \\
  2 & 0 & 4 & 5 \\
  2 & 1 & 5 & 0 \\
  2 & 2 & 0 & 1 \\
  2 & 3 & 1 & 2 \\
  2 & 4 & 2 & 3 \\
  2 & 5 & 3 & 4 \\
  3 & 0 & 3 & 3 \\
  3 & 1 & 4 & 4 \\
  3 & 2 & 5 & 5 \\
  3 & 3 & 0 & 0 \\
  3 & 4 & 1 & 1 \\
  3 & 5 & 2 & 2 \\
  4 & 0 & 2 & 4 \\
  4 & 1 & 3 & 5 \\
  4 & 2 & 4 & 0 \\
  4 & 3 & 5 & 1 \\
  4 & 4 & 0 & 2 \\
  4 & 5 & 1 & 3 \\
  5 & 0 & 1 & 2 \\
  5 & 1 & 2 & 3 \\
  5 & 2 & 3 & 4 \\
  5 & 3 & 4 & 5 \\
  5 & 4 & 5 & 0 \\
  5 & 5 & 0 & 1 \\
\end{matrix}
&
\begin{matrix}
  6 & 6  & 0 & 3 \\
  6 & 7  & 1 & 4 \\
  6 & 8  & 2 & 5 \\
  6 & 9  & 3 & 0 \\
  6 & 10 & 4 & 1 \\
  6 & 11 & 5 & 2 \\
  7 & 6  & 5 & 4 \\
  7 & 7  & 0 & 5 \\
  7 & 8  & 1 & 0 \\
  7 & 9  & 2 & 1 \\
  7 & 10 & 3 & 2 \\
  7 & 11 & 4 & 3 \\
  8 & 6  & 4 & 2 \\
  8 & 7  & 5 & 3 \\
  8 & 8  & 0 & 4 \\
  8 & 9  & 1 & 5 \\
  8 & 10 & 2 & 0 \\
  8 & 11 & 3 & 1 \\
  9 & 6  & 3 & 0 \\
  9 & 7  & 4 & 1 \\
  9 & 8  & 5 & 2 \\
  9 & 9  & 0 & 3 \\
  9 & 10 & 1 & 4 \\
  9 & 11 & 2 & 5 \\
  10& 6  & 2 & 1 \\
  10& 7  & 3 & 2 \\
  10& 8  & 4 & 3 \\
  10& 9  & 5 & 4 \\
  10& 10 & 0 & 5 \\
  10& 11 & 1 & 0 \\
  11& 6  & 1 & 5 \\
  11& 7  & 2 & 0 \\
  11& 8  & 3 & 1 \\
  11& 9  & 4 & 2 \\
  11& 10 & 5 & 3 \\
  11& 11 & 0 & 4 \\
\end{matrix}
\end{array}.
\end{equation*}
Due to page limitation, disjoint subarrays $\mb{T}'_6$ and
$\mb{T}''_6$ of $\mb{T}_6$  are
juxtaposed.
Let $(X_i,i\in N_4)$ be uniformly distributed on the rows of
$\mb{T}_v$, $v=2,6$.
It can be checked that for any nonempty $A\subseteq N_4$,
\begin{equation}
  \label{eq:3}
  H(X_A)=
  \begin{cases}
    \log v \quad &\text{ if } A=\{3\} \text{ or } \{4\}\\
    \log 2v \quad & \text{ if } A=\{1\} \text{ or }  \{2\}\\
    2\log v\quad &\text{ if } A=\{3,4\},\\
    2\log v+\log2 \quad &\text{ o.w.}
  \end{cases}
\end{equation}
It can be checked that the entropy function is in $F$.
Then by \eqref{c750}, \eqref{c751} and  \eqref{eq:3}, we have $a=\log
v$ and $b=\log 2$. The proof is accomplished.
\end{proof}





\begin{theorem}
	\label{rk25}
	For $F=(U_{2,4}, U^{4}_{1,1})$, $\h=(a,b)\in F$ is
	\begin{itemize}
		\item  entropic if $a=\log v$ for integer $v\neq 2,6$ or $a=\log6,b\geq \log2$; and
		\item  non-entropic if $a\neq \log v$ for some integer $v>0$ or $a=\log2$.
	\end{itemize}
\end{theorem}
\begin{figure}[H]
    \centering\mrv{\includegraphics{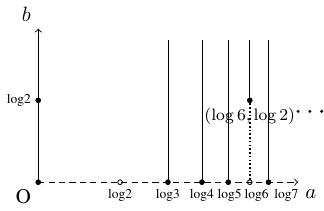}}
	\captionsetup{justification=centering}
	\caption{ The face $( U_{2,4}  ,U^{4}_{1,1})$ }
	\label{fig12}
\end{figure}
  \begin{proof}
	
	If $\mathbf{ h} \in  F$ is entropic, its characterizing random vector
	$(X_i,i\in N_4)$ satisfies the following information equalities,
	\begin{align}
		H(  X_{N_4})&=H(X_{N_4-i})  ,\ i\in \{1,2,3\}, \nonumber \\
		H(X_{ij })&=H(X_{i})+H(X_{j}),i,j\in N_4, \nonumber \\
		H(X_{i\cup K})+H(X_{  j\cup K})&=H(X_{ K})+H(X_{  ij
			\cup K }),\nonumber\\ &\qquad\qquad \quad |K|=2,   K\subseteq N_4.   \nonumber
	\end{align}
  For $ (x_i,i\in N_4) \in
\mathcal{X}_{N_4}$ with $p(x_{1234})>0$, above information equalties
imply that the probability mass function satisfies
	\begin{align}
		p(x_{1234})&=p(x_{1},x_{2},x_{4}) \label{c500}\\
		&=p(x_{1},x_{3},x_{4}) \label{c501} \\ 
		&=p(x_{2},x_{3},x_{4}), \label{c502}\\
		p(x_1,x_2)&=p(x_1)p(x_2), \label{c503}\\
		p(x_1,x_3)&=p(x_1)p(x_3), \label{c504}\\
		p(x_1,x_4)&=p(x_1)p(x_4), \label{c505}\\
		p(x_2,x_3)&=p(x_2)p(x_3), \label{c506}\\
		p(x_2,x_4)&=p(x_2)p(x_4), \label{c507}\\
		p(x_3,x_4)&=p(x_3)p(x_4), \label{c508}\\
		p(x_{1},x_{2},x_{3})p(x_{1},x_{2},x_{4})&=p(x_{1},x_{2})p(x_{1234}),\label{c509}\\
		p(x_{1},x_{2},x_{3})p(x_{1},x_{3},x_{4})&=p(x_{1},x_{3})p(x_{1234}),\label{c510} \\
		p(x_{1},x_{2},x_{4})p(x_{1},x_{3},x_{4})&=p(x_{1},x_{4})p(x_{1234})\label{c511}\\
		p(x_{1},x_{2},x_{3})p(x_{2},x_{3},x_{4})&=p(x_{2},x_{3})p(x_{1234}),\label{c512}\\
		p(x_{1},x_{2},x_{4})p(x_{2},x_{3},x_{4})&=p(x_{2},x_{4})p(x_{1234}),\label{c513} \\
		p(x_{1},x_{3},x_{4})p(x_{2},x_{3},x_{4})&=p(x_{3},x_{4})p(x_{1234}).\label{c514}
	\end{align}
By \eqref{c500}, canceling $p(x_1,x_2,x_4)$ and $p(x_1,x_2,x_3,x_4)$ on either side of \eqref{c511}, 
 we obtain
 \begin{equation}
 	p(x_1,x_3,x_4)=p(x_1,x_4).
 \end{equation}
Together with \eqref{c501}, we have 
\begin{equation}
 p(x_1,x_2,x_3,x_4)=p(x_1,x_4) \label{c1012}
\end{equation}
By the same argument, we obtain 
\begin{align}
	p(x_1,x_2,x_3,x_4)&=p(x_2,x_4), \label{c1013}\\
	p(x_1,x_2,x_3,x_4)&=p(x_3,x_4). \label{c1014}
\end{align}
	Restricting $\h$ on $M=\{1,2,3\}$, we obtain $\h'=a\br'$, where $\br'$ is the rank function of $U_{2,3}$ on $M$. 
	 Thus the characterizing random vector $(X_1,X_2,X_3)$ of $\h'$ is uniformly distributed on the rows of a $\voa(U_{2,3},v)$ $\mb{T}$ for a positive integer $v$, and so $a$ can only take the value of $\log v$. Note that
	 $a=\log v$, $v\neq 2,6$, on the ray $U_{2,4}$, and the whole ray $U_{1,1}^{4}$ are  entropic. 
	 By Lemma \ref{lem}, $\h=(\log v,b)$ is entropic for positive integers $v$ with $v\neq 2,6$ and $b\geq 0$.
	
	 Now we only need to consider $\h=(\log v,b)$ for $v=2,6$, $b\geq 0$.
%
Assume $\h=(\log2,b)$ is entropic. Note that up to isomorphism, there exists only one $\voa(U_{2,3},2)$.
 Without loss of generality, let the characterizing random vector $(X_1,X_2,X_3)$ of $\h'$ be uniformly distributed on the rows of $\mb{T}$ as follows:
		  \begin{equation*}
		 	\mb{T}\ =\ 
		 	\begin{matrix}
		 		0&0 & 0 \\
		 		0&1 & 1  \\     
		 		1&0 & 1 \\  
		 		1&1 & 0
		 		\end{matrix}
		 \end{equation*}
	 As $p_{X_1X_2X_3}(0,0,0)>0$, there exists $x_4\in\mathcal{X}_4$ such that $p_{X_1X_2X_3X_4}(0,0,0,x_4)>0$. 
	 Note that $X_1$ and $X_4$ are independent by \eqref{c505}, we obtain
	 \begin{equation}
	 	 p_{X_1X_4}(1,x_4)=p_{X_1}(1)p_{X_4}(x_4)>0,
	 \end{equation}
which implies that either $p_{X_1X_2X_3X_4}(1,0,1,x_4)>0$ or $p_{X_1X_2X_3X_4}(1,1,0,x_4)>0$. Since $p_{X_1X_2X_3X_4}(x_1,x_2,x_3,x_4)=p_{X_2X_4}(x_2,x_4)$ by \eqref{c1013}, 
\begin{align}
	p_{X_1X_2X_3X_4}(0,0,0,x_4)=p_{X_2X_4}(0,x_4)
\end{align}
However,
\begin{align}
	&p_{X_2X_4}(0,x_4)=\sum_{x_1,x_3}	p_{X_1X_2X_3X_4}(x_1,0,x_3,x_4) \\
	&=p_{X_1X_2X_3X_4}(0,0,0,x_4)+p_{X_1X_2X_3X_4}(1,0,1,x_4),
\end{align}
which implies $p_{X_1X_2X_3X_4}(1,0,1,x_4)=0$ contradicting $p_{X_1X_2X_3X_4}(1,0,1,x_4)>0$. Similarly, we can show that by \eqref{c1014}, $p_{X_1X_2X_3X_4}(1,1,0,x_4)>0$ will also lead a contradiction.
	 
	As for $\h=(\log6,b)$, we will show an inner bound on the entropy region within these polymatroids, i.e, those with $b\geq 0$.  
%
%
%
Let $\mb{T}'$ be the array as follows.

Written down within a page, the first to $36$th and $37$th to $72$th rows of $\mb{T'}$ are juxtaposed. Let $(X_i,i\in N_4)$ be uniformly distributed on the rows of $\mb{T}'$. Then it can be checked that such construction is $(\log6,\log2)$. Then by the fact that the whole ray $U^{4
}_{1,1}$ are entropic and Lemma  \ref{lem}, all $\h=(\log6,b)$ with $b\geq\log2$ are entropic.
\end{proof}

\noindent{\bf{Remark}} In this theorem, we give an inner bound on the
face $(U_{2,4}, U^{4}_{1,1})$. Entropy functions on this face
corresponds to a pair orthogonal squares, one is Latin and the other is
multi-symbol Latin. A square is called a \emph{multi-symbol Lain square
  of order $v$ with symbol set size $v'\ge v$} if it is a $v\times v$
square with a set of symbols with size $v'$ and each cell contains
one or more symbols, and each symbol appears in each row and each
column exactly once. Such kind of pair of squares can be obtained by
splitting the symbols of one square of a pair of orthogonal Latin
squares for $v\neq 2,6$. We proved that such a pair does not exist for
$v=2$. For $v=6$, we gave a pair with $v'=12$. We conjecture that this
inner bound is tight.

\subsection{\rv{Entropy functions on faces with both extreme rays containing
rank $2$ integer polymatroids}}
\label{rank2}
In this subsection, we characterize entropy functions on three
$2$-dimensional faces $(U_{2,4}, \mc{W}^{12}_{2})$, $(U_{2,4}, U^{123}_{2,3} )$, and
$(\hat{U}_{2,5}^{1},\mc{W}_{2}^{12})$ of $\Gamma_4$ with extreme rays
both containing rank $2$ integer
polymatroids. Some other faces in this family $(U_{2,3}^{123},
U_{2,3}^{124})$ and $(\mc{W}_{2}^{12},U_{2,3}^{134})$
have already been characterized in Part I of this \rv{series} of two
papers, while  $(\hat{U}_{2,5}^{1}, U^{234}_{2,3})$ will be
characterized in Subsection \ref{latindecom} as Latin square decomposition will be
used. 

			 \begin{equation*}
	\mb{T}'\ =\ 
	\begin{matrix}
0&0&0&0 \\
0&1&5&7 \\
0&2&3&2 \\
0&3&4&1 \\
0&4&2&4 \\
0&5&1&5 \\
1&0&1&8 \\
1&1&0&2 \\
1&2&4&5 \\
1&3&3&4 \\
1&4&5&0 \\
1&5&2&7 \\
2&0&2&2 \\
2&1&4&6 \\
2&2&1&0 \\
2&3&5&9 \\
2&4&3&5 \\
2&5&0&4 \\
3&0&3&1 \\
3&1&2&5 \\
3&2&5&4 \\
3&3&1&2 \\
3&4&0&6 \\
3&5&4&0 \\
4&0&4&4 \\
4&1&3&0 \\
4&2&2&1 \\
4&3&0&5 \\
4&4&1&3 \\
4&5&5&2 \\
5&0&5&5 \\
5&1&1&4 \\
5&2&0&7 \\
5&3&2&0 \\
5&4&4&2 \\
5&5&3&3 
\end{matrix}
\qquad\qquad
\begin{matrix}
0&0&0&9 \\
0&1&5&11 \\
0&2&3&8 \\
0&3&4&3 \\
0&4&2&10 \\
0&5&1&6 \\
1&0&1&10 \\
1&1&0&3 \\
1&2&4&9 \\
1&3&3&6 \\
1&4&5&1 \\
1&5&2&11 \\
2&0&2&3 \\
2&1&4&8 \\
2&2&1&11 \\
2&3&5&10 \\
2&4&3&7 \\
2&5&0&1 \\
3&0&3&11 \\
3&1&2&9 \\
3&2&5&3 \\
3&3&1&7 \\
3&4&0&8 \\
3&5&4&10 \\
4&0&4&7 \\
4&1&3&10 \\
4&2&2&6 \\
4&3&0&11 \\
4&4&1&9 \\
4&5&5&8 \\
5&0&5&6 \\
5&1&1&1 \\
5&2&0&10 \\
5&3&2&8 \\
5&4&4&11 \\
5&5&3&9 
\end{matrix}
\end{equation*}

\begin{theorem}
	\label{rk26}
	For $F=(U_{2,4},  \mc{W}^{12}_{2})$, $\h=(a,b)\in F$ is entropic if and only if $a+b=\log v$ for integer $v>0$, and there  exists a   $v^2\times 4 $ array $\mb{T}$ such that $\mb{T}(1,3,4)$ and $\mb{T}(2,3,4)$ are $\voa(U_{2,3},v)$, and
	\begin{equation}
		a=H(\bm\alpha) -\log v \nonumber,
	\end{equation} 
where  $\bm\alpha=(\alpha_{x_1,x_2}>0:x_1,x_2\in \mathbb{I}_v)$ and $\alpha_{x_1,x_2}$ denotes the times of the row $(x_1,x_2)$ that occurs in $\mb{T}(1,2)$.
\end{theorem}
\begin{figure}[H]
	\centering
	\begin{tikzpicture}[scale = 2]
		\draw [->,densely dashed] (0,0)--(1.6,0) node[below right] { $a$};
		\draw [->,densely dashed] (0,0)--(0,1.6) node[above left] {$b$};
		\foreach \y in{0.693,1.098,1.386}
		\draw [black, dashed,thin] (\y,0)--(0,\y);
		\fill (1.6,0.8) circle (0.4pt);  
		\fill (1.7,0.8) circle (0.4pt);  
		\fill (1.8,0.8) circle (0.4pt);  
		\fill (0.8,1.6) circle (0.4pt);  
		\fill (0.8,1.7) circle (0.4pt);  
		\fill (0.8,1.8) circle (0.4pt); 
		\node[below left] at (0,0) {O};   
		\foreach \x in{0,1.098,1.386}
		{\draw[fill] (\x,0) circle (.02);
		\draw[fill] (0,\x) circle (.02);}
		\draw (0.693,0) circle (.02);
		\draw[fill] (0,0.693) circle (.02);
		\draw[fill] (0.3466,1.0397) circle (.02);
		\draw[fill] (1.0397,0.34667) circle (.02);
		\draw[fill] (0.6931,0.6931) circle (.02);
		\node[right,font=\fontsize{8}{6}\selectfont] at (0.3466,1.0397) {(0.5,1)}; 
		\node[right,font=\fontsize{8}{6}\selectfont] at (0.6931,0.6931) {(1,1)}; 
		\node[right,font=\fontsize{8}{6}\selectfont] at (1.0397,0.34667) {(1.5,0.5)}; 
		\node[below,font=\fontsize{8}{6}\selectfont] at (0.693,0) {log2};  
		\node[below,font=\fontsize{8}{6}\selectfont] at (1.098,0) {log3};
		\node[below,font=\fontsize{8}{6}\selectfont] at (1.386,0) {log4};
		\node[left,font=\fontsize{8}{6}\selectfont] at (0,0.693) {log2};
		\node[left,font=\fontsize{8}{6}\selectfont] at (0,1.098) {log3};
		\node[left,font=\fontsize{8}{6}\selectfont] at (0,1.386) {log4};
	\end{tikzpicture}
	\caption{The face $(U_{2,4} , W^{12}_{2}) $ }
	\label{fig14}
\end{figure}
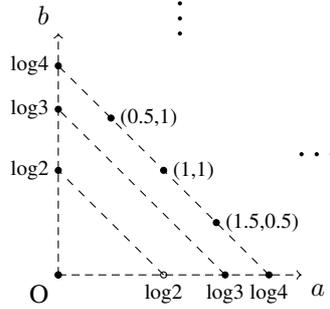

\begin{proof}
	Let $\h=a\br_1+b\br_2$,
	where $\br_1$ and $\br_2$ are the rank functions of the matroids on
	the two extreme rays of the face, respectively.
	Restricting $\h$ on  $M=\{1,3,4\}$ or $\{2,3,4\}$, we obtain $\h'=(a+b)\br'$, where $\br'$ is the rank function of $U_{2,3}$ on $M$. As $\mb{X}=(X_i, i\in N_4)$ is
	the characterizing random vector of $\h$, $(X_1,X_3,X_4)$ and $(X_2,X_3,X_4)$ are uniformly distributed on the rows of some $\voa(U_{2,3},v)s$, which implies that $a+b=\log v$ where $v=|\mathcal{X}_{1}|=|\mathcal{X}_{2}|=|\mathcal{X}_{3}|=|\mathcal{X}_{4}|$.  Let $\mb{T}$ be a $v^2\times 4 $ array such that both $\mb{T}(1,3,4)$ and $\mb{T}(2,3,4)$ are $\voa(U_{2,3},v)s$. 	If $\mathbf{ h} \in  F $ is entropic,
	$\mb{X}$  satisfies the following information equalities
	\begin{align}
		H(X_1|X_2,X_3,X_4)=H(X_2|X_1,X_3,X_4)=0,
	\end{align}
which implies that $\mb{X}$ must be uniformly distributed on such a constructed $\mb{T}$. 
%
 Thus the probability mass of each pair $(x_1,x_2)\in \mathcal{X}_{1}\times\mathcal{X}_{2} $ is 
 \begin{equation}
 	p(x_1,x_2)=\sum_{x_3,x_4:\atop p(x_1,x_2,x_3,x_4)>0 } p(x_1,x_2,x_3,x_4)=\frac{\alpha_{x_1,x_2}}{v^2}, 
 \end{equation}
where $\alpha_{x_1,x_2}$ denotes the times of the row $(x_1,x_2)$ that occurs in $\mb{T}(1,2)$.  Hence, $\bm{\alpha}=(\alpha_{x_1,x_2}>0:x_1,x_2\in \mathbb{I}_v)$ forms a partition of $v^2$. Then
	\begin{align}
		H(X_1,X_2)=H(\bm{\alpha}). \label{r1}
	\end{align}
 As $\h\in F$ and $(X_i, i\in N_4)$ is
	its characterizing random vector, we have 
	\begin{align}
		H(X_1)&=a+b=\log v, \label{r2}\\
		H(X_1,X_2)&=2a+b. \label{r3}
	\end{align}
	By \eqref{r1}-\eqref{r3}, we conclude that 
	\begin{equation}
		a=H(\bm{\alpha}) -\log v .
	\end{equation}

Now we prove the ``if '' part. Since there exists a $v^2\times 4 $ array $\mb{T}$ satisfying the sufficient condition, let $(X_i,i\in N_4)$ be uniformly distributed on the rows of $\mb{T}$. It can be checked that $(X_i,i\in N_4)$ characterizes $(a,b)$. 
\end{proof}	

\noindent \textbf{Remark} This theorem studies the relations of two
Latin squares (of first order) of the same size from the perspective of entropy functions, which
generalize both orthogonal Latin squares and two identical Latin
squares. For any pair of two $v\times v$ Latin squares $S_1$ and
$S_2$, it correspond to a $v^2\times 4$ $\mb{T}$ array with
each row $(i,j, s_1, s_2)$, where $i,j\in \mathbb{I}_v$ and $s_1$ and $s_2$ are
the symbols in $S_1(i,j)$ and $S_2(i,j)$, respectively. On one hand, when $S_1$ and
$S_2$ are orthogonal, entropy function $\h$ of a random vector
distributed on the rows of $\mb{T}$ is in the extreme ray containing
$U_{2,4}$; on the other hand, when they are identical, $\h$ is in the extreme ray
containing $\mc{W}^{12}_2$. When they are neither orthogonal nor identical,
$\h$ locates in the face, but not in the extreme ray.

\begin{theorem}
	\label{rk14}
	For $F=(U_{2,4}, U^{123}_{2,3} )$, $\mathbf{h}=(a,b)\in F$ is
        entropic if and only if  $a+b=\log{v}, a=H(\bm{\alpha})$ and $(a,b)\neq (\log 2,0),(\log 6,0)$, where integer $v>0$ and $\bm{\alpha}$ is
	a partition of $v$.
\end{theorem}
\begin{figure}[H]
	\centering
	\begin{tikzpicture}[scale = 2]
		\draw [->,densely dashed] (0,0)--(2,0) node[below right] { $a$};
		\draw [->,,densely dashed] (0,0)--(0,2) node[above left] {$b$};
		\foreach \y in{0.693,1.098,1.386,1.609,1.791}
		\draw [black, dashed,thin] (\y,0)--(0,\y);
		
		\fill (1.6,0.8) circle (0.4pt);  
		\fill (1.7,0.8) circle (0.4pt);  
		\fill (1.8,0.8) circle (0.4pt);  
		\fill (0.8,1.6) circle (0.4pt);  
		\fill (0.8,1.7) circle (0.4pt);  
		\fill (0.8,1.8) circle (0.4pt);
		\fill (1.6,1.6) circle (0.4pt);  
		\fill (1.7,1.7) circle (0.4pt);  
		\fill (1.8,1.8) circle (0.4pt); 
		
		\draw[fill] (0.636,0.462) circle (.02);
		\draw[fill] (0.562,0.824) circle (.02);
		\draw[fill] (0.693,0.693) circle (.02);
		\draw[fill] (1.039,0.346) circle (.02);
		\node[below left] at (0,0) {O};   
		\foreach \x in{0,0.693,1.098,1.386}
		\foreach \y in{0.693,1.098,1.386}
		\draw[fill] (\x,\y) circle (.02);
		\foreach \x in{1.098,1.386}
		\draw[fill] (\x,0) circle (.02);
		\draw (0.693,0)  circle (.02);
		\draw[fill] (1.609,0)  circle (.02);
		\draw (1.791,0)  circle (.02);
		\foreach \x in{1.609,1.791}
		\draw[fill] (0,\x) circle (.02);
			\draw[fill] (1.332,0.277) circle (.02);
			\draw[fill] (1.06,0.554) circle (.02);
			\draw[fill] (0.95,0.66) circle (.02);
			\draw[fill] (0.67,0.94) circle (.02);
			\draw[fill] (0.5,1.1) circle (.02);
			\draw[fill] (1.56,0.23) circle (.02);
			\draw[fill] (1.33,0.46) circle (.02);
			\draw[fill] (1.098,0.693) circle (.02);
			\draw[fill] (1.242,0.549) circle (.02);
			\draw[fill] (1.01,0.78) circle (.02);
			\draw[fill] (0.693,1.098) circle (.02);
			\draw[fill] (0.8676,0.9242) circle (.02);
			\draw[fill] (0.636,1.155) circle (.02);
			\draw[fill] (0.45,1.34) circle (.02);
		    \draw[fill] (0,0) circle (.02);
		\node[below,font=\fontsize{7}{6}\selectfont] at (0.693,0) {log2};  
		\node[below,font=\fontsize{7}{6}\selectfont] at (1.098,0) {log3};
		\node[below,font=\fontsize{7}{6}\selectfont] at (1.386,0) {log4};
		\node[below,font=\fontsize{7}{6}\selectfont] at (1.62,0) {log5};
		\node[below,font=\fontsize{7}{6}\selectfont] at (1.85,0) {log6};
		\node[left,font=\fontsize{7}{6}\selectfont] at (0,0.693) {log2};
		\node[left,font=\fontsize{7}{6}\selectfont] at (0,1.098) {log3};
		\node[left,font=\fontsize{7}{6}\selectfont] at (0,1.386) {log4};
		\node[left,font=\fontsize{7}{6}\selectfont] at (0,1.62) {log5};
		\node[left,font=\fontsize{7}{6}\selectfont] at (0,1.85) {log6};
	\end{tikzpicture}	
	\caption{The face $(U_{2,4}, U^{123}_{2,3})$ }
	\label{fig9}
\end{figure}
  \begin{proof}
	For entropic $\mathbf{ h} \in F$, its characterizing random vector
	$(X_i,i\in N_4)$ satisfies the following information equalities,
	\begin{align}
		H(  X_{N_4})&=H(X_{N_4-i})  ,\ i\in N_4 \nonumber \\
		H(X_{ij })&=H(X_{i})+H(X_{j}) ,i,j\in  N_4,  \nonumber \\
		H(X_{i\cup K})+H(X_{  j\cup K})&=H(X_{ K})+H(X_{  ij
			\cup K }), \nonumber\\& \qquad\quad|K|=2, K\subseteq\{1,2,3\}.   \nonumber
	\end{align}
  For $ (x_i,i\in N_4) \in
\mathcal{X}_{N_4}$ with $p(x_{1234})>0$, above information equalties
imply that the probability mass function satisfies
	\begin{align}
		p(x_{1234})&=p(x_{1},x_{2},x_{3}) \label{c300}\\
		&=p(x_{1},x_{2},x_{4}) \label{c301} \\ 
		&=p(x_{1},x_{3},x_{4}) \label{c302}\\
		&= p(x_{2},x_{3},x_{4}), \label{c303}\\
		p(x_1,x_2)&=p(x_1)p(x_2),\label{c304} \\
		p(x_1,x_3)&=p(x_1)p(x_3), \label{c305}\\
		p(x_1,x_4)&=p(x_1)p(x_4), \label{c306}\\
		p(x_2,x_3)&=p(x_2)p(x_3), \label{c307}\\
		p(x_2,x_4)&=p(x_2)p(x_4), \label{c308}\\
		p(x_3,x_4)&=p(x_3)p(x_4), \label{c309}\\
		p(x_{1},x_{2},x_{3})p(x_{1},x_{2},x_{4})&=p(x_{1},x_{2})p(x_{1234}),\label{c310}\\
		p(x_{1},x_{2},x_{3})p(x_{1},x_{3},x_{4})&=p(x_{1},x_{3})p(x_{1234}),\label{c311}\\
		p(x_{1},x_{2},x_{3})p(x_{2},x_{3},x_{4})&=p(x_{2},x_{3})p(x_{1234}).\label{c312}
	\end{align}
	By \eqref{c300}, canceling $p(x_1,x_2,x_3)$ and $p(x_1,x_2,x_3,x_4)$ on both side of \eqref{c310}, we have 
	\begin{equation}
		p(x_1,x_2,x_4)=p(x_1,x_2). \label{c313}
	\end{equation}
	Equating \eqref{c304} and \eqref{c313} implies
	\begin{equation}
		p(x_1,x_2,x_4)=p(x_1)p(x_2). \label{c314}
	\end{equation}
	By the same argument,
	\begin{align}
		p(x_1,x_3,x_4)=p(x_1)p(x_3), \label{c315} \\
		p(x_2,x_3,x_4)=p(x_2)p(x_3). \label{c316}
	\end{align}
	By \eqref{c301}-\eqref{c303}, $p(x_1,x_2,x_4)=p(x_1,x_3,x_4)=p(x_2,x_3,x_4)$, together with \eqref{c314}-\eqref{c316}, we have
	\begin{equation}
		p(x_1)=p(x_2)=p(x_3). \label{c317}
	\end{equation}
	By \eqref{c304}, \eqref{c305} and \eqref{c307}, $X_1$, $X_2$ and  $X_3$ are pairwise independent. By Lemma \ref{lem1}, $X_i$ are  uniformly distributed on $\mathcal{X}_{i} $ for $i=1,2,3$, and
	\begin{align}
     H(X_1)=H(X_2)=H(X_3)=\log v,
	\end{align}
   where $v=|\mathcal{X}_{1}|=|\mathcal{X}_{2}|=|\mathcal{X}_{3}|$.
	As $\h\in F$, $(X_i, i\in N_4)$ is
	its characterizing random vector, we have 
	\begin{align}
		H(X_1)&=a+b,\label{c318} \\
		H(X_4)&=a, \label{c320}
	\end{align}
	which implies that $a+b=\log v$.
	Note that $X_1$ and $X_2$ are uniformly distributed, then $p(x_1)=p(x_2)=\frac{1}{v}$. By \eqref{c301} and \eqref{c314}, we have
	\begin{equation}
		p(x_{1234})=p(x_1,x_2,x_4)=p(x_1)p(x_2)=\frac{1}{v^2}.   \label{c322}
	\end{equation}
 As 
\begin{equation}
	p(x_1,x_4)=\sum_{x'_2,x'_3: \atop p(x_1,x'_2,x'_3,x_4)>0} p(x_1,x'_2,x'_3,x_4),  \label{c321}
\end{equation} 
	and the \rv{choice} of $(x_2,x_3)\in\mc{X}_2\times\mc{X}_3$ can be
        arbitrary for a fixed $(x_1,x_4)\in\mc{X}_1\times\mc{X}_4$, replacing $p(x_1,x'_2,x'_3,x_4)$ by $\frac{1}{v^2}$ in \eqref{c321}, we obtain
	\begin{align}
		p(x_1,x_4)=\sum_{x'_2,x'_3: \atop p(x_1,x'_2,x'_3,x_4)>0} \frac{1}{v^2}=\frac{\alpha(x_1,x_4)}{v^2}, \label{c323}
	\end{align}
	where $\alpha(x_1,x_4)\triangleq|\{(x'_2,x'_3)\in\mathcal{X}_{2}\times\mathcal{X}_{3}: p(x_1,x’_2,x'_3,x_4)>0\}|$.
	By \eqref{c306}, we have 
	\begin{equation}
		p(x_1,x_4)=p(x_1)p(x_4)=\frac{1}{v}p(x_4).  \label{c324}
	\end{equation}
	In light of \eqref{c323} and \eqref{c324}, we obtain
	\begin{equation}
		p(x_4)=\frac{\alpha(x_4)}{v} \label{c325}
              \end{equation}
              where 
              $\alpha(x_4)=\alpha(x_1,x_4)$ for any
              $x_1\in\mc{X}_1$.
	Together with \eqref{c320}, it can be seen that 
	\begin{align}
		a=H(X_4)=H(\bm{\alpha}),\label{4.55}
	\end{align}
	where  $\bm{\alpha}\triangleq(\alpha(x_4),x_4\in\mc{X}_4)$ is a number partition of $v$. 
	
	So far, we have proved that
        \begin{align*}
          \{\h=(a,b)\in F: \ & a+b=\log v,\quad v\in \SetZ^+ \\
          & a=H(\bm{\alpha}), \quad \bm{\alpha} \text{ is a partition of } v\}
        \end{align*}
        forms an outer bound on the entropic region in $F=(U_{2,4}, U^{123}_{2,3})$. To prove the theorem,
        we now only need to check whether this outer bound is
        tight. We will see in the following that all $\h$ in it are
        entropic except for $(\log2,0)$ and $(\log6,0)$.

        \begin{itemize}
        \item  For any positive integer $v\neq 2,6$, let
        $\bm{\alpha}=(\alpha_1,\alpha_2,\cdots,\alpha_t)$
        be a
        partition of \rv{$v$}, and $\{A_i,i=1,\dots,t\}$ be a partition of
        $\mathbb{I}_v$ with $|A(i)|=\alpha_{i}$. Let $\mb{T}$ be a
        $\voa(U_{2,4},v)$, and $\mb{T'}$ be a $v^2\times4$ array such that $\mb{T}'(i)=\mb{T}(i)$ for $i=1,2,3$ and  each entry in $\mb{T}'(4)$ be $j$ if the corresponding entry in $\mb{T}(4)$ is in $A_j$. Then let $(X_i,i\in N_4)$ be uniformly distributed on the rows of $\mb{T'}.$ It can be checked that $(X_i,i\in N_4)$ characterizes $(a,b)$.
	
	\begin{example}
		Let $\h=(a,b)\in F$ with $a+b=\log 3$ and
                $a=H((1,2))$. Let
		 \begin{equation*}
		  \mb{T}=\
		  \begin{matrix}
		0 & 0 & 0 & 0 \\
		0 & 1 & 1 & 2 \\     		
		0 & 2 & 2 & 1 \\		
		1 & 0 & 2 & 2 \\ 		
		1 & 1 & 0 & 1 \\		
		1 & 2 & 1 & 0 \\		
		2 & 0 & 1 & 1 \\		
		2 & 1 & 2 & 0 \\
		2 & 1 & 0 & 2 \\
		\end{matrix}
	\hspace{0.5cm}
	\text{ and }\hspace{0.5cm} \mb{T}'=\
	\begin{matrix}
	0 & 0 & 0 & 0 \\
0 & 1 & 1 & 1 \\     
0 & 2 & 2 & 1 \\
1 & 0 & 2 & 1 \\ 
1 & 1 & 0 & 1 \\
1 & 2 & 1 & 0 \\
2 & 0 & 1 & 1 \\
2 & 1 & 2 & 0 \\
2 & 1 & 0 & 1 \\
	\end{matrix}
	\end{equation*} 
   Note that $\mb{T}$ is a $\voa(U_{2,4},3)$ and $\bm{T}'$ is
   constructed as above with $A_1=\{0\}$ and $A_2=\{1,2\}$. Let $(X_i,i\in N_4)$ be uniformly
   distributed on the rows of $\mb{T'}.$ We can see that
   $H(X_4)=H((1,2))=a$, and $\h$ is the entropy function of $(X_i,i\in N_4)$.
 	\end{example}  
   \end{itemize}  
For $v=2$ and $6$, as there is no such $\voa(U_{2,4},2)$
 or  $\voa(U_{2,4},6)$  \cite[Proposition 2]{CCB21}, the above
construction is invalid, and so we have to discuss them separately.
        \begin{itemize}
      \item 
        For $v=2$, there are only two
        partitions $(2)$ and $(1,1)$, which correspond to $(0,\log 2)$
        and $(\log 2, 0)$, respectively. For $(0,\log 2)$, it is
        entropic as $\voa(U_{2,3},2)$ is constructible, while $(\log
        2, 0)$ is non-entropic as $\voa(U_{2,4},2)$ is not
        constructible. 

\item For $v=6$, let 
 \begin{equation*}
 \mrv{
	\mb{T}_{\rm Eu} =\ 
	\begin{matrix}
	0 & 0 & 0 & 0 \\
	0 & 1 & 1 & 5 \\     
	0 & 2 & 2 & 3 \\
	0 & 3 & 3 & 4 \\ 
	0 & 4 & 4 & 2 \\
	0 & 5 & 5 & 1 \\
	1 & 0 & 1 & 1 \\
	1 & 1 & 2 & 0 \\
	1 & 2 & 5 & 4 \\
	1 & 3 & 4 & 3 \\
	1 & 4 & 0 & 5 \\     
	1 & 5 & 3 & 2 \\
	2 & 0 & 2 & 2 \\ 
	2 & 1 & 3 & 4 \\
	2 & 2 & 0 & 1 \\
	2 & 3 & 1 & 5 \\
	2 & 4 & 5 & 3 \\
	2 & 5 & 4 & 0 
    \end{matrix}
        \qquad\qquad
    \begin{matrix}
	3 & 0 & 3 & 3 \\
	3 & 1 & 5 & 2 \\     
	3 & 2 & 4 & 5 \\
	3 & 3 & 2 & 1 \\ 
	3 & 4 & 1 & 0 \\
	3 & 5 & 0 & 4 \\
	4 & 0 & 4 & 4 \\
	4 & 1 & 0 & 3 \\
	4 & 2 & 1 & 2 \\
	4 & 3 & 5 & 0 \\
	4 & 4 & 3 & 1 \\     
	4 & 5 & 2 & 5 \\
	5 & 0 & 5 & 5 \\ 
	5 & 1 & 4 & 1 \\
	5 & 2 & 3 & 0 \\
	5 & 3 & 0 & 2 \\
	5 & 4 & 2 & 4 \\
	5 & 5 & 1 & 3 \\
	\end{matrix}
    }
\end{equation*}

Note that both $\mb{T}_{\rm Eu}(\{1,2,3\})$ and $\mb{T}_{\rm
  Eu}(\{1,2,4\})$ are $\voa(U_{2,3},6)$s. However, $\mb{T}_{\rm
  Eu}$ is not a $\voa(U_{2,4},6)$, as only
$34$ different pairs occur in $\mb{T}_{\rm
          Eu}(\{2,4\})$, while
 $(1,5)$ and 
        $(3,4)$ each appear twice in $\mb{T}_{\rm
          Eu}(\{3,4\})$. 
\footnote{
The array $\mb{T}_{\rm
  Eu}$ is constructed from the following two Latin squares
	\begin{table}[H]
		\renewcommand{\arraystretch}{1.5}
		\centering
		\label{table:3} 
		\begin{tabular}{|c|c|c|c|c|c|} 		   
		\hline  0& 1 &2& 3&4 & 5\\
		\hline 1& 2 &5&4 &0 &3 \\
		\hline 2& 3 &0&1 & 5&4\\
		\hline  3& 5 &4& 2& 1&0 \\
		\hline 4& 0 &1& 5& 3&2 \\
		\hline 5& 4 &3& 0& 2&1\\
		\hline
	\end{tabular} 
\hspace{0.5 cm}
\begin{tabular}{|c|c|c|c|c|c|} 		   
\hline  0& 5 &3& 4&2 &1 \\
\hline 1& 0 &4&3 &5 &2 \\
\hline 2& 4 &1& 5&3 &0\\
\hline  3& 2 &5& 1&0 &4 \\
\hline 4& 3 &2& 0&1 &5 \\
\hline 5& 1 &0& 2&4 &3\\
\hline
		\end{tabular} 
	\end{table}
\noindent discovered by Euler in \cite{euler1782recherches} in 1782, in a manner
how we construct $\voa(U_{2,4},v)$ from two $v\times v$ orthogonal Latin
squares. That is, for each row of $\mb{T}_{\rm Eu}$, the first two entries are the row and column indices of the two
squares, and the third and forth entries are the symbols of the first
square and the second square, respectively. 
}


Let $\bm{\alpha}=(1,1,1,1,2)$.
Let 
$A_i=\{i-1\},i=1,2,3,4$ and $A_5=\{4,5\}$. For any partition $\bm{\beta}=(\beta_1,\beta_2,\ldots,\beta_t)$ of $6$
other than $(1,1,1,1,1,1)$, it is coarser than $\bm{\alpha}$. Let
$\{B_i,i=1,2,\ldots,t\}$ be a partition of $N_6$ with
$|B_i|=\beta_i$ such that each $B_i$ is a union of some $A_i$s. Let $\mb{T}'_{\rm
          Eu}$ be a $36\times 4$ array such that $\mb{T}'_{\rm
          Eu}(i)=\mb{T}_{\rm
          Eu}(i)$, $i=1,2,3$ and each entry in $\mb{T}'_{\rm
          Eu}(4)$ is $j$ if the corresponding entry in $\mb{T}_{\rm
          Eu}(4)$ is in $B_j$. Let $\mb{X}=(X_i,i\in N_4)$
        distributed on the rows of $\mb{T}'_{\rm
  Eu}$. The entropy function $\h$ of $\mb{X}$ satisfies that $a+b=\log 6$ and
$a=H(\bm{\beta})$. 
      \end{itemize}
The theorem is proved.
\end{proof}

\noindent \textbf{Remark} A \emph{frequency square} induced by
partition $\bm{\alpha}=(\alpha_1,\ldots, \alpha_t)$ of integer $v$ is a 
$v\times v$ square $S$ with symbols $k\in \mathbb{I}_t$, each appearing in
each row and each column of $S$ $\alpha_k$ times. Note that when
$\bm{\alpha}$ is the all-$1$ partition of $v$, $S$ reduces to a Latin square.
A Latin square $S_1$
of order $v$ (and of the first kind) and a frequency square $S_2$ induced by partition
$\bm{\alpha}=(\alpha_1,\ldots, \alpha_t)$ of integer $v$ are called
\emph{orthogonal} if each pair $(s_1,s_2)\in \mathbb{I}_v\times \mathbb{I}_t$
  appears $\alpha_i$, where $s_2$ is the symbol that appears $\alpha_i$
  times in each row and column of $S_2$.
By characterizing entropy functions on the face
$(U_{2,4}, U^{123}_{2,3})$, 
this theorem also studies arrays corresponding to
orthogonal two squares with one Latin square $S_1$ and one frequency
square $S_2$. When $\bm{\alpha}$ is the all-$1$ partition of $v$, it reduces to orthogonal of two Latin squares and then corresponding entropy functions are on the extreme ray
containing $U_{2,4}$. On the other
hand, when $\bm{\alpha}$ is the trivial partition of a single $v$, it
reduces to orthogonal squares of one Latin square of the first kind and
the other of the zeroth kind, and entropy functions are in the extreme
ray containing $U^{123}_{2,3}$. It is interesting that although
$\voa(U_{2,4},6)$ does not exist, arrays ``between''
$\voa(U^{123}_{2,3},6)$ and $\voa(U_{2,4},6)$ exist.




\begin{theorem}
	\label{rk20}
	For   $F=(\hat{U}_{2,5}^{1},\mc{W}_{2}^{12})$, $\h=(a,b)\in F$ is entropic if and only if $a+b= \log v$ for some positive $v$  and $a=\frac{1}{v}\sum_{i=0}^{v-1} H(\bm{\alpha_i})$, where $ \bm{\alpha_i}\in\mathcal{P}(v),i\in \mathbb{I}_v$. \rv{Here, $\mathcal{P}(v)$ denotes the family of all partitions of $v$.}
\end{theorem}
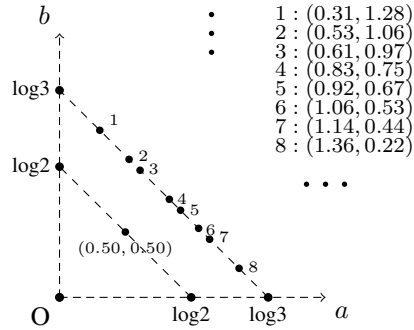
\begin{figure}[H]
	\centering
	\begin{tikzpicture}[scale = 2.5]
		\draw [->,densely dashed] (0,0)--(1.4,0) node[below right] { $a$};
		\draw [->,densely dashed] (0,0)--(0,1.4) node[above left] {$b$};
		\foreach \y in{0.693,1.098}
		\draw [black, dashed,thin] (\y,0)--(0,\y);
		\fill (1.3,0.6) circle (0.4pt);  
		\fill (1.4,0.6) circle (0.4pt);  
		\fill (1.5,0.6) circle (0.4pt);  
		\fill (0.8,1.3) circle (0.4pt);  
		\fill (0.8,1.4) circle (0.4pt);  
		\fill (0.8,1.5) circle (0.4pt); 
		\node[below left] at (0,0) {O};   
		\foreach \x in{0,1.098}
		{\draw[fill] (\x,0) circle (.02);
			\draw[fill] (0,\x) circle (.02);}
		\draw[fill] (0.693,0) circle (.02);
		\draw[fill] (0,0.693) circle (.02);
		\fill (0.3466,0.3466) circle (0.02); 
		\node[below,font=\fontsize{6}{6}\selectfont] at (0.3466,0.3466) {$(0.50,0.50)$}; 
		\node[below,font=\fontsize{8}{6}\selectfont] at (0.693,0) {log2};  
		\node[below,font=\fontsize{8}{6}\selectfont] at (1.098,0) {log3};
		\node[left,font=\fontsize{8}{6}\selectfont] at (0,0.693) {log2};
		\node[left,font=\fontsize{8}{6}\selectfont] at (0,1.098) {log3};
		\fill (0,1.098) circle (0.4pt); 
		\fill (0.212,0.886) circle (0.02);
		\node[right,font=\fontsize{6}{6}\selectfont] at (0.212,0.94) {$1$};  
		\fill (0.366,0.732) circle (0.02); 
		\node[right,font=\fontsize{6}{6}\selectfont] at (0.366,0.75) {$2$}; 

		\fill (0.424,0.674) circle (0.02); 
		\node[right,font=\fontsize{6}{6}\selectfont] at (0.42,0.68) {$3$}; 
		\fill (0.578,0.52) circle (0.02);
		\node[right,font=\fontsize{6}{6}\selectfont] at (0.57,0.52) {$4$}; 
			\fill (0.637,0.462) circle (0.02); 
		\node[right,font=\fontsize{6}{6}\selectfont] at (0.637,0.462) {$5$};
		
		\fill (0.732,0.366) circle (0.02); 
		\node[right,font=\fontsize{6}{6}\selectfont] at (0.72,0.36){$6$};
		
		\fill (0.79,0.308) circle (0.02);
			\node[right,font=\fontsize{6}{6}\selectfont] at (0.79,0.32){$7$};
			
		\node[right,font=\fontsize{6}{6}\selectfont] at (0.945,0.16){$8$};
		\fill (0.945,0.154) circle (0.02); 
		\fill (1.099,0) circle (0.02);
		\node[font=\fontsize{8}{6}\selectfont] at (1.5,1.5){$1:(0.31,1.28)$};
		\node[font=\fontsize{8}{6}\selectfont] at (1.5,1.4){$2:(0.53,1.06)$};
		\node[font=\fontsize{8}{6}\selectfont] at (1.5,1.3){$3:(0.61,0.97)$};
		\node[font=\fontsize{8}{6}\selectfont] at (1.5,1.2){$4:(0.83,0.75)$};
		\node[font=\fontsize{8}{6}\selectfont] at (1.5,1.1){$5:(0.92,0.67)$};
		\node[font=\fontsize{8}{6}\selectfont] at (1.5,1){$6:(1.06,0.53)$};
		\node[font=\fontsize{8}{6}\selectfont] at (1.5,0.9){$7:(1.14,0.44)$};
		\node[font=\fontsize{8}{6}\selectfont] at (1.5,0.8){$8:(1.36,0.22)$};

	\end{tikzpicture}
	\caption{The face $(\hat{U}_{2,5}^{1},\mc{W}_{2}^{12})$}
	\label{fig18}
\end{figure}

\begin{proof}
	If $\mathbf{ h} \in  F$ is entropic, its characterizing random vector
	$(X_i,i\in N_4)$ satisfies the following information equalities,
	\begin{align}
		H(  X_{N_4})&=H(X_{N_4-i}), \quad i\in N_4, \nonumber \\
		H(X_{ij })&=H(X_{i})+H(X_{j}),i,j\in\{2,3,4\}, \nonumber \\
		H(X_{  12 })+H(X_{1i })&=H(X_{ 1})+H(X_{  12i} ),\quad
                                         i\in\{3,4\}, \nonumber \\
		H(X_{i\cup K})+H(X_{  j\cup K})&=H(X_{ K})+H(X_{  ij
			\cup K }), \nonumber\\ &\qquad\qquad\qquad |K|=2, K\neq\{1,2\}.   \nonumber
	\end{align}
  For $ (x_i,i\in N_4) \in
\mathcal{X}_{N_4}$ with $p(x_{1234})>0$, above information equalties
imply that the probability mass function satisfies
	\begin{align}
		p(x_{1234})&=p(x_{1},x_{2},x_{3}) \label{c270}\\
		&=p(x_{1},x_{2},x_{4}) \label{c271} \\ 
		&=p(x_{1},x_{3},x_{4}) \label{c272}\\
		&= p(x_{2},x_{3},x_{4}), \label{c273}\\
		p(x_2,x_3)&=p(x_2)p(x_3), \label{c274}\\
		p(x_2,x_4)&=p(x_2)p(x_4), \label{c275}\\
		p(x_3,x_4)&=p(x_3)p(x_4), \label{c276}\\
		p(x_1,x_2)p(x_1,x_3)&=p(x_1)p(x_1,x_2,x_3), \label{c277}\\
		p(x_1,x_2)p(x_1,x_4)&=p(x_1)p(x_1,x_2,x_4), \label{c278}\\
		p(x_{1},x_{2},x_{3})p(x_{1},x_{3},x_{4})&=p(x_{1},x_{3})p(x_{1234}),\label{c279}\\
		p(x_{1},x_{2},x_{4})p(x_{1},x_{3},x_{4})&=p(x_{1},x_{4})p(x_{1234}).\label{c280}\\	p(x_{1},x_{2},x_{3})p(x_{2},x_{3},x_{4})&=p(x_{2},x_{3})p(x_{1234}),\label{c281}\\
		p(x_{1},x_{2},x_{4})p(x_{2},x_{3},x_{4})&=p(x_{2},x_{4})p(x_{1234}),\label{c282}\\
		p(x_{1},x_{3},x_{4})p(x_{2},x_{3},x_{4})&=p(x_{3},x_{4})p(x_{1234}).\label{c283}
	\end{align}
	Restricting $\h=(a,b)$ on $\{2,3,4\}$, we obtain
        $\h'=(a+b)\br'$, where $\br'$ is the rank function of
        $U_{2,3}$ on $\{2,3,4\}$. Thus the characterizing random
        vector $(X_2,X_3,X_4)$ of $\h'$ is uniformly distributed on
        the rows of a $\voa(U_{2,3},v)$ for a positive integer $v$,
        and so $a+b$ can only take the value of $\log v$. Now let
        $\mb{T}$ be a $v^2\times 4$ array with $\mb{T}(2,3,4)$ a
        $\voa(U_{2,3},v)$. By \eqref{c273}, $X_1$ is a function of
        $(X_2,X_3,X_4)$, which implies that $(X_i,i\in N_4)$ must be
        distributed on the rows of a such constructed $\mb{T}$.
	
	According to \eqref{c272}, canceling  $p(x_1,x_3,x_4)$ on the left side and $p(x_1,x_2,x_3,x_4)$ on the right side of \eqref{c279}, we have
	\begin{align}
		p(x_1,x_2,x_3)= p(x_1,x_3). \label{c286}  
	\end{align}
	Then canceling	$p(x_1,x_3)$ and $p(x_1,x_2,x_3)$ in \eqref{c277}, we obtain
	\begin{equation}
		p(x_1,x_2)=p(x_1),
	\end{equation}
	which implies $X_2$ is a function of $X_1$. Then for each $j\in\mc{X}_1$, there exists a unique $i\in\mc{X}_2=\mathbb{I}_v$ such that $(j,i)$ forms a row in $\mb{T}(1,2)$. Let $\beta_{i,j}$ denote the times of the row $(j,i)$ that occurs in $\mb{T}(1,2)$. Note that each $i\in \mathbb{I}_v$ occurs in $\mb{T}(2)$ exactly $v$ times by the definition of  a $\voa(U_{2,3},k)$. So for each $i\in \mathbb{I}_v$, $\bm{\beta_i}=(\beta_{i,j}>0,j\in \mc{X}_1)$ forms a partition of $v$.
	 We assume that there exist $t_i$ different $j$ such that $\beta_{i,j}>0$ for $i\in \mathbb{I}_v$. Then, $\bm{\beta_i}$ can be written as $\mrv{\bm{\alpha_i}=(\alpha_{i,0},\dots,\alpha_{i,t_i-1}) } $. Then	
	\begin{align}
		H(X_1)=H( &\frac{ \alpha_{0,0}}{v^2},\frac{ \alpha_{0,1}}{v^2},\dots,\frac{ \alpha_{ 0,t_1-1}}{v^2},  \nonumber \\
		&\frac{ \alpha_{1,0}}{v^2},\frac{ \alpha_{1,1}}{v^2},\dots,\frac{ \alpha_{ 1,t_2-1}}{v^2}, \nonumber \\
		& ...\nonumber,\\
		&\frac{ \alpha_{v-1,0}}{v^2},\frac{ \alpha_{v-1,1}}{v^2},\dots,\frac{ \alpha_{ v-1,t_v-1}}{v^2}) \nonumber    \\
		=-\sum_{i=0}^{v-1}& \sum_{j=0}^{t_i-1} \frac{ \alpha_{i,j}}{v^2}\log\frac{ \alpha_{i,j}}{v^2}. \label{c287}
	\end{align}
	As $\h\in F$, $(X_i, i\in N_4)$ is
	its characterizing random vector, we have 
	\begin{align}
		H(X_1)&= 2a+b, \label{c288} \\
		H(X_2)&=H(X_3)=H(X_4)=a+b. \label{c289}  
	\end{align}
	Since $a+b=\log v$, we have
	\begin{align}
		a&=H(X_1)-(a+b)\label{c290}   \\
		&=-\sum_{i=0}^{v-1} \sum_{j=0}^{t_i-1} \frac{ \alpha_{i,j}}{v^2}\log\frac{ \alpha_{i,j}}{v^2}-\log v  \label{c291}  \\
		&=-\sum_{i=0}^{v-1} \sum_{j=0}^{t_i-1} \frac{ \alpha_{i,j}}{v^2}\log\frac{ \alpha_{i,j}}{v} =\frac{1}{v}\sum_{i=0}^{v-1} H(\bm{\alpha_i}). \label{c292}
	\end{align}

Now we prove the ``if''  part. For any $\h=(a,b)$ satisfying $a+b= \log v$  and $a=\frac{1}{v}\mrv{\sum_{i=0}^{v-1}} H(\bm{\alpha_i})$, where $\bm{\alpha_i}=\mrv{(\alpha_{i,0},\dots,\alpha_{i,t_i-1})} \in \mathcal{P}(v),\mrv{i\in \mathbb{I}_v}$, let $\mb{T}$ be a $v^2\times 4$ array such that $\mb{T}(2,3,4)$ is a $\voa(U_{2,3},v)$. In the rows where $i$ occurs in the  $\mb{T}(2)$, let the entry $(\sum_{m=0}^{i-1}t_m)+j$ occur  $\mrv{\alpha_{i,j}}$ times in $\mb{T}(1)$ for $i\in \mathbb{I}_v$ and $j\in \mathbb{I}_{t_i}$ where we set $t_{-1}=0$. Then let $(X_i,i\in N_4)$ be uniformly distributed on the rows of $\mb{T}.$ It can be checked that $(X_i,i\in N_4)$ characterizes $(a,b)$. The proof is accomplished.
\end{proof}

\noindent \textbf{Remark} Similar to the first two faces in this
Subsection, the characterization of $(\hat{U}_{2,5}^{1},\mc{W}_{2}^{12})$ can also be considered as two
squares of order $v$, where $S_2$ is a Latin square and $S_1$ is a
square whose symbols can be obtained from splitting symbols in $S_2$,
that is, each symbol $i$ splits into $t_i$ symbols with each occurring
$\alpha_{i,j}$ times according to the partition $\bm{\alpha}_i$.

\subsection{Entropy functions on faces involving Latin square decompositions}
\label{latindecom}

In this subsection, we introduce three types of decompositions of a
$\voa(U_{2,3},v)$, which characterize three $2$-dimensional faces,
respectively. As discussed in Subsection \ref{jlvoaolh}, each $\voa(U_{2,3},v)$
corresponds to a Latin square, the three types of $\voa(U_{2,3},v)$
decompositions correspond to three types of Latin square decomposition.

                  \begin{definition}
                    	Given $A,B\subseteq \mathbb{I}_v$ and a
                $\voa(U_{2,3},v)$ $\mb{T}$, an $|A||B|\times 3$ subarray $\mb{T}'$ of
                $\mb{T}$ is called induced by $A$ and $B$ if rows in
                $\mb{T}'(1,2)$ are exactly those pairs in $A\times B$.
                  \end{definition}

	\begin{definition}
		Given $A,B\subseteq \mathbb{I}_v$ with $|A||B|=v$ and a
                $\voa(U_{2,3},v)$ $\mb{T}$,
                \begin{itemize}
                \item a subarray $\mb{T}'$ of $\mb{T}$ induced by $A$
                  and $B$ is called a \emph{unit
                      subarray} of  $\mb{T}$ if each $e\in \mathbb{I}_v$
                  occurs exactly once in $\mb{T}'(3)$.
                  \item $\{\mb{T}_i,i\in \mathbb{I}_v\}$ is called an
                    \emph{uniform decomposition} of
                    a $\voa(U_{2,3},v)$ $\mb{T}$ if
                    \begin{itemize}
                    \item each $\mb{T}_i$ induced by $A_i$ and $B_i$ is a unit subarray of $\mb{T}$ and
                      \item $\biguplus\limits_{i\in \mathbb{I}_v}A_i\times B_i=\mathbb{I}^2_v$. 
                    \end{itemize}
                \end{itemize}
	\end{definition}

\begin{example}
\label{unidecomplatin}
	Here is an example of uniform decomposition of a $\voa(U_{2,3},4)$ $\mb{T}$.

\begin{figure*}
	\begin{equation*}
		\mb{T}\ =\ 
	\begin{matrix}
		0&0&0\\
		0&1&3\\
		0&2&1\\
		0&3&2\\
		1&0&1\\
		1&1&2\\
		1&2&0\\
		1&3&3\\
		2&0&2\\
		2&1&0\\
		2&2&3\\
		2&3&1\\	
		3&0&3\\
		3&1&1\\
		3&2&2\\
		3&3&0\\
	\end{matrix}
	\qquad\mb{T}_0\ =\ 
	\begin{matrix}
		0&0&0\\
		0&1&3\\
		0&2&1\\
		0&3&2
	\end{matrix}
	\qquad\mb{T}_1\ =\ 
	\begin{matrix}
		1&0&1\\
		1&1&2\\
		1&2&0\\
		1&3&3\\
	\end{matrix}
	\qquad\mb{T}_2\ =\ 
	\begin{matrix}
		2&0&2\\
		2&1&0\\
		3&0&3\\
		3&1&1
	\end{matrix}
	\qquad\mb{T}_3\ =\ 
	\begin{matrix}
		2&2&3\\
		2&3&1\\
		3&2&2\\
		3&3&0
	\end{matrix}
		\end{equation*}
 \end{figure*}
Note that in this exmaple, $\mb{T}_i$ is induced by $A_i$ and $B_i$,
where $i\in \mathbb{I}_4$ and 
\begin{itemize}
\item $A_0=\{0\}$, $B_0=\mathbb{I}_4$,
\item $A_1=\{1\}$, $B_1=\mathbb{I}_4$,
\item $A_2=\{2,3\}$, $B_2=\{0,1\}$ and 
\item $A_3=\{2,3\}$, $B_3=\{2,3\}$. 
\end{itemize}
\end{example}



\begin{theorem}
	\label{rk28}
	For $F=(\mc{W}_{2}^{12}, \mc{W}_{2}^{13})$, $\h=(a,b)\in F$ is
        entropic if and only if there exists a uniform decomposition $\{\mb{T}_0,\dots,\mb{T}_{v-1}\}$ of a $\voa(U_{2,3},v)$ $\mb{T}$ such that 
	\begin{align}
		a =\log v -\frac{1}{v}\sum_{i=0}^{v-1}\log |B_i|  \text{
          and }
		  b =\log v -\frac{1}{v}\sum_{i=0}^{v-1}\log |A_i|,  \nonumber
	\end{align}
where the subarray $\mb{T}_i$ of $\mb{T}$ are induced by $A_i$ and $B_i$  for $i\in \mathbb{I}_v$.
\end{theorem}

\begin{figure}[H]
	\centering
	\begin{tikzpicture}[scale = 2]
		\draw [->,densely dashed] (0,0)--(1.6,0) node[below right] { $a$};
		\draw [->,densely dashed] (0,0)--(0,1.6) node[above left] {$b$};
		\foreach \y in{0.693,1.098,1.386}
		\draw [black, dashed,thin] (\y,0)--(0,\y);
		\fill (1.6,0.8) circle (0.4pt);  
		\fill (1.7,0.8) circle (0.4pt);  
		\fill (1.8,0.8) circle (0.4pt);  
		\fill (0.8,1.6) circle (0.4pt);  
		\fill (0.8,1.7) circle (0.4pt);  
		\fill (0.8,1.8) circle (0.4pt); 
		\node[below left] at (0,0) {O};   
		\foreach \x in{0,0.693,1.098,1.386}
		{\draw[fill] (\x,0) circle (.02);
			\draw[fill] (0,\x) circle (.02);}
		\draw[fill] (0,0.693) circle (.02);
		\draw[fill] (0.3466,1.0397) circle (.02);
		\draw[fill] (1.0397,0.34667) circle (.02);
		\draw[fill] (0.6931,0.6931) circle (.02);
		\node[right,font=\fontsize{8}{6}\selectfont] at (0.3466,1.0397) {$(0.5,1.5)$}; 
		\node[right,font=\fontsize{8}{6}\selectfont] at (0.6931,0.6931) {$(1,1)$}; 
		\node[right,font=\fontsize{8}{6}\selectfont] at (1.0397,0.34667) {$(1.5,0.5)$}; 
		 
		 \node[below,font=\fontsize{8}{6}\selectfont] at (0.6931,0) {log2};
		\node[below,font=\fontsize{8}{6}\selectfont] at (1.098,0) {log3};
		\node[below,font=\fontsize{8}{6}\selectfont] at (1.386,0) {log4};
		\node[left,font=\fontsize{8}{6}\selectfont] at (0,0.693) {log2};
		\node[left,font=\fontsize{8}{6}\selectfont] at (0,1.098) {log3};
		\node[left,font=\fontsize{8}{6}\selectfont] at (0,1.386) {log4};
	\end{tikzpicture}
	\caption{The face $ ( \mc{W}_{2}^{12}, \mc{W}_{2}^{13}) $ }
	\label{fig15}
\end{figure} 

\begin{proof}
	If $\mathbf{h} \in  F$ is entropic, its characterizing random vector
	$(X_i,i\in N_4)$ satisfies the following information equalities,
	\begin{align}
		H(  X_{N_4})&=H(X_{N_4-i})  ,\ i\in N_4 \nonumber \\
		H(X_{ij })&=H(X_{i})+H(X_{j}), \nonumber\\& \qquad\qquad\{i,j\}\neq \{1,2\},\{1,3\} \nonumber \\
		H(X_{12})+H(X_{13})&=H(X_1)+H(X_{123}),\nonumber \\ 
		H(X_{i\cup K})+H(X_{  j\cup K})&=H(X_{ K})+H(X_{  ij
			\cup K }),\nonumber\\ &\qquad|K|=2,   K\neq \{1,2\},\{1,3\}.   \nonumber
	\end{align}
 For $ (x_i,i\in N_4) \in
\mathcal{X}_{N_4}$ with $p(x_{1234})>0$, above information equalties
imply that the probability mass function satisfies
	\begin{align}
		p(x_{1234})&=p(x_{1},x_{2},x_{3}) \label{c600}\\
		&=p(x_{1},x_{2},x_{4}) \label{c601} \\ 
		&=p(x_{1},x_{3},x_{4}) \label{c602}\\
		&=p(x_{2},x_{3},x_{4}) \label{c612}\\
		p(x_1,x_4)&=p(x_1)p(x_4), \label{c603}\\
		p(x_2,x_3)&=p(x_2)p(x_3), \label{c604}\\
		p(x_2,x_4)&=p(x_2)p(x_4), \label{c605}\\
		p(x_3,x_4)&=p(x_3)p(x_4), \label{c606}\\
		p(x_{1},x_{2})p(x_{1},x_{3})&=p(x_{1})p(x_{1},x_{2},x_{3}),\label{c611}\\
		p(x_{1},x_{2},x_{4})p(x_{1},x_{3},x_{4})&=p(x_{1},x_{4})p(x_{1234}),\label{c607}\\
		p(x_{1},x_{2},x_{3})p(x_{2},x_{3},x_{4})&=p(x_{2},x_{3})p(x_{1234}),\label{c608} \\
		p(x_{1},x_{2},x_{4})p(x_{1},x_{3},x_{4})&=p(x_{2},x_{4})p(x_{1234})\label{c609}\\
		p(x_{1},x_{3},x_{4})p(x_{2},x_{3},x_{4})&=p(x_{3},x_{4})p(x_{1234}).\label{c610}
	\end{align}
	Restricting $\h$ on $\{2,3,4\}$, we obtain $\h'=(a+b)\br'$,
        where $\br'$ is the rank function of $U_{2,3}$ on
        $\{2,3,4\}$. Thus the characterizing random vector
        $(X_2,X_3,X_4)$ of $\h'$ is uniformly distributed on the rows
        of a $\voa(U_{2,3},v)$ for a positive integer $v$, and so
        $a+b$ can only take the value of $\log v$.  

By \eqref{c600}-\eqref{c612}, \eqref{c607} and \eqref{c609}, we obtain $p(x_1,x_4)=p(x_2,x_4)$. Then with \eqref{c603} and \eqref{c605}, we have  $p(x_1)p(x_4)=p(x_2)p(x_4)$. Therefore, we obtain $p(x_1)=p(x_2)=\frac{1}{v}$, which implies that $X_1$ is uniformly distributed on $\mathcal{X}_{1}$ and $H(X_1)=\log v$. 
	
	Since $p(x_1,x_2,x_3,x_4)=p(x_2,x_3,x_4)$, $(X_i,i\in N_4$)
        must be uniformly distributed on the rows of a $v^2\times 4$
        array $\mb{T}$ such that $\mb{T}(2,3,4)$ is a
        $\voa(U_{2,3},v)$, and the first entry of each row in $\mb{T}$ is uniquely
        determined by the remaining three entries. Assume $\mathcal{X}_{1}=\mathbb{I}_v$. Let $A_i$ and
        $B_i$ denote the set of all $j_1$ such that $(i,j_1)$ appears
        on the row of $\mb{T}(1,2)$, and the set of $j_2$ such that  $(i,j_2)$ appears on the row of $\mb{T}(1,3)$ for $i\in \mathbb{I}_v$, respectively. For any $j_1\in A_i$ and  $j_2\in B_i$, we have 
	\begin{align}
		 p_{X_1,X_2}(i,j_1)>0 \text{ and } p_{X_1,X_3}(i,j_2)>0.
	\end{align}
	Together with \eqref{c611}, we obtain
	\begin{align}
		p_{X_1,X_2,X_3}(i,j_1,j_2)>0.
	\end{align}
Note that $(j_1,j_2)$ occurs exactly once on the rows of $\mb{T}(2,3)$ due to the definition of  $\voa(U_{2,3},v)$, 
$(i,j_1,j_2)$ will appear  exactly once on the rows of $\mb{T}(1,2,3)$, and so  
\begin{equation}
	p_{X_1,X_2,X_3}(i,j_1,j_2)=\frac{1}{v^2}. \label{c1007}
\end{equation}
The probability mass of $(i,j_1)$, $j_i\in A_i, i\in \mathbb{I}_v$
\begin{align}
	p_{X_1,X_2}(i,j_1)=\sum_{j_2\in B_i}p_{X_1,X_2,X_3}(i,j_1,j_2)=\frac{|B_i|}{v^2}. \label{c1005}
\end{align}
 Similarly, the probability mass of $(i,j_2)$, $j_2\in B_i, i\in \mathbb{I}_v$  
 \begin{align}
 	p_{X_1,X_3}(i,j_2)=\sum_{j_1\in A_i}p_{X_1,X_2,X_3}(i,j_1,j_2)=\frac{|A_i|}{v^2}.\label{c1006}
 \end{align}
Since $p(x_1)=\frac{1}{v}$, together with \eqref{c611} and \eqref{c1007}-\eqref{c1006}, we obtain
\begin{equation}
	|A_i|\times |B_i|=v, \label{c1008}
\end{equation}
which implies that $i$ will occur $v$ times in $\mb{T}(1)$. Note $X_1$
is independent of $X_4$ and $|\mathcal{X}_4|=v$, $(i,j_3)$ for $j_3\in
N_4$ appears exactly once on the rows of $\mb{T}(1,4)$. Thus $\mb{T}$
can be decomposed into $v$ arrays based on the entry $i$ occuring on
the rows of $\mb{T}(1)$, that is, the entries on the rows that $i\in
\mb{T}(1)$ appears of $\mb{T}$ forms a $v\times 4$ array $\mb{T}_{i}$
for $i\in \mathbb{I}_v$. We can check that $\{\mb{T}_i(2,3,4),i\in \mathbb{I}_v\}$ is a uniform decomposition of $\mb{T}(2,3,4)$, and the entries of $\mb{T}_i(2)$,$\mb{T}_i(3)$ and $\mb{T}_i(4)$ are from $A_i$, $B_i$ and $\mathbb{I}_v$, respectively.
The entropy of $(X_1,X_2)$
	\begin{align}
	H(X_1,X_2)&=H(\underbrace{\dfrac{|B_0|}{v^2},\dots,\dfrac{|B_0|}{v^2}}_{|A_0|},\dots, \underbrace{\dfrac{|B_{v-1}|}{v^2},\dots,\dfrac{|B_{v-1}|}{v^2}}_{|A_{v-1}|} )  \label{c614}\\
	&= 2\log v -\frac{1}{v}\sum_{i=0}^{v-1 }\log |B_i|. \label{c615}
\end{align}
	As $\h\in F$, $(X_i, i\in N_4)$ is its characterizing random vector, we have 
	\begin{align}
		H(X_1)&=a+b \label{c616} \\
		H(X_1,X_2)&=2a+b   \label{c617}
	\end{align}
	Note that $a+b=\log v$, we have
	\begin{align}
		a&=2\log v -\frac{1}{v}\sum_{i=0}^{v-1}\log |B_i|  -\log k  \label{c618}\\
		&=\log v -\frac{1}{v}\sum_{i=0}^{v-1}\log |B_i|.  \label{c619}
	\end{align}
	By the same argument,
	\begin{align}
		b =\log v -\frac{1}{v}\sum_{i=0}^{v-1}\log |A_i|.  \label{c620}
	\end{align}

  As for the ``if'' part, let $(X_2,X_3,X_4)$ be uniformly distributed
  on the rows of the  $\voa(U_{2,3},v)$ $\mb{T}$. Let $X_1$ be $i$ if $(x_2,x_3,x_4)$ appears on the rows of $\mb{T}_i$. Then $(X_i,i\in N_4)$ characterizes $(a,b)$. The proof has been completed.
\end{proof}

\noindent \textbf{Remark}  Theorem \ref{rk28} establishes a correspondence between the
$2$-dim face characterization problem and the uniform decomposition
problem of a $\voa(U_{2,3},v)$.
When \rv{$v$} is prime, $\voa(U_{2,3},v)$ can be decomposed into  \rv{$v$} uniform
subarrays where either $|A_i|=1$ and $|B_i|=v$ for $i\in \mathbb{I}_v$, or
$|A_i|=v$ and $|B_i|=1$ for $i\in \mathbb{I}_v$. These correspond to the
polymatroids $(0,\log v)$ and  $(\log v,0)$, respectively. While for a
composite $v$, the uniform decomposition of a $\voa(U_{2,3},v)$ \rv{will}
be more complicated. In Example \ref{unidecomplatin}, the uniform decomposition
corresponds to the entropy function $(0.5,1.5)$ on the face $F$.

	\begin{definition}
	Given $A,B\subseteq \mathbb{I}_v$ with $|A|=|B|=v'\le v$ and a
		$\voa(U_{2,3},v)$ $\mb{T}$,
		\begin{itemize}
			\item  a subarray $\mb{T}'$ of $\mb{T}$
                          induced by $A$ and $B$ is called a
                          \emph{suborder $\voa$} of
                            $\mb{T}$ if $\mb{T}'$ is a $\voa(U_{2,3}, v')$.
	  \item  $\{\mb{T}_i,i\in \mathbb{I}_t\}$ is called a
            \emph{suborder decomposition} of $\mb{T}$ if
		\begin{itemize}
		\item  each $\mb{T}_i$ induced by $A_i$ and $B_i$ is a suborder $\voa$ of $\mb{T}$ and 
		\item $\biguplus\limits_{i\in \mathbb{I}_t}A_i\times B_i= \mathbb{I}^2_v$. 
		\end{itemize}  
		\end{itemize}
	\end{definition}

\begin{example}
\label{suborder}
	Given a $\voa(U_{2,3},4)$ $\mb{T}$ in the following, it can be
        seen that $\{\mb{T}_0,\mb{T}_1,\ldots,\mb{T}_6\}$ forms a
        suborder $\voa$ decomposition of $\mb{T}$.

\begin{equation*}
\mrv{
\begin{array}{@{}l@{\qquad}l@{}}
\mrv{\mathbf T}=
\begin{matrix}
0&0&0\\
0&1&3\\
0&2&1\\
0&3&2\\
1&0&1\\
1&1&2\\
1&2&0\\
1&3&3\\
2&0&2\\
2&1&0\\
2&2&3\\
2&3&1\\
3&0&3\\
3&1&1\\
3&2&2\\
3&3&0
\end{matrix}
& \hspace{1cm}
\begin{array}{@{}l@{\qquad}l@{\qquad}l@{\qquad}l@{}}
\mathbf T_0=\begin{matrix}
0&0&0\\0&2&1\\1&0&1\\1&2&0
\end{matrix}
&
\mathbf T_1=\begin{matrix}
0&1&3\\0&3&2\\1&1&2\\1&3&3
\end{matrix}
&
\mathbf T_2=\begin{matrix}
2&0&2\\2&2&3\\3&0&3\\3&2&2
\end{matrix}
&
\mathbf T_3=\begin{matrix}2&1&0\end{matrix}
\\ \noalign{\vskip 30pt}
\mathbf T_4=\begin{matrix}2&3&1\end{matrix}
&
\mathbf T_5=\begin{matrix}3&1&1\end{matrix}
&
\mathbf T_6=\begin{matrix}3&3&0\end{matrix}
&
{}
\end{array}
\end{array}
}
\end{equation*}
where
\begin{itemize}
\item $A_0=\{0,1\}$, $B_0=\{0,2\}$,
\item $A_1=\{0,1\}$, $B_1=\{1,3\}$,
\item $A_2=\{2,3\}$, $B_1=\{0,2\}$,
\item $A_3=\{2\}$, $B_3=\{1\}$,
\item $A_4=\{2\}$, $B_4=\{3\}$,
\item $A_5=\{3\}$, $B_3=\{1\}$ and
\item $A_6=\{3\}$, $B_4=\{3\}$.
\end{itemize}
\end{example}

\begin{theorem}
	\label{rk27}
	For $F=(\hat{U}_{2,5}^{1},  U^{234}_{2,3})$, $\h=(a,b)\in F$ is entropic if and only if $a+b=\log v$ for some positive $v$ and there exists a suborder decomposition $\{\mb{T}_0,\mb{T}_1,\ldots,\mb{T}_{t-1}\}$ of a $\voa(U_{2,3},v)$ $\mb{T}$ such that 
	\begin{equation}
	a=\frac{1}{2} H(\dfrac{|A_i|^2}{k^2}:i\in \mathbb{I}_t), \nonumber
\end{equation}
where subarray $\mb{T}_i$ of $\mb{T}$ are induced by $A_i$ and $B_i$ for $i\in \mathbb{I}_t$.
\end{theorem}
\begin{figure}[H]
	\centering
	\begin{tikzpicture}[scale = 2]
		\draw [->,densely dashed] (0,0)--(1.6,0) node[below right] { $a$};
		\draw [->,densely dashed] (0,0)--(0,1.6) node[above left] {$b$};
		\foreach \y in{0.693,1.098,1.386}
		\draw [black, dashed,thin] (\y,0)--(0,\y);
		\fill (1.6,0.8) circle (0.4pt);  
		\fill (1.7,0.8) circle (0.4pt);  
		\fill (1.8,0.8) circle (0.4pt);  
		\fill (0.8,1.6) circle (0.4pt);  
		\fill (0.8,1.7) circle (0.4pt);  
		\fill (0.8,1.8) circle (0.4pt); 
		\node[below left] at (0,0) {O};   
		\foreach \x in{0,0.693,1.098,1.386}
		{\draw[fill] (\x,0) circle (.02);
		\draw[fill] (0,\x) circle (.02);}
		\draw[fill] (0,0.693) circle (.02);
		\draw[fill] (1.0397,0.34667) circle (.02);
		\draw[fill] (0.6931,0.6931) circle (.02);
		\draw[fill] (0.8664,0.5199) circle (.02);
		\draw[fill] (1.213,0.1733) circle (.02);
		\node[below,font=\fontsize{8}{6}\selectfont] at (0.693,0) {log2};  
		\node[below,font=\fontsize{8}{6}\selectfont] at (1.098,0) {log3};
		\node[below,font=\fontsize{8}{6}\selectfont] at (1.386,0) {log4};
		\node[left,font=\fontsize{8}{6}\selectfont] at (0,0.693) {log2};
		\node[left,font=\fontsize{8}{6}\selectfont] at (0,1.098) {log3};
		\node[left,font=\fontsize{8}{6}\selectfont] at (0,1.386) {log4};
		\node[right,font=\fontsize{8}{6}\selectfont] at (0.6931,0.6931) {(1,1)};
		\node[right,font=\fontsize{8}{6}\selectfont] at (0.8664,0.5199) {(1.25,0.75)};
		\node[right,font=\fontsize{8}{6}\selectfont] at (1.0397,0.34667) {(1.5,0.5)};
		\node[right,font=\fontsize{8}{6}\selectfont] at  (1.213,0.1733) {(1.75,0.25)};
		
	\end{tikzpicture}
	\caption{The face $(\hat{U}_{2,5}^{1},  U^{234}_{2,3} )$}
	\label{fig16}
\end{figure} 

\begin{proof}
	If $\mathbf{ h} \in  F$ is entropic, its characterizing random vector
	$(X_i,i\in N_4)$ satisfies the following information equalities,
	\begin{align}
		H(  X_{N_4})&=H(X_{N_4-i}),\ i\in N_4, \nonumber \\
		H(X_{ij })&=H(X_{i})+H(X_{j}),i,j\in \{2,3,4\}, \nonumber \\
		H(X_{  1i })+H(X_{1j })&=H(X_{ 1})+H(X_{ 1ij} ),i,j\in \{2,3,4\}, \nonumber\\
		H(X_{i\cup K})+H(X_{  j\cup K})&=H(X_{ K})+H(X_{  ij
			\cup K }),\nonumber\\&\qquad\qquad\qquad |K|=2,   K\subseteq \{2,3,4\}.   \nonumber
	\end{align}
	  For $ (x_i,i\in N_4) \in
	\mathcal{X}_{N_4}$ with $p(x_{1234})>0$, above information equalties
	imply that the probability mass function satisfies
	\begin{align}
		p(x_{1234})&=p(x_{1},x_{2},x_{3}) \label{c517}\\
		&=p(x_{1},x_{2},x_{4}) \label{c518} \\ 
		&=p(x_{1},x_{3},x_{4}) \label{c519}\\
		&=p(x_{2},x_{3},x_{4}), \label{c520}\\
		p(x_2,x_3)&=p(x_2)p(x_3), \label{c521}\\
		p(x_2,x_4)&=p(x_2)p(x_4), \label{c522}\\
		p(x_3,x_4)&=p(x_3)p(x_4), \label{c523}\\
		p(x_1,x_2)p(x_1,x_3)&=p(x_1)p(x_1,x_2,x_3), \label{c524}\\
		p(x_1,x_2)p(x_1,x_4)&=p(x_1)p(x_1,x_2,x_4), \label{c525}\\
		p(x_1,x_3)p(x_1,x_4)&=p(x_1)p(x_1,x_3,x_4), \label{c526}\\
		p(x_{1},x_{2},x_{3})p(x_{2},x_{3},x_{4})&=p(x_{2},x_{3})p(x_{1234}),\label{c527}\\
		p(x_{1},x_{2},x_{4})p(x_{2},x_{3},x_{4})&=p(x_{2},x_{4})p(x_{1234}),\label{c528} \\
		p(x_{1},x_{3},x_{4})p(x_{2},x_{3},x_{4})&=p(x_{3},x_{4})p(x_{1234}).
		\label{c529}
	\end{align}
		Restricting $\h$ on $\{2,3,4\}$, we obtain
                $\h'=(a+b)\br'$, where $\br'$ is the rank function of
                $U_{2,3}$ on $\{2,3,4\}$. Thus the characterizing
                random vector $(X_2,X_3,X_4)$ of $\h'$ is uniformly
                distributed on the rows of a $\voa(U_{2,3},v)$ for a
                positive integer $v$, and so $a+b$ can only take the
                value of $\log v$. By \eqref{c520},
                $p(x_1,x_2,x_3,x_4)=p(x_2,x_3,x_4)$, which implies
                that $(X_i,i\in N_4)$ must be uniformly distributed on
                the rows of a $v^2\times 4$ array $\mb{T}$ with
                $\mb{T}(2,3,4)$ a $\voa(U_{2,3},v)$. Assume
                $\mathcal{X}_{1}=\mathbb{I}_t$. Let $A_i$ and $B_i$
denote the set of all $j_1$ such that $(i,j_1)$ appears
        on the row of $\mb{T}(1,2)$, and the set of $j_2$ such that  $(i,j_2)$ appears on the row of $\mb{T}(1,3)$ for $i\in \mathbb{I}_t$, respectively. 
                For any $j_1\in A_i$ and  $j_2\in B_i$, we have 
		\begin{align}
			p_{X_1,X_2}(i,j_1)>0,p_{X_1,X_3}(i,j_2)>0.
		\end{align}
		Together with \eqref{c524}, we obtain
		\begin{align}
			p_{X_1,X_2,X_3}(i,j_1,j_2)>0.
		\end{align}
		Note that $(j_1,j_2)$ occurs exactly once on the rows of $\mb{T}(2,3)$ due to the definition of  $\voa(U_{2,3},v)$, 
		$(i,j_1,j_2)$ will appear  exactly once on the rows of $\mb{T}(1,2,3)$, and so  
		\begin{equation}
			p_{X_1,X_2,X_3}(i,j_1,j_2)=\frac{1}{v^2}. 
		\end{equation}
		The probability mass of $(i,j_1)$ for $i\in \mathbb{I}_t,j_1\in A_i $
		\begin{align}
			p_{X_1,X_2}(i,j_1)=\sum_{j_2\in B_i}p_{X_1,X_2,X_3}(i,j_1,j_2)=\frac{|B_i|}{v^2}. \label{c1111}
		\end{align}
		Similarly, the probability mass of $(i,j_2)$ for $i\in \mathbb{I}_t,j_2\in B_i $  
		\begin{align}
			p_{X_1,X_3}(i,j_2)=\sum_{j_1\in A_i}p_{X_1,X_2,X_3}(i,j_1,j_2)=\frac{|A_i|}{v^2}.\label{c1112}
		\end{align}
	 By \eqref{c517}-\eqref{c519}, $p(x_1,x_2,x_3)=p(x_1,x_2,x_4)=p(x_1,x_3,x_4)$. Equating the left side of \eqref{c524}-\eqref{c526}, we obatin
		\begin{align}
			p(x_1,x_2)p(x_1,x_3)=p(x_1,x_2)p(x_1,x_4)=p(x_1,x_3)p(x_1,x_4),
		\end{align}
		which implies 
		\begin{equation}
			p(x_1,x_2)=p(x_1,x_3)=p(x_1,x_4). \label{c1015}
		\end{equation}
		Together with \eqref{c1111} and \eqref{c1112}, we obtain
		\begin{align}
		|A_i|=|B_i|.
		\end{align}
		Assume $E_i$ denotes the set of $j_3$ satisfying $(i,j_3)$ appears on the row of $\mb{T}(1,4)$ for $i\in \mathbb{I}_t$. By symmetry, we conclude 
		\begin{equation}
			|A_i|=|B_i|=|E_i|.
		\end{equation}
		Note that $\mb{T}(2,3,4)$ is a $\voa(U_{2,3},v)$, the
                subarray $\mb{T}_i$ of $\mb{T}(2,3,4)$ induced by
                $A_i$ and $B_i$ is a $\voa(U_{2,3},|A_i|)$. Therefore,
                each $i\in N_t$ determines a suborder $\voa$
                $\mb{T}_i$ of $\mb{T}(2,3,4)$, which implies
                $\{\mb{T}_i,i\in \mathbb{I}_t\}$ is a suborder $\voa$ decomposition of $\mb{T}(2,3,4)$. The probalities mass of $i\in \mathbb{I}_t$
		\begin{align}
			p_{X_1}(i)=\sum_{j_1\in A_i}p_{X_1,X_2}(i,j_1)\\ =\frac{|A_i||B_i|}{v^2}=\frac{|A_i|^2}{v^2}.
		\end{align}
	Thus the entropy of $X_1$ is equal to 
	\begin{equation}
			H(X_1)= H(\dfrac{|A_i|^2}{v^2}:i\in \mathbb{I}_t) \label{c530}
	\end{equation}
	 As $\h\in F$ and $(X_i, i\in N_4)$ is its characterizing
         random vector, restricting $\h$ on $\{1\}$, we have 
	\begin{align}
		H(X_1)&=2a.  \label{c531}
	\end{align}
	which implies
	\begin{equation}
		a=\frac{1}{2} H(\dfrac{|A_i|^2}{v^2}:i\in \mathbb{I}_t) .
              \end{equation}
              
 For the ``if'' part, let $(X_2,X_3,X_4)$ be uniformly distributed
  on the rows of a $\voa(U_{2,3},v)$ $\mb{T}'$. Let
  $\{\mb{T}_0,\ldots,\mb{T}_{t-1}\}$ be a suborder $\voa$ decomposition of
  $\mb{T}'$. Let $X_1$ be $i$ if $(x_2,x_3,x_4)$ appear on the rows of
  $\mb{T}_i$. Then the entropy function $\h$ of $(X_i,i\in N_4)$ is in
  $F$. The proof has been completed.
\end{proof}
\noindent \textbf{Remark} 
  Theorem \ref{rk27} establishes a correspondence between the face $(\hat{U}_{2,5}^{1}, U^{234}_{2,3})$
  characterization and suborder decomposition problem.
	It is obvious that $\voa(U_{2,3},v)$ is inherently a suborder
        $\voa$ of itself. On the other hand, any $\voa(U_{2,3},v)$ can
        be  decomposed into $v^2$ suborder VOA $\voa(U_{2,3},1)$. These two cases correspond to the polymatroids $(0, \log v)$ and $(\log v, 0)$, respectively.  However, listing all the $\voa$ decompositions of a $\voa(U_{2,3},v)$ can be challenging.

%


\begin{definition}
  For a $v^2\times 4$ array $\mb{T}$, if
  \begin{itemize}
  \item $\mb{T}(2,3,4)$ is a $\voa(U_{2,3},v)$,
  \item entries in $\mb{T}(1)$ is from $\mathbb{I}_t$ with $v\le t\le v^2$, and 
  \item for each $i=2,3,4$, each row in $\mb{T}(1,i)$ occurs exactly
     once,
  \end{itemize}
we call $\mb{T}$ a  $\{1\}$-partial $\voa(U_{2,4})$.
\end{definition}


\begin{example}
	Let
	\begin{equation*}
		\mb{T}=
		\begin{matrix}
		0 &0 &0 &0 \\
		2&0 &1 &1 \\	
		1 &0 &2 &2 \\	
  	    3 &1 &0 &2 \\	
		1 &1 &1 &0 \\	
		0 &1 &2 &1 \\	
		1 &2 &0 &1 \\
		4&2 &1 &2 \\
		2 &2 &2 &0 
		\end{matrix}
              \end{equation*}
      It can be seen that $\mb{T}$ is a $\{1\}$-partial $\voa(U_{2,4})$, and the entries of $\mb{T}(1)$ is from $\mathbb{I}_5$.
\end{example}
\begin{theorem}
	\label{rk29}
	For $F=(\hat{U}_{2,5}^{1},  U_{2,4})$, $\h=(a,b)\in F$ is entropic if and only if $a+b=\log v$ and there exists
        a $\{1\}$-partial $\voa(U_{2,4},v)$
        such that 
	\begin{equation}
	a=H(\dfrac{\alpha_0}{v^2},\dfrac{\alpha_1}{v^2},\dots, \dfrac{\alpha_{t-1}}{v^2})-\log v,   \nonumber
\end{equation}
	where $\alpha_i$ denotes the times of the entry $i\in \mathbb{I}_t$ that occurs in $\mb{T}(1)$. 
\end{theorem}
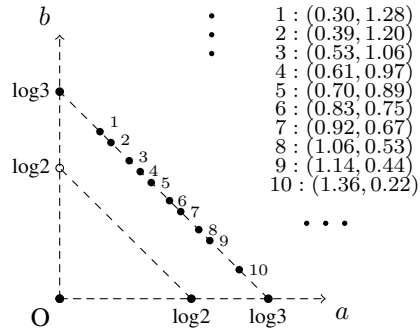
\begin{figure}[H]
	\centering
	\begin{tikzpicture}[scale = 2.5]
		\draw [->,densely dashed] (0,0)--(1.4,0) node[below right] { $a$};
		\draw [->,densely dashed] (0,0)--(0,1.4) node[above left] {$b$};
		\foreach \y in{0.693,1.098}
		\draw [black, dashed,thin] (\y,0)--(0,\y);
		\fill (1.3,0.4) circle (0.4pt);  
		\fill (1.4,0.4) circle (0.4pt);  
		\fill (1.5,0.4) circle (0.4pt);  
		\fill (0.8,1.3) circle (0.4pt);  
		\fill (0.8,1.4) circle (0.4pt);  
		\fill (0.8,1.5) circle (0.4pt); 
		\node[below left] at (0,0) {O};   
		\foreach \x in{0,1.098}
		{\draw[fill] (\x,0) circle (.02);
			\draw[fill] (0,\x) circle (.02);}
		\draw[fill] (0.693,0) circle (.02);
		\draw (0,0.693) circle (.02);
		\node[below,font=\fontsize{8}{6}\selectfont] at (0.693,0) {log2};  
		\node[below,font=\fontsize{8}{6}\selectfont] at (1.098,0) {log3};
		\node[left,font=\fontsize{8}{6}\selectfont] at (0,0.693) {log2};
		\node[left,font=\fontsize{8}{6}\selectfont] at (0,1.098) {log3};
		\fill (0,1.098) circle (0.4pt); 
		\fill (0.212,0.886) circle (0.02);
		\node[right,font=\fontsize{6}{6}\selectfont] at (0.212,0.94) {$1$};  
     	\fill (0.366,0.732) circle (0.02); 
     		\node[right,font=\fontsize{6}{6}\selectfont] at (0.366,0.75) {$3$}; 
        \fill (0.27,0.828) circle (0.02);
        \node[right,font=\fontsize{6}{6}\selectfont] at (0.27,0.84) {$2$}; 
    	\fill (0.424,0.674) circle (0.02); 
    	\node[right,font=\fontsize{6}{6}\selectfont] at (0.42,0.68) {$4$}; 
        \fill (0.578,0.52) circle (0.02);
        \node[right,font=\fontsize{6}{6}\selectfont] at (0.57,0.52) {$6$}; 
        \fill (0.732,0.366) circle (0.02); 
        \node[right,font=\fontsize{6}{6}\selectfont] at (0.732,0.366){$8$};
        \fill (0.482,0.616) circle (0.02);
        \node[right,font=\fontsize{6}{6}\selectfont] at (0.482,0.616) {$5$}; 
    	\fill (0.637,0.462) circle (0.02); 
    	\node[right,font=\fontsize{6}{6}\selectfont] at (0.637,0.462) {$7$};
        \fill (0.79,0.308) circle (0.02);
         \node[right,font=\fontsize{6}{6}\selectfont] at (0.79,0.308){$9$};
        \fill (0.945,0.154) circle (0.02); 
         \node[right,font=\fontsize{6}{6}\selectfont] at (0.945,0.154){$10$};
         \fill (1.099,0) circle (0.02);
            \node[font=\fontsize{8}{6}\selectfont] at (1.5,1.5){$1:(0.30,1.28)$};
           \node[font=\fontsize{8}{6}\selectfont] at (1.5,1.4){$2:(0.39,1.20)$};
           \node[font=\fontsize{8}{6}\selectfont] at (1.5,1.3){$3:(0.53,1.06)$};
           \node[font=\fontsize{8}{6}\selectfont] at (1.5,1.2){$4:(0.61,0.97)$};
           \node[font=\fontsize{8}{6}\selectfont] at (1.5,1.1){$5:(0.70,0.89)$};
           \node[font=\fontsize{8}{6}\selectfont] at (1.5,1){$6:(0.83,0.75)$};
           \node[font=\fontsize{8}{6}\selectfont] at (1.5,0.9){$7:(0.92,0.67)$};
           \node[font=\fontsize{8}{6}\selectfont] at (1.5,0.8){$8:(1.06,0.53)$};
           \node[font=\fontsize{8}{6}\selectfont] at (1.5,0.7){$9:(1.14,0.44)$};
           \node[font=\fontsize{8}{6}\selectfont] at (1.5,0.6){$10:(1.36,0.22)$};
	\end{tikzpicture}
	\caption{The face $(\hat{U}_{2,5}^{1},  U_{2,4} )$}
	\label{fig17}
\end{figure} 
\begin{proof}
	If $\mathbf{ h} \in  F$ is entropic, its characterizing random vector
	$(X_i,i\in N_4)$ satisfies the following information equalities,
	\begin{align}
		H(  X_{N_4})&=H(X_{N_4-i})  ,\ i\in N_4 \nonumber \\
		H(X_{ij })&=H(X_{i})+H(X_{j}),i,j\in\{2,3,4\} \nonumber \\
		H(X_{i\cup K})+H(X_{  j\cup K})&=H(X_{ K})+H(X_{  ij
			\cup K }), |K|=2\nonumber.
	\end{align}
	 For $ (x_i,i\in N_4) \in
	\mathcal{X}_{N_4}$ with $p(x_{1234})>0$, above information equalties
	imply that the probability mass function satisfies
	\begin{align}
		p(x_{1234})&=p(x_{1},x_{2},x_{3}) \label{c532}\\
		&=p(x_{1},x_{2},x_{4}) \label{c533} \\ 
		&=p(x_{1},x_{3},x_{4}) \label{c534}\\
		&=p(x_{2},x_{3},x_{4}) \label{c535}\\
		p(x_2,x_3)&=p(x_2)p(x_3), \label{c536}\\
		p(x_2,x_4)&=p(x_2)p(x_4), \label{c537}\\
		p(x_3,x_4)&=p(x_3)p(x_4), \label{c538}\\
		p(x_{1},x_{2},x_{3})p(x_{1},x_{2},x_{4})&=p(x_{1},x_{2})p(x_{1234}),\label{c539}\\
		p(x_{1},x_{2},x_{3})p(x_{1},x_{3},x_{4})&=p(x_{1},x_{3})p(x_{1234}),\label{c540} \\
		p(x_{1},x_{2},x_{4})p(x_{1},x_{3},x_{4})&=p(x_{1},x_{4})p(x_{1234}).\label{c541} \\
		p(x_{1},x_{2},x_{3})p(x_{2},x_{3},x_{4})&=p(x_{2},x_{3})p(x_{1234}),\label{c542}\\
		p(x_{1},x_{2},x_{4})p(x_{2},x_{3},x_{4})&=p(x_{2},x_{4})p(x_{1234}),\label{c543} \\
		p(x_{1},x_{3},x_{4})p(x_{2},x_{3},x_{4})&=p(x_{3},x_{4})p(x_{1234}).\label{c544} 
	\end{align}
By \eqref{c532}, canceling $p(x_1,x_2,x_3)$ and  $p(x_1,x_2,x_3,x_4)$ on either side of \eqref{c539}, we have
\begin{align}
	p(x_1,x_2,x_4)=p(x_1,x_2). 
\end{align}
Together with \eqref{c533}, we obtain
\begin{equation}
  p(x_1,x_2,x_3,x_4)=p(x_1,x_2). \label{c1016} 
\end{equation}
	By the same argument, we have
	\begin{align}
		p(x_1,x_2,x_3,x_4)&=p(x_1,x_3)=p(x_1,x_4) \label{c545} 
	\end{align}
Restricting $\h$ on $\{2,3,4\}$, we obtain $\h'=(a+b)\br'$, where
$\br'$ is the rank function of $U_{2,3}$ on $\{2,3,4\}$. Thus the
characterizing random vector $(X_2,X_3,X_4)$ of $\h'$ is uniformly
distributed on the rows of a $\voa(U_{2,3},v)$ for a positive integer
$k$, and so $a+b$ can only take the value of $\log v$.

By \eqref{c535}, $p(x_1,x_2,x_3,x_4)=p(x_2,x_3,x_4)$, which implies
that $(X_i,i\in N_4)$ must  be  uniformly distributed on the rows of a
$v^2\times 4$ array $\mb{T}$ with $\mb{T}(2,3,4)$ a
$\voa(U_{2,3},v)$. Note that $p(x_1,x_2,x_3,x_4)=p(x_1,x_2)$ by
\eqref{c1016}, each row of $\mb{T}(1,2)$ occurs exactly once in
$\mb{T}(1,2)$. Similarly, by \eqref{c545}, each row of  $\mb{T}(A)$
occurs exactly once in $\mb{T}(A)$ for $A=\{1,3\}$ and
$M=\{1,4\}$. Hence, $\mb{T}$ is a $\{1\}$-partial $\voa(U_{2,4})$.
Recall that $(X_i,i\in N_4)$ is uniformly distributed on the rows of $\mb{T}$, the probability of each row $(x_1,x_2,x_3,x_4)$ of $\mb{T}$ is
\begin{equation}
	p(x_1,x_2,x_3,x_4)=\frac{1}{v^2}.
      \end{equation}
      Then 
	\begin{align}
		H(X_1)=H(\dfrac{\alpha_0}{v^2},\dfrac{\alpha_1}{v^2},\dots, \dfrac{\alpha_{t-1}}{v^2}),  \label{c547}
	\end{align}
	where $\alpha_i$ denotes the times of $i\in \mathbb{I}_t$ that occurs in $\mb{T}(1)$.  As $\h\in F$ and $(X_i, i\in N_4)$ is its characterizing random vector, we have 
	\begin{align}
		H(X_1)&=2a+b.  \label{c548}  \\
		H(X_2)&=H(X_3)=H(X_4)=a+b   \label{c549}
	\end{align}
	Note that $a+b=\log v$, we obtain \rv{that}
	\begin{equation}
		a=H(\dfrac{\alpha_0}{v^2},\dfrac{\alpha_1}{v^2},\dots, \dfrac{\alpha_{t-1}}{v^2})-\log v.   \label{c550}
	\end{equation}

       To prove the ``if'' part of the theorem, let $\mb{T}$ be a $\{1\}$-partial $\voa(U_{2,4})$.
      Let $(X_i,i\in N_4)$ be uniformly distributed on the rows of $\mb{T}$. Then $(X_i,i\in N_4)$ characterizes $(a,b)$.  The proof is accomplished.
\end{proof}

\subsection{Entropy functions on the face $(\hat{U}_{3,5}^{4},  U_{2,4})$}
\label{u35}
In this subsection, we characterize entropy functions on the face 
$(\hat{U}_{3,5}^{4},  U_{2,4})$, which is a face with one extreme ray
containing a rank 3 integer polymatroid and another containing a rank
2 matroid.
\begin{theorem}
	\label{rk30}
	For $F=(\hat{U}_{3,5}^{4},  U_{2,4})$, $\h=(a,b)\in F$  is
	\begin{itemize}
		\item  entropic if 
		\begin{itemize}
			\item[$\ast$] $a+b=\log v$ for integer $v\neq2,6$;
			\item[$\ast$]  $(a,b)=(\log2,0)$; or
			\item[$\ast$]   $a+b=\log6,a\geq\log2$; and
		\end{itemize}
		\item non-entropic if
		\begin{itemize}
			\item[$\ast$] $a+b\neq \log v$ for some integer $v>0$;
			\item[$\ast$]  $a+b=\log2$, $a<\log2$; or
			\item[$\ast$]  $(a,b)=(0,\log6)$.
		\end{itemize} 
	\end{itemize}
\end{theorem}

\begin{figure}[H]
	\centering\mrv{
	\begin{tikzpicture}[scale = 2]
		\draw [->,densely dashed] (0,0)--(2.2,0) node[below right] { $a$};
		\draw [->,densely dashed] (0,0)--(0,2.2) node[above left] {$b$};
		\foreach \y in{1.098,1.386,1.609,1.945}
		\draw [black,thin] (\y,0)--(0,\y);
		\draw [black,thin] (1.791,0)--(0.693,1.098);
		\draw [thick,densely dotted] (0.693,1.098)--(0,1.791);
		\fill (1.6,0.8) circle (0.4pt);  
		\fill (1.7,0.8) circle (0.4pt);  
		\fill (1.8,0.8) circle (0.4pt);  
		\fill (0.8,1.6) circle (0.4pt);  
		\fill (0.8,1.7) circle (0.4pt);  
		\fill (0.8,1.8) circle (0.4pt);
		\fill (1.6,1.6) circle (0.4pt);  
		\fill (1.7,1.7) circle (0.4pt);  
		\fill (1.8,1.8) circle (0.4pt); 
			\draw[fill] (0,0) circle (.02);
		\draw[fill] (0.693,1.098) circle (.02);
    	\node[right,font=\fontsize{7}{6}\selectfont] at (0.693,1.098) {$(\log2,\log3)$};  
		\node[below left] at (0,0) {O};   
		\foreach \x in{1.098,1.386,1.945}
		\draw[fill] (0,\x) circle (.02);
		\draw(0,0.6993) circle (.02);

		\foreach \x in{0.693,1.098,1.386,1.609,1.791,1.945}
		\draw[fill] (\x,0) circle (.02);
		\draw (0,1.791) circle (.02);
		\node[below,font=\fontsize{7}{6}\selectfont] at (0.693,0) {log2};  
		\node[below,font=\fontsize{7}{6}\selectfont] at (1.098,0) {log3};
		\node[below,font=\fontsize{7}{6}\selectfont] at (1.386,0) {log4};
		\node[below,font=\fontsize{7}{6}\selectfont] at (1.62,0) {log5};
		\node[below,font=\fontsize{7}{6}\selectfont] at (1.85,0) {log6};
        \node[below,font=\fontsize{7}{6}\selectfont] at (2.1,0) {log7};
		\node[left,font=\fontsize{7}{6}\selectfont] at (0,0.693) {log2};
		\node[left,font=\fontsize{7}{6}\selectfont] at (0,1.098) {log3};
		\node[left,font=\fontsize{7}{6}\selectfont] at (0,1.386) {log4};
		\node[left,font=\fontsize{7}{6}\selectfont] at (0,1.62) {log5};
		\node[left,font=\fontsize{7}{6}\selectfont] at (0,1.85) {log6};
        \node[left,font=\fontsize{7}{6}\selectfont] at (0,2) {log7};
        
	\end{tikzpicture}}	
	\caption{The face $ (\hat{U}_{3,5}^{4},  U_{2,4}) $ }
	\label{fig13}
\end{figure}
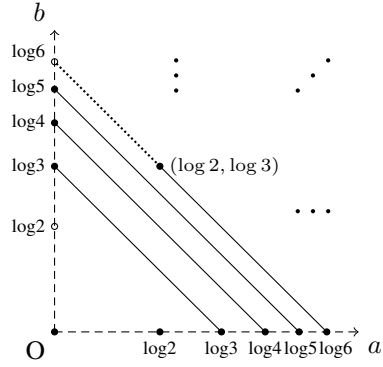
\begin{proof}
	
	If $\mathbf{ h} \in F$ is entropic, its characterizing random vector
	$(X_i,i\in N_4)$ satisfies the following information equalities,
	\begin{align}
		H(  X_{N_4})&=H(X_{N_4-i})  ,\ i\in N_4, \nonumber \\
		H(X_{ij })&=H(X_{i})+H(X_{j}),i,j\in N_4,\nonumber \\
		H(X_{i\cup K})+H(X_{  j\cup K})&=H(X_{ K})+H(X_{  ij
			\cup K }),\nonumber\\&\qquad\quad |K|=2,\{4\}\subseteq K \subseteq N_4.\nonumber
	\end{align}
 For $ (x_i,i\in N_4) \in
\mathcal{X}_{N_4}$ with $p(x_{1234})>0$, above information equalities
imply that the probability mass function satisfies
	\begin{align}
		p(x_{1234})&=p(x_{1},x_{2},x_{3}) \label{c551}\\
		&=p(x_{1},x_{2},x_{4}) \label{c552} \\ 
		&=p(x_{1},x_{3},x_{4}) \label{c553}\\
		&=p(x_{2},x_{3},x_{4}), \label{c554}\\
		p(x_1,x_2)&=p(x_1)p(x_2), \label{c555}\\
		p(x_1,x_3)&=p(x_1)p(x_3), \label{c556}\\
		p(x_1,x_4)&=p(x_1)p(x_4), \label{c557}\\
		p(x_2,x_3)&=p(x_2)p(x_3), \label{c558}\\
		p(x_2,x_4)&=p(x_2)p(x_4), \label{c559}\\
		p(x_3,x_4)&=p(x_3)p(x_4), \label{c560}\\
		p(x_{1},x_{2},x_{3})p(x_{1},x_{2},x_{4})&=p(x_{1},x_{4})p(x_{1234}),\label{c561}\\
		p(x_{1},x_{2},x_{3})p(x_{1},x_{3},x_{4})&=p(x_{2},x_{4})p(x_{1234}),\label{c562} \\
		p(x_{1},x_{2},x_{4})p(x_{1},x_{3},x_{4})&=p(x_{3},x_{4})p(x_{1234}).\label{c563} 
	\end{align}
	By \eqref{c551}, canceling $p(x_1,x_2,x_3)$ and $p(x_{1234})$  on either side of \eqref{c561}, we have
	\begin{equation}
		p(x_1,x_2,x_4)=p(x_1,x_4).
	\end{equation}
Together with \eqref{c552}, we obtain
\begin{equation}
	p(x_1,x_2,x_3,x_4)=p(x_1,x_4). \label{c1001} 
\end{equation}
By the same argument, 
	\begin{align}
		p(x_1,x_2,x_3,x_4)&=p(x_2,x_4)=p(x_3,x_4).     \label{c564} 
	\end{align}
In light of  \eqref{c557}, \eqref{c559} and \eqref{c560}, together with \eqref{c1001} and \eqref{c564},
\begin{equation}
	p(x_1)p(x_4)=p(x_2)p(x_4)=p(x_3)p(x_4),
\end{equation}
	which impies  $p(x_1)=p(x_2)=p(x_3)$. Since  $X_1$, $X_2$, and $X_3$ are pairwise independent by \eqref{c555}, \eqref{c556} and \eqref{c558}, by Lemma \ref{lem1}, we obtain that $X_i$ are uniformly distributed on   $\mathcal{X}_{i}$ for $i=1,2,3$, and so $H(X_1)=H(X_2)=H(X_3)=\log v$ where $v=|\mathcal{X}_{1}|=|\mathcal{X}_{2}|=|\mathcal{X}_{3}|$. As $\h\in F$, $(X_i, i\in N_4)$ is its characterizing random vector, we have 
	\begin{align}
		H(X_1)&=H(X_2)=H(X_3)=a+b,   \label{c566} \\
		H(X_4)&=2a+b,     \label{c569}\\
		H(X_2,X_4)&=3a+2b, \label{c1003} \\
		H(X_1,X_2,X_3)&=3a+2b\label{c1002}.
	\end{align}
	Thus  $a+b$ must take the value of $\log v$.
	
	Now we give all the construction of $(a,b)$ on the region $a+b=\log v$ for $v\neq 2,6$.
	Let $p_i>0$, $i\in \mathbb{I}_v$ such that $H(p_0,p_1,\dots,p_{v-1})=a$. Let $\mb{T}_0$ be a $\voa(U_{2,4},v)$. Let $\mb{T}_i$, $i= 1,\dots,v-1$ be a $v^2\times4$ array such that 
	 \begin{align}
	 \mb{T}_i(1,2)&=\mb{T}_0(1,2),\\
	  \mb{T}_i(3;j)&=\mb{T}_0(3;j)+i \mod v  \text{ for } j\in N_{k^2}.
	 	\\
	 \mb{T}_i(4;j)&=\mb{T}_0(4;j)+vi \text{ for } j\in N_{k^2}, \label{z1}
	 \end{align}
	It can be seen that each of such constructed $\mb{T}_i$ is a $\voa(U_{2,4},v)$. Let $(X_i,i\in N_4)$ be distributed on the rows of  $\mb{T}\triangleq\left[ 
	\begin{matrix}
		\mb{T}_0,\\
		\dots\\
		\mb{T}_{v-1}
	\end{matrix} \right]$
such that the probability mass of each row of $\mb{T}_i$ is $\frac{p_i}{v^2}$.
	Now we show the entropy function of $(X_i,i\in N_4)$ is $(a,b)$. By \eqref{z1}, $|\mathcal{X}_4|=v^2$ and each entry $j\in\mathcal{X}_4$ occurs only in one of the arrays $\mb{T}_0(4)$, $ \dots$, $\mb{T}_{v-1}(4)$, which implies that for any subset $A$ such that $\{4\}\subseteq A \subseteq N_4$,  
    \rv{each $x_A\in \mathcal{X}_A$ appears $\Delta_A$ times, where}
\[
\mrv{\Delta_A := v^{\mb{r}_{U_{2,4}}(N_4)-\mb{r}_{U_{2,4}}(A)},}
\]
\rv{and $\mb{r}_{U_{2,4}}$ is the rank function of $U_{2,4}$.}
So   
\begin{align}
H(X_A)
&=H\Big(
\underbrace{\frac{p_0\Delta_A}{v^2},\dots,\frac{p_0\Delta_A}{v^2}}_{ v^{\mb{r}_{U_{2,4}}(A)}},
\dots,\nonumber  \\&\qquad
\underbrace{\frac{p_{v-1}\Delta_A}{v^2},\dots,\frac{p_{v-1}\Delta_A}{v^2}}_{v^{\mb{r}_{U_{2,4}}(A)}}
\Big) \\
&=\mrv{H(p_0,p_1,\dots,p_{v-1})+ \log v\cdot\mb{r}_{U_{2,4}}(A)} \\
&=\mrv{a+ (a+b)\cdot\mb{r}_{U_{2,4}}(A)\\
&=a\cdot(1+\mb{r}_{U_{2,4}}(A))+b\cdot\mb{r}_{U_{2,4}}(A)} .
\end{align}

Then it is clear that $H(X_1)=H(X_2)=\log{v}$ and
$H(X_1,X_2)=2\log{v}$. Since each $j\in \mathcal{X}_3$ appears exactly
$v$ times in each $\mb{T}_i$, $i=0,\dots v-1$, we obtain
\begin{align}
	H(X_3)&=H(\frac{v(p_0+p_1+\dots,+p_{v-1})}{v^2}, \dots, \nonumber\\ &\frac{v(p_0.+p_1+\dots,+p_{v-1}) }{v^2})\\&=\log{v}.
\end{align}
Each $(i,j)\in \mathcal{X}_1\times \mathcal{X}_3$  or $\mathcal{X}_2\times \mathcal{X}_3$ appears exactly once in $\mb{T}_0$, $\dots$, and $\mb{T}_{v-1} $, which implies 
\begin{align}
	H(X_1,X_3)&=H(X_2,X_3)=H(\frac{p_0+p_1+\dots,+p_{v-1}}{v^2}, \dots,\nonumber \\ &\frac{p_0+p_1+\dots,+p_{v-1} }{v^2})\\&=2\log{v}.
\end{align}
It remains to verify $H(X_1,X_2,X_3)$. Each $(i,j,k)\in  \mathcal{X}_{123}$ appears once in the rows of $\mb{T}(1,2,3)$, and so
\begin{align}
	H(X_1,X_2,X_3)&=H(\underbrace{\frac{p_0}{k^2},\frac{p_0}{v^2},\dots, \frac{p_0}{v^2}}_{v^2},\dots,\nonumber \\ &\qquad\underbrace{\frac{p_{v-1}}{v^2},\frac{p_{v-1}}{v^2},\dots,\frac{p_{v-1}}{v^2}}_{v^2}) \\
	&=H(p_0,\dots,p_{v-1})+ 2\log{v}\\&=a+2(a+b)=3a+2b.
\end{align}
	
Now we show that all polymatroids with $a+b=\log 2$ are non-entropic
except for $(\log2,0)$. Assume $\h=(a,b)$ is entropic with
$a+b=\log2$ and $(X_i,i\in N_4)$ is its characterizing random
vector. By the discussion above, assume without loss of generality
that $X_1, X_2$ and $X_3$ are all uniformly distributed on $\mathbb{I}_2$. Let
$\mb{T}_0$ be the array consisting of rows being all $3$-tuples with
entries in $\mathbb{I}_2$, that is,
	\begin{equation*}
		\begin{matrix}
			0&0&0\\
			0&0&1\\
			0&1&0\\
			0&1&1\\
			1&0&0\\
			1&0&1\\
			1&1&0\\
			1&1&1
		\end{matrix}
	\end{equation*} 
It can be seen that $(X_1,X_2,X_3)$ must be distributed on the rows of
$\mb{T}_0$. Let the probatility of the $i$th-row of $\mb{T}_0$ be $p_i$.  
	Since  $X_1$, $X_2$, and $X_3$ are uniform and pairwise
        independent,  each of
       $(x_1,x_2)$, $(x_1,x_3)$ and $(x_2,x_3)$ has probability
       $\frac{1}{4}$, and so
	\begin{align}
		p_1+p_2=\dfrac{1}{4}, p_3+p_4=\dfrac{1}{4}, p_5+p_6=\dfrac{1}{4}, p_7+p_8=\dfrac{1}{4}, \nonumber \\
		p_1+p_5=\dfrac{1}{4}, p_2+p_6=\dfrac{1}{4}, p_3+p_7=\dfrac{1}{4}, p_4+p_8=\dfrac{1}{4}, \nonumber\\ 
		p_1+p_3=\dfrac{1}{4}, p_2+p_4=\dfrac{1}{4}, p_5+p_7=\dfrac{1}{4}, p_6+p_8=\dfrac{1}{4}. \nonumber
	\end{align} 
	Solving above equations, we obtain
	\begin{align}
		p_1=p_4=p_6=p_7,  \label{c567}    \\
		p_2=p_3=p_5=p_8.  \label{c568}  
	\end{align}
	Assume that either \eqref{c567} or \eqref{c568} vanishes, then $\mb{T}_0$ degenerates a $\voa(U_{2,3},2)$ $\mb{T}_1$ and  $(X_1,X_2,X_3)$ is uniformly distributed on the rows of $\mb{T}_1$. Together with \eqref{c566} and \eqref{c1002},
	\begin{equation}
		a=0,b=\log2,
	\end{equation}
	which contradicts the fact that $\h=(0,\log2)$ is
        non-entropic. Hence, both \eqref{c567} and \eqref{c568} must be positive.
	By \eqref{c551}, $X_4$ is a function of $(X_1,X_2,X_3)$, which
        implies that $(X_i,i\in N_4)$ must be distributed on a $\mb{T}$ such that $\mb{T}(1,2,3)=\mb{T}_0$. By \eqref{c1001} and \eqref{c564},
 for each $x_4\in\mathcal{X}_4, x_j\in\mathcal{X}_j$, $(x_j,x_4)$
 appear exactly once on the row of $\mb{T}(j,4)$ for $j=1,2,3$. Additionally, $X_4$ is independent of $X_i,i=1,2,3$ by \eqref{c557}, \eqref{c559}, and \eqref{c560}. There
 exists a unique $\mb{T}$ satisfying the above information equalities
 up to symmetry, and 
 \begin{equation*}
   \mb{T}\ =\
		\begin{matrix}
   0&0&0&0\\
   0&0&1&1\\
   0&1&0&2\\
   0&1&1&3\\
   1&0&0&3\\
   1&0&1&2\\
   1&1&0&1\\
   1&1&1&0
		\end{matrix}
	\end{equation*} 
	Calculating the entropy of $X_1$, $X_2$ and $(X_1,X_2)$, we obtain
	\begin{align}
		H(X_2)&=\log2, \\
		H(X_4)&=\log4,\\
		H(X_2,X_4)&=H(p_1,p_2,\dots,p_8).
	\end{align}
	Together with \eqref{c569}-\eqref{c1003}, we have
	\begin{align}
		&a=\log2,b=0,  \label{c571}\\
		&p_1=p_2=\dots=p_8=\dfrac{1}{8}.  \label{c572}
	\end{align}
	Thus the entropy function of $(X_i,i\in
        N_4)$ must be $(\log2,0)$, which implies that all polymatroids on the region $a+b=\log2$ are non-entropic expect for $(\log2,0)$.

	Now we show an inner bound on the entropy region on $F$ that
        $a+b=\log 6$ and $\log2\leq a\leq\log6$. Let $p_i>0$,
        $i=0,1,2$, and $p_0+p_1+p_2=1$ and $H(p_0,p_1,p_2)=a-\log2$. Let $\mb{T}^{(k)}_0,k=1,2$ be arrays  defined as follows.

\begin{equation*}
	\mb{T}^{(1)}_0\ = \
	\begin{matrix}
		0&0 & 0& 0\\
		0&1 & 1 &5 \\     
		0	&2 & 2 &3 \\
		0&3 & 3 &10 \\ 
		0	&4 & 4 &2 \\
		0	&5 & 5&1  \\
		1	&	0 & 1 &1 \\
		1	&	1 & 2 &0\\
		1	&	2 & 5 &4 \\
		1	&	3 & 4 &3 \\
		1	&	4 & 0 &5 \\      
		1	&	5 & 3 &2\\
		2	&		0 & 2 &2 \\ 
		2	&	1 & 3&11  \\
		2&		2 & 0&1  \\
		2	&		3 & 1&4  \\
		2	&		4 & 5&3\\
		2	&		5 & 4&0  \\
		3	&		0 & 3 &3 \\
		3	&	1 & 5 &2 \\     
		3	&	2 & 4&5  \\
		3	&	3 & 2&1 \\ 
		3	&	4 & 1 &0 \\
		3	&	5 & 0 &4 \\
		4	&	0 & 4 &4 \\
		4	&	1 & 0&3  \\
		4	&	2 & 1&2  \\
		4	&	3 & 5 &0 \\
		4	&	4 & 3&1 \\     
		4	&	5 & 2&5 \\
		5	&	0 & 5&5 \\ 
		5	&	1 & 4 &1\\
		5	&	2 & 3 &0 \\
		5	&	3 & 0 &2 \\
		5	&	4 & 2  &4\\
		5	&	5 & 1 &3 
	\end{matrix}
	\qquad\qquad\mb{T}^{(2)}_0\ =\ 
	\begin{matrix}
		0&0 & 2&6 \\
		0&1 & 3 &4 \\     
		0	&2 & 0&9  \\
		0&3 & 1&11  \\ 
		0	&4 & 5 &8 \\
		0	&5 & 4 &7 \\
		1	&	0 & 3&7  \\
		1	&	1 & 0&6 \\
		1	&	2 & 4  &10\\
		1	&	3 & 5 &9 \\
		1	&	4 & 2 &11 \\     
		1	&	5 & 1 &8\\
		2	&		0 & 0 &8 \\ 
		2	&	1 & 1 &10 \\
		2&		2 & 2 &7 \\
		2	&		3 & 3 &5 \\
		2	&		4 & 4&9\\
		2	&		5 & 5 &6 \\
		3	&	0 & 1 &9 \\
		3	&	1 & 4&8  \\     
		3	&	2 & 5&11  \\
		3	&	3 & 0 &7\\ 
		3	&	4 & 3 &6 \\
		3	&	5 & 2 &10 \\
		4	&	0 & 5 &10 \\
		4	&	1 & 2 &9 \\
		4	&	2 & 3 &8 \\
		4	&	3 & 4&6  \\
		4	&	4 & 1 &7\\     
		4	&	5 & 0&11 \\
		5	&	0 & 4&11 \\ 
		5	&	1 & 5&7 \\
		5	&	2 & 1 &6 \\
		5	&	3 & 2 &8 \\
		5	&	4 & 0 &10 \\
		5	&	5 & 3  &9
	\end{matrix}
\end{equation*}

   Let $\sigma_2$=
	$\left(
	\begin{array}{l}
		\text{0 1 2 3 4 5} \\
		\text{3 2 4 5 1 0} \\
	\end{array}
	\right)$ and $\sigma_3$=
	$\left(
	\begin{array}{l}
		\text{0 1 2 3 4 5} \\
		\text{1 0 5 4 3 2} \\
	\end{array}
	\right)$ be two permutations on $\mathbb{I}_6$. Let $\mb{T}^{(k)}_i$, $i=1,2,k=1,2$, be the $36\times 4$ arrays such that
	\begin{align}
		\mb{T}^{(k)}_i(1,2)&=\mb{T}^{(k)}_1(1,2), \\
		\mb{T}^{(k)}_i(3
		;j)&=\sigma_i(\mb{T}^{(k)}_1(3;j)) \text{ for } j\in N_{36}\\
				\mb{T}^{(k)}_i(4;j)&=\mb{T}^{(k)}_1(4;j)+12i \text{ for } j\in N_{36}, 
	\end{align}

 Let 
\begin{equation*}
	\mb{T}=\left[
	\begin{matrix}
		\mb{T}^{(1)}_0\\
		\mb{T}^{(2)}_0\\
		\mb{T}^{(1)}_1\\
		\mb{T}^{(2)}_1\\
		\mb{T}^{(1)}_2\\
		\mb{T}^{(2)}_2
	\end{matrix}
	\right]
\end{equation*}
Let $(X_i,i\in N_4)$ be distributed on the rows of $\mb{T}$ such that
the probability mass of each row of $\mb{T}^{(k)}_i$ is equal to $\frac{p_i}{72}$ for $i=0,1,2,k=1,2$.

Now we show the entropy function of $(X_i,i\in N_4)$ is $(a,b)$. Since entry $j+12i$, $j\in \mathbb{I}_{12}$ occurs $6$ times in $\left[ 
\begin{matrix}
	\mb{T}^{(1)}_i(4) \\ \mb{T}^{(2)}_i(4)
\end{matrix} \right]$
for $i=0,1,2$.

\begin{align}
	H(X_4)&=H(\underbrace{\frac{p_0}{12},\frac{p_0}{12},\dots,\frac{p_0}{12}}_{12},\dots,\\&\qquad\underbrace{\frac{p_2}{12},\frac{p_2}{12},\dots,\frac{p_2}{12}}_{12}) \\
	&=H(p_0,p_1,p_2)+ \log{12}\\&=a-\log2+ \log12\\&=2a+b.
\end{align}
Since each entry $j\in \mathbb{I}_6$ occurs $12$ times in  $\left[ 
\begin{matrix}
	\mb{T}^{(1)}_i(1) \\ \mb{T}^{(2)}_i(1)
\end{matrix} \right]$, 
$i=0,1,2$, which implies 
\begin{align}
	p_{X_1}(j)=\frac{12p_0}{72}+\frac{12p_1}{72}+\frac{12p_2}{72}=\frac{1}{6},
\end{align}
and so
\begin{align}
	H(X_1)=\log6=a+b.
\end{align} 
Similarly,
\begin{align}
	H(X_2)=H(X_3)=\log6=a+b.
\end{align}
Each row of $\mb{T}(1,4)$ appears exactly once in $\mb{T}(1,4)$, which implies $H(X_2,X_3|X_1,X_4)=0$. Hence,
\begin{align}
	H(X_1,X_4)&=H(X_1,X_2,X_3,X_4) \\
	&=H(\underbrace{\frac{p_0}{72},\frac{p_0}{72},\dots,\frac{p_0}{72}}_{72},\dots,\\&\qquad\underbrace{\frac{p_2}{72},\frac{p_2}{72},\dots,\frac{p_2}{72}}_{72}) \\
	&=H(p_0,p_1,p_2)+ \log{72}\\&=a-\log2+ \log72\\
    &=3a+2b.
\end{align}
By the same argument, we obtain
\begin{align}
&H(X_{1234})=H(X_1,X_2,X_3)=H(X_1,X_2,X_4)\\
&=H(X_1,X_3,X_4)=H(X_2,X_3,X_4)\\
	&=H(X_1,X_4)=H(X_2,X_4)=H(X_3,X_4)\\&=3a+2b.
\end{align}
Each row of $\mb{T}(1,2)$ appears exactly twice in $\left[ 
\begin{matrix}
	\mb{T}^{(1)}_i(1,2) \\ \mb{T}^{(2)}_i(1,2)
\end{matrix} \right]$,
$i=0,1,2$, which implies that 
\begin{align}
	p_{X_1X_3}(x_1,x_2)=\frac{2p_0}{72}+\frac{2p_1}{72}+\frac{2p_2}{72}=\frac{1}{36},
\end{align}
and so
\begin{align}
	H(X_1,X_2)=\log36=2a+2b.
\end{align}
By the same argument,
\begin{align}
		H(X_1,X_2)=H(X_1,X_3)=H(X_2,X_3)=2a+2b.
\end{align}
 Then $(X_i,i\in N_4)$ characterizes $(a,b)$. The proof is accomplished.
\end{proof}

\subsection{Discussion}

In this section, we characterized $10$ types of $2$-dimensional faces
of $\Gamma_4$. For a face $F$ with one of its two extremes rays containing a rank-1 matroid, we can see that entropy region $F^*$ has a similar shape to the entropy region on  $(U_{2,3},U^{12,3}_{1,2})$ characterized by Mat{\'u}{\v{s}} in \cite{matus2005piecewise} or $(U_{2,3},U^{1,3}_{1,1})$ by Chen and Yeung in \cite{chen2012characterizing}. The characterization of such faces are generalized to faces of general $n$ and both extreme rays containing matroids in \cite{hechen2026}.

As each $\mvoa(P, v)$ with $M$ a rank 2 or greater integer
polymatroid corresponds to a type of orthogonal Latin hypercubes, the
characterization of a face with both extreme rays containing an integer polymatroid exceeding 1,
in Subsection \ref{rank2}-\ref{u35}, breeds a new combinatorial
structure which can be considered as an intermediate form of the two
types of orthogonal Latin hypercubes. Specifically, in Subsection
\ref{latindecom}, three faces are characterized by the Latin square
decompositions, which can be considered as three new types of Latin
square substructures.

\bibliography{reference}
\bibliographystyle{IEEEtran}

%
%
%

\end{document}